\documentclass[english]{iopart}

\usepackage{graphicx}
\usepackage{url}
\usepackage{amssymb}

\newcounter{fig}
\begin{document}

\title[Ising model, modular forms and Calabi-Yau]
{\Large The Ising model: from elliptic curves to
modular forms and Calabi-Yau equations}
\vskip .3cm 
 {\bf 10th December 2010} 

\author{
A. Bostan$^\P$,
S. Boukraa$^\dag$, S. Hassani$^\S$,
 M. van Hoeij\footnote[8]{Supported by NSF 1017880}$^\ddag$, \\
J.-M. Maillard$^\pounds$,
 J-A. Weil$^{||}$  and N. Zenine$^\S$}
\address{$^\P$ \ INRIA Paris-Rocquencourt, 
Domaine de Voluceau, B.P. 105
78153 Le Chesnay Cedex, France} 
\address{\dag LPTHIRM and D\'epartement d'A{\'e}ronautique,
 Universit\'e de Blida, Algeria}
\address{\ddag Florida State University, Department of Mathematics,
1017 Academic Way, Tallahassee, FL 32306-4510 USA}
\address{\S  Centre de Recherche Nucl\'eaire d'Alger, 
2 Bd. Frantz Fanon, BP 399, 16000 Alger, Algeria}
\address{$^\pounds$ LPTMC, UMR 7600 CNRS, 
Universit\'e de Paris, Tour 23,
 5\`eme \'etage, case 121, 
 4 Place Jussieu, 75252 Paris Cedex 05, France} 
\address{$||$ \  XLIM, Universit\'e de Limoges, 
123 avenue Albert Thomas,
87060 Limoges Cedex, France} 
 
\ead{alin.bostan@inria.fr,  boukraa@mail.univ-blida.dz,  hoeij@mail.math.fsu.edu,
 maillard@lptmc.jussieu.fr, jacques-arthur.weil@unilim.fr, njzenine@yahoo.com}

\begin{abstract}
We show that almost all the linear differential operators 
factors obtained in the analysis of the 
$\, n$-particle contributions
$\, {\tilde \chi}^{(n)}$'s of the 
susceptibility of the Ising
 model for $\, n \le 6$, 
 are linear differential operators
 ``{\em associated with elliptic curves}''. Beyond the simplest
differential operators factors which are homomorphic to symmetric 
powers of the second order operator associated
 with the complete elliptic integral $\, E$, 
the second and third order
differential operators 
$\, Z_2$, $\, F_2$, $\, F_3$,
  $\, {\tilde L}_3$ can actually be interpreted as
{\em  modular forms} of the elliptic curve of the Ising model.
A last order-four globally nilpotent linear 
differential operator is not reducible to this 
elliptic curve, modular forms scheme. 
This operator is shown to actually correspond to
a natural generalization of this 
elliptic curve, modular forms scheme, with
the emergence of a Calabi-Yau
equation, corresponding to a selected $_4F_3$ 
hypergeometric function. This hypergeometric function 
 can also be seen as a Hadamard product  
of the complete elliptic integral $\, K$,
with a remarkably simple algebraic pull-back
 (square root extension), 
 the corresponding Calabi-Yau fourth-order differential 
 operator having 
 a symplectic differential Galois group
 $\, SP(4, \, \mathbb{C})$. 
The mirror maps and higher order Schwarzian ODEs,
associated with this Calabi-Yau ODE, present all 
 the nice physical and mathematical ingredients we had 
 with elliptic curves and modular forms, in particular 
 an exact (isogenies) representation of the generators of the 
renormalization group,
 extending the modular group $SL(2, \, \mathbb{Z})$ 
to a $GL(2, \, \mathbb{Z})$ symmetry group.

\end{abstract}

\vskip .5cm


\noindent {\bf AMS Classification scheme numbers}: 34M55, 
47E05, 81Qxx, 32G34, 34Lxx, 34Mxx, 14Kxx 


\section{Introduction}
\label{pre}
\vskip .1cm

In a previous paper~\cite{bo-gu-ha-je-ma-ni-ze-08} some
 massive computer calculations have been performed 
on the susceptibility of the square
 Ising model and on the $\, n$-particle 
($n$-fold integrals) contributions  $ \, \tilde{\chi}^{(n)}$ of the 
susceptibility~\cite{wu-mc-tr-ba-76,nickel-99,nickel-00}. In three
 more recent papers~\cite{High,bernie2010,Khi6} the linear differential
operators for $ \, \tilde{\chi}^{(5)}$ and $ \, \tilde{\chi}^{(6)}$ were
 carefully analysed. In particular, 
it was found that the minimal order linear differential operator 
for $ \, \tilde{\chi}^{(5)}$  can be reduced
to a  minimal order linear differential operator $L_{29}$ of order 29 
for the linear combination
\begin{eqnarray}
\label{thePhi5}
\Phi^{(5)} \,=\,\,\,\, \tilde{\chi}^{(5)}\,\, 
-{1 \over 2}\, \tilde{\chi}^{(3)} \,\,
 + {\frac{1}{120}}\, \tilde{\chi}^{(1)}.
\end{eqnarray}
This specific linear combination series being annihilated
by an ODE of lower order, one has, thus, 
the occurrence of a {\em direct sum} structure.
It was found~\cite{High,bernie2010} that this linear differential
operator $L_{29}$ can be factorised 
as a product of an order-five, an order-twelve,
 an order-one, and an order eleven linear differential operators 
 \begin{eqnarray}
\label{63}
L_{29} \,=\,\, \,\,\, L_5 \cdot L_{12} \cdot \tilde{L}_1 \cdot L_{11},
\end{eqnarray}
where  the order-one linear differential  
operator $\tilde{L}_1$ 
has a rational solution and
where the order-eleven linear differential operator 
 has a direct-sum decomposition
\begin{eqnarray}
\label{L11}
L_{11} \,=\,\,\,\,\,
 (Z_2 \cdot N_1) \oplus V_2 \oplus 
(F_3 \cdot F_2 \cdot L_1^s), 
\end{eqnarray}
where $\, L_1^s$ and $\, N_1$ are order-one (globally 
nilpotent\footnote[5]{That is of the form
 (\ref{globallyone}) for order-one operators (see below). 
$\, L_1^s$ has the simple rational solution $w^2/(1-4\, w)^2$.}) 
operators,  
$\, Z_2$ is the second order operator occurring
 in the factorization of the linear differential 
operator~\cite{ze-bo-ha-ma-04}
 associated with $ \, \tilde{\chi}^{(3)}$
and seen to correspond\footnote[1]{As well 
as the second order operator 
occurring in Ap\'ery's proof of the irrationality 
of $\, \zeta(3)$.} to a 
{\em modular form of weight
 one}~\cite{bo-bo-ha-ma-we-ze-09}, 
$\, V_2$ is a second order operator
 equivalent\footnote[3]{In the sense
of the equivalence of linear 
differential operators~\cite{Homomorphisms1,Homomorphisms}
(corresponding to the ``Homomorphisms'' command in Maple).
We refer to this (classical) notion of equivalence of linear 
differential operators everywhere in the paper. In the literature
the wording ``operators of the same type'' is also
 used~\cite{Singer}.} to
the second order operator  associated with 
 $ \, \tilde{\chi}^{(2)}$ (or
equivalently to the complete elliptic integral $E$),
 $\, F_2$ and $\, F_3$ are remarkable second order 
 and third order {\em globally nilpotent} linear
 differential operators~\cite{High,bo-bo-ha-ma-we-ze-09}. 

The order-five linear differential operator
 $L_5$ was shown to be equivalent  to the
 {\em symmetric fourth power} of (the second
 order operator~\footnote[2]{This second
 order operator does play a central role 
in the Gauss-Manin, or Picard-Fuchs, origin of the
 (sigma form of the) Painlev\'e equations occurring
 for the two-point correlation
functions of the Ising model~\cite{Painleve,Fuchs}.}) 
$L_E$ corresponding to the complete
 elliptic integral $\, E$.
These operators were actually obtained
 in  {\em exact arithmetics}~\cite{High,bernie2010}.
The  order-twelve linear differential operator $\, L_{12}$
has been shown to be {\em irreducible} and has been proved to
not be a symmetric product of differential operators of
smaller orders (see~\cite{bernie2010} for details).

Similar calculations were actually
 achieved~\cite{Khi6} on $\, \tilde{\chi}^{(6)}$.
For the linear combination
\begin{eqnarray}
\label{thePhi6}
\Phi^{(6)}\,\,  = \,\, \,\, \, 
\tilde{\chi}^{(6)}\,\,  -\frac23 \tilde{\chi}^{(4)}\,\, 
+\frac{2}{45}\tilde{\chi}^{(2)}, 
\end{eqnarray}
 one obtains a linear differential operator 
of (minimal) order $\, 46$
which has the following factorization 
\begin{eqnarray}
\label{L46}
L_{46} \,\,  =\,\,\,\, \, L_6  \cdot L_{23}  \cdot L_{17},
\end{eqnarray}
with
\begin{eqnarray}
\label{L17}
&&L_{17} \,\, =\, \,\,\,\,
\tilde{L}_5 \oplus  L_{3} \oplus 
\, (L_4 \cdot \tilde{L}_3 \cdot L_2), \\
&&\tilde{L}_5 \,= \, \,\,
\left( D_x\, -{1 \over x} \right) \oplus \,
\Bigr( L_{1,3} \cdot 
 \left( L_{1,2} \oplus L_{1,1} \oplus D_x \right)\Bigr).
 \nonumber 
\end{eqnarray}
where $\, D_x$ denotes\footnote[2]{In this paper
we will use the notations $\, D_t$ 
(resp. $\, D_z$) for  $\,d/dt$ (resp. $\, d/dz$).} $\,d/dx$, 
and where the $\, L_{1,n}$'s ($n\, = \, 1, 2, 3$)
 are  order-one linear differential operators
 (see Appendix A in~\cite{Khi6})
and $\,L_2$, $\,L_3$ and $\,L_6$ are~\cite{Khi6} respectively 
equivalent (homomorphic) to $L_E$, to the
 symmetric square of $L_E$ and to the
 {\em symmetric fifth power} of $L_E$.

The factorization (if any) of the order twenty-three
linear differential operator $L_{23}$ 
 is beyond our current computational
resources (see~\cite{Khi6} for details).

\vskip .1cm

$\bullet$ {\em Understanding the elementary factors: 
the modular form challenges}
\vskip .1cm
Among these various globally nilpotent
 factors~\cite{bo-bo-ha-ma-we-ze-09}, 
 of large order operators, one discovers order-one
linear differential operators which, because they are
globally nilpotent, are all of the form 
\begin{eqnarray}
\label{globallyone}
 D_x \,\,  - {{1} \over {N}} \cdot {{d \ln(R(x))} \over {dx}}, 
\end{eqnarray}
thus having $\, N$-th root of rational functions solutions.
One also discovers operators 
of various orders which are
 equivalent to symmetric powers of $\, L_E$,
the second order operator corresponding to complete elliptic integral 
of the first (or second kind), $E$ or $\, K$,
 like $\, V_2$ in (\ref{L11}), or
$\, L_5$ in (\ref{63}), or  $\, L_{3}$ in (\ref{L17}),  
or $\, L_6$ in (\ref{L46}), a remarkable 
second order operator $\, Z_2$
having a  {\em modular form
 interpretation}~\cite{bo-bo-ha-ma-we-ze-09},
and a miscellaneous set of  operators  $\, F_2$,  $\, F_3$, 
$\, \tilde{L}_3$, $\, L_4$,  $\, L_{12}$ and $\, L_{23}$. 

These last linear differential operators are {\em not}
 equivalent to symmetric powers of $\, L_E$,
and are still waiting for a {\em modular form}
 interpretation, if we think that all
the globally nilpotent factors of these operators 
of the ``Ising class''~\cite{Borwein} {\em should have 
an interpretation in terms of elliptic curves 
in a modern sense}\footnote[1]{
Before Wiles' recent results, only elliptic
 curves with the property known as
 "{\em complex multiplication}" 
had been shown to be parametrised by
 modular functions (by Shimura in 1961).}. These
linear differential operators, emerging in the analysis of
 $\, \tilde{\chi}^{(5)}$, namely
the second order operator $\, F_2$
and the third order operator $\, F_3$, 
 are waiting for such a modular form interpretation,
 as well as $\, \tilde{L}_3$ and  $\, L_4$
 emerging in the analysis of
 $\, \tilde{\chi}^{(6)}$. The two remaining operators 
$\, L_{12}$ and $\, L_{23}$ are too involved, and of too large order,
for seeking for a possible modular form
 interpretation (up to equivalence).
\vskip .1cm 
The purpose of this paper is to provide a mathematical interpretation
of the $\, F_2$,  $\, F_3$, 
$\, \tilde{L}_3$, $\, L_4$ elementary ``bricks'' 
of the $n$-fold integrals $\, \tilde{\chi}^{(5)}$ and
$\, \tilde{\chi}^{(6)}$, in order to {\em mathematically 
understand the Ising model}. In this paper we will,
as much as possible, use the same notations as in
our previous
 papers~\cite{bo-gu-ha-je-ma-ni-ze-08,High,Khi6,ze-bo-ha-ma-04,bo-bo-ha-ma-we-ze-09}.
The linear differential operators, or equations, are in 
the $\, x\, = \, w$ variable for the high temperature 
 differential operators and in the  $\, x\, = \, w^2$
 variable for the low temperature 
 differential operators, where  $\,w\, = \, \, (1+s^2)/s/2$
and where $\, s \, = \, \sinh(2\, K)$ with
 the standard notations for the Ising model.

\section{Modular form recalls}
\label{lodrecall}
\vskip .1cm
\subsection{Modular form recalls : $\, Z_2$ and Ap\'ery modular
linear differential operator}
\label{Z2}
\vskip .1cm

Let us introduce as in~\cite{bo-bo-ha-ma-we-ze-09} 
the order-two Heunian operator
which has the following solution $\, Heun(8/9, 2/3, 1, 1, 1, 1; t)$:
\begin{eqnarray}
\label{lets}
 {\cal H} \,\,  = \,\,  \, \, \,
 D_t^2 \, \,  \,
+\Bigl( {{1} \over {t}} \, \, 
  +{{1} \over {t-1}} \, + {{9} \over {9\, t\, -8}} \Bigr) \cdot D_t
 \, \,  \,
 + \, 3\,{\frac {3\,t-2}{ (9\,t-8)\, (t-1)\, t  }}.\nonumber 
\end{eqnarray}
A simple change of variable
 (see equation (43) in~\cite{bo-bo-ha-ma-we-ze-09}):
\begin{eqnarray}
\label{simplechange}
t \, \, = \,\,  \, \, \,{\frac {-8\, x}{ (1-4\,x) 
 \, (1-x) }},
\end{eqnarray}
 transforms $\,{\cal H}$  into the order-two 
linear differential operator\footnote[1]{Do note
 two minus sign misprints
in the numerator of the $\, D_x$ coefficient 
equation (44) of~\cite{bo-bo-ha-ma-we-ze-09}: 
$\,-10\,x +19\,{x}^{2}$ 
must be replaced by $\,+10\,x -19\,{x}^{2}$, see (\ref{Heunx}).}:
\begin{eqnarray}
\label{Heunx}
&&  {\cal H}_{x}\, = \, \, \, \, D_x^2 \\
&&\, \,\qquad   +\, {\frac {1 +10\,x -19\,{x}^{2}-92\,{x}^{3} 
 +12\,{x}^{4}\,+224\,{x}^{5}-64\,{x}^{6}  }
{ ( 1+3\,x+4\,{x}^2)  \, (1-2\,x)  \, (1+2\,x)\, 
  (1-4\,x)
  \, (1-x) \cdot x  }} \cdot D_x 
\nonumber \\
&&\qquad \, \, + \, 6\,{\frac { (1+7\,x+4\,{x}^{2}) 
 \, (1-2\,x)^2}{ (1+ 3\,x+4\,{x}^2) \,
\, (1-4\,x)^{2} \,(1-x)^2 \cdot x}}.
\nonumber
\end{eqnarray}

We found~\cite{bo-bo-ha-ma-we-ze-09} 
that the second order linear differential operator 
$\, Z_2$, occurring as a factor of the linear differential
operator annihilating $ \, \tilde{\chi}^{(3)}$, 
is homomorphic~\cite{Homomorphisms1,Homomorphisms}
 to the operator (\ref{Heunx}).
Recall~\cite{Maier1} that
 the {\em fundamental weight-1 modular form}\footnote[2]{
The modular form $\, h_6$ 
is also combinatorially significant: the perimeter generating 
function of the three-dimensional staircase polygons~\cite{Prell} 
can be expressed~\cite{bo-bo-ha-ma-we-ze-09}
 in terms~\cite{Maier1} of $\, h_6$. The modular form $\, h_6$ 
also occurs~\cite{bo-bo-ha-ma-we-ze-09} 
in Ap\'ery's study of $\, \zeta(3)$. }
  $\, h_N$ for the modular group $\, \Gamma_0(N)$
for $\, N\, = \, 6$,  can 
be expressed as a simple {\em Heun} function, 
$\, Heun(9/8, 3/4, 1, 1, 1, 1, - t/8)$, or 
as a hypergeometric function:
\begin{eqnarray}
\label{46}
&&{\frac { 2\,\sqrt {3}}{ (  (t+6)^{3}
 \, (t^3+18\,t^2+84\,t+24)^3 )^{1/12}}} \\
&& \qquad \quad \times \, \, 
 _2F_1 \Bigl([{{1} \over {12}}, {{5} \over {12}}];[1];  
\, 1728\,\, {\frac { (t+9)^2 \, (t+8)^3 \, t}{
 \,(t+6)^3 \, ({t}^{3}+18\,{t}^{2}+84\,t+24)^3}} \Bigr), 
\nonumber 
\end{eqnarray}
which is solution of the 
order-two linear differential operator:
\begin{eqnarray}
\label{h6}
D_t^2 \, \, \, \,  + \Bigl(  {{1} \over { t+8 }} \, 
+ {{1} \over {t}} \, +\, {{1} \over { t+9 }}  \Bigr)
 \cdot  D_{t} \, \, 
\,  +{\frac {t+6}{ (t+8) \, (t+9)\cdot  t}}.
\end{eqnarray}
simply related to $\,{\cal H}$.
Therefore, after some changes of variables, one
 can see the (selected)
solution of $\, Z_2$
as a hypergeometric function ({\em up to a Hauptmodul pull-back})
 corresponding to
a {\em weight-1 modular form}\footnote[3]{The simplest
 weight-1 modular form
is $\, _2F_1([1/12,5/12],[1], \, {\hat J}) \, = \, 
\, 12^{1/4} \, \eta(\tau)^2 \, {\hat J}^{-1/12}$
where $\, \hat{J}$ is the Hauptmodul, $\, \eta$ 
is the Dedekind eta function, 
and $\, \tau$ is the ratio of periods (see (4.6) 
in~\cite{Maier4}). } (namely $h_6$ in~\cite{Maier}). 

To sum-up $\, {\cal H}_{x}$, given by (\ref{Heunx}), has 
the following solution:
\begin{eqnarray}
\label{bingo}
&& {\cal S} \, = \,\, \,
 \Bigl( \Omega \cdot  {\cal M}_x \Bigr)^{1/12} 
\times  \, \, \,
 _2F_1 \Bigl([{{1} \over {12}}, {{5} \over {12}}];[1];
 \, \, {\cal M}_x\Bigr), 
\quad \quad \, \, \, \,\, \,\hbox{where:}  
\nonumber  \\
&&\Omega \, = \, \, \, \, 
{\frac {1}{1728}}\,{\frac { (1-4\,x)^6 
\, (1-x)^{6}} {x \cdot (1 \,+ 3\,x\, +4\,x^2)^2 \,
 (1+ 2\,x)^6}}, \\
&&  {\cal M}_x\, = \, \,  \,  \, \,
1728\,{\frac {x \cdot (1 \, +3\,x\, +4\, x^2)^2
 \, (1\, +2\,x)^{6}
 \, (1-4\,x) ^{6}
 \, (1-x) ^{6}}{ 
 (1\, +7\,x+4\,{x}^2) ^{3} \cdot  P^{3}  }},  \nonumber \\
&& P \, = \, \,  \, \,1\, +237\,x\, +1455\,{x}^{2}+ 4183\,{x}^{3}
+5820\,{x}^{4}+3792\,{x}^{5} +64\,{x}^{6}. 
\nonumber 
\end{eqnarray}
 
The solution of the operator $\, Z_2$,
 in terms of hypergeometric functions,
 can now be understood 
from this hypergeometric function ({\em up to a  modular invariant 
pull-back}) structure.

\subsection{Recall on modular forms: Hauptmoduls
 and fundamental modular curve}
\label{recallmodumar}
\vskip .1cm

In several papers~\cite{broglie,bo-ha-ma-ze-07b} we underlined, 
 for Yang-Baxter integrable models with a canonical {\em genus-one} 
parametrization~\cite{Automorphisms,Baxterization} (elliptic functions
 of modulus $\, k$), that the {\em exact}
generators of the {\em renormalization group}
 must necessarily identify
 with various isogenies~\cite{Barcau} which amounts to multiplying, or 
dividing, $\tau$ the ratio of the two periods of the elliptic curves, 
by an integer~\cite{Renorm}.  The simplest example is the {\em Landen
 transformation}~\cite{bo-ha-ma-ze-07b} 
which corresponds to multiplying ({\em or dividing} because of the 
modular group symmetry $\tau \,\longleftrightarrow \, 1/\tau$, 
i.e. the exchange 
 of the two periods of the elliptic curve),
 the ratio of the two periods:  
\begin{eqnarray} 
\label{Landen}
k \, \quad  \longleftrightarrow \, \quad k_L \, = \, \, 
{{2 \sqrt{k}} \over {1+k}},  \qquad \qquad 
  \tau \, \, \longleftrightarrow \, \, 2\, \tau.
\end{eqnarray} 
The other transformations\footnote[1]{See 
for instance (2.18) in~\cite{Canada}.} 
correspond to 
$\tau \,\leftrightarrow \, N \cdot \tau$,
for various integers $\, N$.  However, in 
the natural variables of the model
 (as $e^{K}, \, \tanh(K), \, k\, = \, s^2 \,$
$ = \, \, \sinh^2(2\, K)$,
 but not the ``transcendental
variables'' like $\tau$ or the nome $\, q$), 
these transformations
 are {\em algebraic} transformations
corresponding, in fact, to the 
{\em fundamental modular curves}. 
For instance, the Landen transformation 
 (\ref{Landen}) corresponds to the 
{\em genus zero fundamental modular curve}
\begin{eqnarray} 
\label{fundmodular1}
&&j^2 \cdot j'^2 \, \, -(j+j') \cdot
 (j^2+1487 \cdot j\, j' \, +j'^2) \nonumber \\
&& \quad  \quad  \quad \,  
 +3 \cdot 15^3 \cdot (16\, j^2\, -4027\, j\, j' \, +16\, j'^2) \\
 &&   \qquad \, \quad  \quad  -12\cdot 30^6\cdot (j+j') \, \, \, +8\cdot 30^9
\,\, \, = \, \, \, \, \, 0 , \nonumber
\end{eqnarray} 
which relates the two $\, j$-functions
\begin{eqnarray} 
j(k) \, = \, \, \, \, 256
\cdot {{(1-k^2+k^4)^3} \over {(1-k^2)^2 \cdot k^4}},
 \, \,  \qquad  j(k_L) \, = \, \, \, \, 
16 \cdot {\frac { (1+14\,{k}^{2}+{k}^{4})^3}{
 (1-{k}^{2})^{4} \cdot {k}^{2} }}.
 \nonumber 
\end{eqnarray}
or to the fundamental modular curve:
\begin{eqnarray} 
\label{fundmodular}
&&5^9\, v^3\, u^3\, -12 \cdot 5^6 \, u^2\, v^2 \cdot (u+v)\, 
+375\, \, u\, v \cdot  (16\, u^2\, +16\, v^2\, -4027\, v\, u)
 \nonumber \\
&&\quad \quad -64\, (v+u)\cdot (v^2+1487\, v\, u\, +u^2) \, \, 
+2^{12}\cdot  3^3 \cdot  u\, v  \, = \, \, 0,
\end{eqnarray} 
which relates the two Hauptmoduls $\, u\, = \, \, 12^{3}/j(k)$,
 $\,\,\, v\, = \, \, 12^{3}/j(k_L)$:

The fact that such transformations correspond to either
 multiplying, or dividing, $\, \tau$
 is associated with the {\em reversibility}\footnote[2]{The 
renormalization group (RG) is introduced,
 in the textbooks, as a semi-group,
the transformations of the RG being {\em non-invertible}. We do 
not try here to address the embarrassing question of
providing a mathematically well-defined definition of the RG. 
 It is known that there are many
 problems with the RG, for instance
 for the (two-parameter) non-zero-temperature
 Ising model 
{\em with a magnetic field}~\cite{GriffithsPearce,Griffiths2,Griffiths3}. 
We just remark here, for a Yang-Baxter integrable model
 like the Ising model without magnetic field, 
that the RG generators, seen as transformations
in the parameter space, identify with the isogenies
of elliptic curves and are thus reversible in the $\, \tau$-space.
}
of these exact representations 
of the {\em renormalization group}~\cite{Renorm}. 
The ``price to pay'' is that these algebraic transformations are 
not one-to-one transformations, they are sometimes 
called ``correspondences'' by some authors.

A simple rational parametrization\footnote[9]{Corresponding to
Atkin-Lehner polynomials and Weber's functions.} of the genus zero
 modular curve (\ref{fundmodular}) reads:
\begin{eqnarray}
\label{parak}
u \, = \, \, u(z) \, = \, \, \,{\frac {1728\,\, z}{ (z+16)^{3}}}, \qquad  
v\, = \, \, \,\,{\frac {1728\,\, z^2}{ (z+256)^3}}
\, = \, \, u\Bigl( {{2^{12}} \over {z}}\Bigr).
\end{eqnarray}
Note that the previously mentioned reversibility
is also associated with the fact that 
the modular curve (\ref{fundmodular})
is invariant by the permutation $\, u \, \leftrightarrow \, v$, 
which, within the previous rational parametrization (\ref{parak}), 
corresponds\footnote[3]{Conversely, and more precisely,
writing $\, 1728 \,z^2/(z+256)^3\,  = \,\, \,1728\, z'/(z'+16)^3$
 gives the
 Fricke-Atkin-Lehner~\cite{Atkin,Fricke} involution $\,z \cdot z'= \,2^{12}$, 
together with the quadratic relation
 $\, z^²  - \, z \, z'^²  -48 \, z \, z'  -4096 \, z'
\, = \,  0$. }
 to the {\em Atkin-Lehner involution}~\cite{Atkin}
 $\, z  \, \leftrightarrow \,2^{12}/z$.

 It has also been
 underlined in~\cite{broglie,bo-ha-ma-ze-07b}
that seeing (\ref{Landen}) as a transformation on {\em complex variables}
(instead of real variables) 
provides, beyond $\, k \, = \, \, 0, \, \, 1$
(the infinite temperature fixed point
 and the critical temperature fixed point),
 two other {\em complex} fixed points which actually correspond to
{\em complex multiplication} for the elliptic curve, and are, actually,  
fundamental new singularities\footnote[5]{Suggesting an 
understanding~\cite{bo-ha-ma-ze-07b,bo-ha-ma-ze-07}
 of the quite rich structure~\cite{bo-ha-ma-ze-07b} 
of the infinite number of singularities of the $\, \chi^{(n)}$'s 
in the complex plane, from a Hauptmodul 
approach~\cite{bo-ha-ma-ze-07b,bo-ha-ma-ze-07}.
Furthermore, the notion of {\em Heegner numbers}
 is closely linked to the {\em isogenies}
mentioned here~\cite{bo-ha-ma-ze-07b}. An exact 
value of the $j$-function
$\, j(\tau)$ corresponding to one of the first Heegner number is,
for instance, $\, j\, = \, \,12^3$.}
discovered on the  $\, \chi^{(3)}$ linear 
ODE~\cite{ze-bo-ha-ma-04,ze-bo-ha-ma-05,ze-bo-ha-ma-05c}.
In general,  within the theory of elliptic curves,  
this underlines the interpretation of
the (generators of) the renormalization group
 as {\em isogenies of elliptic curves}~\cite{Renorm}. 
{\em  Hauptmoduls}\footnote[1]{It 
should be recalled that the {\em mirror
symmetry} found with {\em Calabi-Yau 
manifolds}~\cite{Candelas,Doran,Doran2,Kratten,mirror,LianYau} 
 {\em can be seen as higher order
generalizations of Hauptmoduls}. We can,
 thus, expect generalizations
of this identification of the renormalization and modular structure
when one is not restricted to elliptic curves anymore.},
 {\em modular curves and modular forms} play here a fundamental
role.

Along this modular form line,
let us consider the second order linear differential operator
\begin{eqnarray}
\alpha \, = \,\,\, \, \,\, \, D_z^{2} \,\, \, 
+{\frac { \left( {z}^{2}+56\,z+1024 \right)}
{ z \cdot (z+16)\,  (z+64)  }} 
\cdot D_z\,\, \,
- \,{\frac {240}{ z \cdot (z+16)^2  \,(z+64) }},
\nonumber 
\end{eqnarray}
which has the (modular form) solution:
\begin{eqnarray}
\label{cov}
&&_2F_ 1 \Bigl([{{1} \over {12}},\,  {{5} \over {12}}],\, [1];  
 1728\,{\frac {z}{ (z+16)^3}}\Bigr)
 \,\,\,   \\
&& \quad \quad \, = \, \, \,\, 
2 \cdot \Bigl({\frac {z+256}{z+16}}\Bigr)^{-1/4}
\cdot \, \, \, 
_2F_ 1 \Bigl([{{1} \over {12}},\,  {{5} \over {12}}],\, [1];
  1728\,\,{\frac {{z}^2}{ (z+256)^3}}\Bigr).
 \nonumber 
\end{eqnarray}
In fact the {\em two pull-backs} in the arguments of the 
{\em same} hypergeometric function are {\em actually related
by the fundamental modular curve} 
 (\ref{fundmodular}) (see (\ref{parak})).
Do note that, generically, the
 existence of {\em several pull-backs}
for a hypergeometric function is a {\em quite rare situation}. 
The covariance  (\ref{cov}) is thus the very expression of a
modular form property with respect to
transformation $\tau \, \leftrightarrow \, 2 \, \tau$,
corresponding to the modular curve (\ref{fundmodular}).

This example (see (\ref{cov})) is a simple illustration
 of the special role 
played by selected hypergeometric functions 
{\em having several possible pull-backs} 
(in fact an infinite number). This phenomenon is 
 {\em linked to elliptic curves in a deep and fundamental sense},
 namely the occurrence of {\em modular forms}
 and of isogenies represented by 
algebraic transformations, called by some authors
 ``correspondence'',  between the pull-backs.
These algebraic transformations 
correspond to {\em modular curves} (and to {\em exact algebraic
 representations of the generators of the 
renormalization group}~\cite{Renorm}).
We will say in short, that such hypergeometric functions 
are ``associated with elliptic curves''. 
Simpler examples of {\em isogenies} associated
 with rational transformations, instead of 
``correspondence'' like (\ref{fundmodular}),
 are displayed in~\cite{Renorm}.

\section{Modular form solution of $\, F_2$
 and the corresponding fundamental 
modular curve $\, X_0(2)$}
\label{soluF2}
\vskip .1cm

The exact expressions  of the selected linear differential 
operators $\, Z_2$, $\, F_2$, and $\, F_3$ which emerged~\cite{High} in 
$\, \tilde{\chi}^{(5)}$, can be found in~\cite{jensen}, 
 and the exact expressions  of the selected linear differential 
operators $\, \tilde{L}_3$, 
and $\, L_4$ which emerged~\cite{Khi6} in $\, \tilde{\chi}^{(6)}$,
can be found in~\cite{jensen2}.

In the following we will not detail
 how we have been able to get the
solutions of the linear differential 
operators $\, F_2$, and $\, F_3$, and the
solutions of $\, \tilde{L}_3$, and $\, L_4$.
These details will be displayed
 in forthcoming publications.
These (slightly involved) solutions\footnote[3]{The reader can 
view all these linear differential equations,
as well as their explicit $_2F_1$-type 
solutions, in~\cite{URL}.} are displayed in~\cite{URL}. 
We focus, here, on the structures, and mathematical meaning, 
associated with these solutions.

Actually, the series expansion of the solutions of (the globally
 nilpotent) operator $\, F_2$
gives more than $G$-series~\cite{Andre5,Andre6}: it yields series
 with {\em integer} coefficients, suggesting that 
 $\, F_2$ could also have, like the previous $Z_2$,
 a modular form interpretation.

The solutions of the second order linear differential
operator~\cite{High} $\, F_2$ can, actually, be written in terms of 
hypergeometric functions (the $\rho_i(x)$'s 
are two rational functions) 
\begin{eqnarray}
\rho_1(x)^{1/4}\cdot \, \, 
 _2F_1\Bigl([{{1} \over {4}},\,  {{1} \over {4}}],
[{{1} \over {2}}]; \, \, p(x)\Bigr)
\,\,  + \, \,\rho_2(x)^{1/4}\cdot  \, \, 
 _2F_1\Bigl([{{5} \over {4}},\,  {{5} \over {4}}],
[{{3} \over {2}}]; \, \, p(x)\Bigr),
\nonumber 
\end{eqnarray}
with the pull-back:
\begin{eqnarray}
\label{pul}
p(x) \, = \, \,\,  
-\, {\frac {1}{64}}\,{\frac { (1\, - 4\,x) 
 \left( 1+6\,x+13\,{x}^{2}+4\,{x}^{3} \right)^{2}}
{ \left( 1+2\,x \right)^{3} \cdot x^3 }}. 
\end{eqnarray}
A well-known symmetry of many hypergeometric functions 
amounts to changing the pull-back $\, p(x)$
into\footnote[2]{In addition to
 $\, p(x) \,  \rightarrow \, \,   q(x)  \, = \,\,  \, 1/p(x)$.}
 $\, q(x)  \, = \,\,  \, 1\, -p(x)$:
\begin{eqnarray}
q(x)  \, = \,\,  \, 1\, -p(x) \, = \,\,  \,\,
 {\frac {1}{64}}\,{\frac { (1+4\,x)^2
 \, (1- x)^3 \, (1+3\,x+4\,x^2)  }
{ (1+2\,x)^3\cdot x^3 }
},
\end{eqnarray}
where the selected\footnote[1]{Complex fixed points
of the  Landen transformation, Heegner numbers with
complex multiplication of the elliptic 
curve~\cite{bo-ha-ma-ze-07b}.} singularities
$\,1+3\,x+4\,x^2 \, = \, \, 0$,
 seen~\cite{ze-bo-ha-ma-04} in $\tilde{\chi}^{(3)}$,
and more specifically~\cite{bo-bo-ha-ma-we-ze-09}
 in $Z_2$, clearly occur.  

Alternatively, the solutions of $\, F_2$ can also be written as
 (the $\, R_i$'s are 
rational functions)
\begin{eqnarray}
&&R_1(x)^{1/12}\cdot \, \, 
 _2F_1\Bigl([{{1} \over {12}},\,  {{1} \over {12}}],\,
 [{{2} \over {3}}]; \, \, p_i(x)\Bigr)
\nonumber \\
&&\qquad \qquad  \quad \quad
 \, + \, \,R_2(x)^{1/12}\cdot  \, \, 
 _2F_1\Bigl([{{13} \over {12}},\,  {{13} \over {12}}], \, 
[{{5} \over {3}}]; \, \, p_i(x)\Bigr),
\nonumber 
\end{eqnarray}
with {\em two different possible pull-backs}
 $\, p_1(x)$ and  $\, p_2(x)$
reading respectively:
\begin{eqnarray}
&&\quad \,{\frac { (1\, +8\,x \,  +14\,{x}^{2} \,
 -36\,{x}^{3} -151\,{x}^{4}
 -188\,{x}^{5} -16\,{x}^{6} -64\,{x}^{7})^3}{1728 \cdot (1+2\,x)^6 
 \cdot (1 \, + 4\,x)^{2} \cdot (1\, -x)^{3}
 \cdot (1+3\,x+4\,{x}^{2}) \cdot x^6 }},
 \nonumber \\
&& - \,{\frac { (1 \, +8\,x\, +14\,x^2 \,
- 276\,x^3 -1591\,x^4 \, -3068\,x^5\, -1936\,x^6\, -64\,x^7)^3}
{1728 \cdot  (1 +3\,x +4\,x^2)^2 \cdot (1- x)^6
 \cdot (1\, + 4\,x)^{4} \cdot (1\, +2\, x)^3 \cdot x^3}}.
 \nonumber 
\end{eqnarray}
Note that these two pull-backs 
{\em are not related by a simple Atkin-Lehner involution}~\cite{Atkin}
($p_2(x) \, \ne \, \,p_1(N/x)$,  with $\, N$ some integer).
However,  introducing the rational expression
\begin{eqnarray}
\label{ru}
R(U)\,\, = \,\,\,
- {{1} \over {27}} \,{\frac { (U-4)^{3}}{{U}^{2}}},
\end{eqnarray}
one actually finds that 
\begin{eqnarray}
\label{RU}
 p_1(x) \, = \, \, R\Bigl( {{1} \over {q(x)}} \Bigr),
 \qquad  \quad   p_2(x) \, = \, \, R\Bigl(q(x)  \Bigr).
\end{eqnarray}

 The relation between these two pull-backs
corresponds to the (genus zero) curve
\begin{eqnarray}
\label{alphabeta}
&& 110592 \cdot \alpha^2\, \beta^2  \, \, \,
  -64\cdot (\alpha^3+\beta^3) \, \,\,  
-95232 \cdot (\alpha^2\, \beta\, +\alpha\, \beta^2)
 \nonumber  \\
&&\qquad  +6000\cdot (\alpha^2+\beta^2)\, \,\, 
  -1510125 \cdot \alpha\, \beta \,
 \nonumber  \\
&&\qquad 
 -187500 \cdot (\alpha\, +\beta) \,\,   +1953125
 \,\,\,  \,  =\, \,  \, \, \, \, 0,
\end{eqnarray}
with the simple rational parametrization deduced from (\ref{RU}):
$\, (\alpha(z), \, \, \beta(z)) \,\, \,  = \,\, \, \,  \, \, 
 (R(U),\,  \, \, R(1/U))$.
One immediately finds that (\ref{alphabeta})
is nothing but (\ref{fundmodular}) where $\,( u, \, v)$  have been changed
into $(1/u, \, 1/v)$.
Actually, up to a 
rescaling\footnote[5]{$R(U)\, = \,\, j_2 (-64\, U)/1728 \,$ 
with $ \, \, j_2(t)\, = \,\, (t\, +256)^3/1728/t^2$. }
 of $U$ by a factor $-64$, the parametrization
of the rational curve (\ref{alphabeta}) can be rewritten
in a form where a $ \,\, z \, \leftrightarrow \,2^{12}/z\,\, $
 Atkin-Lehner involution is made explicit:
\begin{eqnarray}
\label{alphazbetaz}
&&(\alpha(z), \, \beta(z)) \,\, \,  = \,\, \, \, \,  
\Bigl({\frac {1}{1728}}\,{\frac { (z+16)^{3}}{z}}, \, \, \,
{\frac {1}{1728}}\,{\frac { (z+256)^{3}}{{z}^{2}}}\Bigr), \\ 
&&\beta(z)\,\,
\,\, = \, \,\, \, 
\alpha\Bigl( {{2^{12}} \over {z}}\Bigr).  
\end{eqnarray}
One recognizes (up to a 1728 normalization factor) 
the {\em fundamental modular curve}
 $\, X_0(2)$
\begin{eqnarray}
\label{X02}
&&A^{2}B^{2}\, \, \, 
- \left( A\, +B \right)  \cdot (A^{2} +1487\,A\, B +B^{2})
 \\
&& \qquad \quad -40773375 \cdot A \, B\,\,\, 
  +162000 \cdot (A^2 \, + B^2)
\nonumber \\
&& \qquad \quad  -8748000000 \cdot (A\, +B)
\,  \, \, +157464000000000 
\,  \,\,  \,\,  = \,\,  \, \,\,  \, 0, 
 \nonumber 
\end{eqnarray}
and its 
well-known rational parametrization:
\begin{eqnarray}
\label{paraX02}
A \, = \, \,\, A(j_2) \, = \, \,
{\frac { \left( 256+j_2 \right)^3}{{ j_2}^2}}, \qquad \quad 
B \, = \, \,\, A\Bigl({{2^{12} } \over {j_2}} \Bigr). 
\end{eqnarray}

\subsection{Dedekind  eta function parametrization}
\label{Dedekind}

It is well known that the {\em Dedekind eta
 function}~\cite{Ramanujan,Elkies}, in power 24, is a cusp automorphic
 form of weight 12, related to the
 {\em discriminant of the elliptic curve}.
Recalling the Weierstrass' modular 
discriminant~\cite{Weierstrass}, defined as
\begin{eqnarray}
\label{discriminant} 
    \Delta(\tau) \,  \,=\,\, \,  \,  (2\pi)^{12} \cdot \eta(\tau)^{24},  
\end{eqnarray}
which is this modular form of weight 12, we get rid of this $\,(2\pi)^{12}$
factor and define
\begin{eqnarray}
\label{Dedekindeta}
 \Delta(q) \, \, =\,\, \, \,\,   \, 
  q \cdot \prod_{n=1}^{\infty} \, (1\, -q^n)^{24}, 
\end{eqnarray}
where $\, q$ is the nome of the elliptic curve, 
that is, the exponential 
of the ratio of the two periods\footnote[3]{In
the literature the ratio of the two periods
 of the elliptic curve often 
encapsulates a $\,2\, i \, \pi$ (or $\,\, i \, \pi$)
 factor and one defines: 
$\,q \,  = \, \,  \exp(2\, i \, \pi \, \tau)$. For some
 Calabi-Yau reason
(see (\ref{defnome}) below) we prefer to write 
$\,q \,  = \, \,  \exp(\tau)$.} of the elliptic curve:
$\,q \,  = \, \,  \exp(\tau)$. 

One can now introduce a ``second layer'' of parametrization 
writing the $\, j$-function as a ratio
 of Dedekind eta function
\begin{eqnarray}
\label{paraj2}
j_2 \, = \, \,\, j_2(q) \, = \, \,\, 
{{\Delta(q)} \over {\Delta(q^2)}}, \qquad \quad 
 A(j_2) \, = \, \, {\frac { \left( 256+j_2  \right)^3}{{ j_2}^2}}. 
\end{eqnarray}
One deduces the alternative parametrization of (\ref{X02})
\begin{eqnarray}
\label{Ded2}
(A, \, B) \, = \, \, (A(j_2(q)), \,A(j_2(q^2)) \, = \, \,
\Bigl(A \Bigl( {{\Delta(q)} \over {\Delta(q^2)}} \Bigr), 
\,A\Bigl({{\Delta(q^2)}\over {\Delta(q^4)}} \Bigr) \Bigr),
\end{eqnarray}
making crystal clear that the fundamental modular curve (\ref{X02})
 is a representation of $\, \tau \, \rightarrow \, 2 \cdot \tau$, or 
$\, q \, \rightarrow \, q^2$
(and in the same
 time\footnote[1]{Thanks to the 
$\tau \, \leftrightarrow \, 1/\tau$ symmetry
 of the modular group corresponding 
to the exchange the two periods of the
 elliptic curve.} $\, \tau \, \rightarrow \, \tau/2$).

The Atkin-Lehner involutive
transformation $j_2 \, \rightarrow \, 2^{12}/j_2$ 
and  transformation $q \, \rightarrow \, q^2$
are {\em actually compatible} thanks to 
the remarkable ``Ramanujan-like'' 
functional identity on Dedekind $\eta$ functions
\begin{eqnarray}
\label{Ramanujan}
&&4096 \cdot \Delta(q) \cdot  \Delta(q^4)^2 
\,\, \, -\Delta(q^2)^3  \,\, 
 \\
&& \qquad  \quad  + (\Delta(q)\, 
+48 \cdot \Delta(q^2)) \cdot \Delta(q)  \cdot  \Delta(q^4)
 \,\,\,\, \,  = \,\,\,\,  \, \, 0, 
 \nonumber
\end{eqnarray}
yielding: 
\begin{eqnarray}
\label{compatible}
A \Bigl( {{\Delta(q^2)} \over {\Delta(q^4)}}  \Bigr) 
\, \, \, \, = \,\,  \,\, \,\, \,
 A \Bigl(  2^{12}  /\Bigl(  {{\Delta(q)} 
\over {\Delta(q^2)}}  \Bigr) \Bigr), 
\end{eqnarray}
making (\ref{paraX02}) and (\ref{Ded2}) compatible. 

There are many other nice functional and differential 
relations on the Dedekind eta functions 
that are shown below.

$\, \bullet$ The  modular functions
 (see (19) in~\cite{ModularForm} page 16)
\begin{eqnarray}
\label{beforecover}
t \, = \, \, \Bigl({{ \eta(6\tau) \, \eta(\tau)} 
\over { \eta(2\tau) \, \eta(3\tau)}}\Bigr)^{12},
 \qquad \qquad 
g \, = \, \,
 {{ \eta(6\tau)^8 \, \eta(\tau)^4}
 \over { \eta(2\tau)^8 \, \eta(3\tau)^4}},
\end{eqnarray}
have the following relation:
\begin{eqnarray}
\label{cover}
t \,\, \,  = \, \, \, \,\, \, 
 g \cdot {{ 1\, -9\, g  } \over {  1-g }}. 
\end{eqnarray}
This is {\em exactly} the covering necessary 
to see  $Z_2$ as a modular
 form (see equation (A.3)
in~\cite{bo-bo-ha-ma-we-ze-09}). 

$\, \bullet$ Differential equations are actually
 satisfied by modular
 forms~\cite{ModularForm,ModularFormTwoVar}.
Introducing the same $\, t$ as in (\ref{beforecover}) and
the following function $F(t)$
\begin{eqnarray}
t \, = \, \,  \Bigl({{\eta(\tau)\eta(6\tau) }
 \over {\eta(2\tau)\eta(3\tau)  }}\Bigr)^{12},
 \qquad \quad 
F  \, = \, \,  {{(\eta(2 \tau)\eta(3\tau))^7 }
 \over {(\eta(\tau) \eta(6\tau))^5  }},
\end{eqnarray}
one has an Ap\'ery's
 third order ODE~\cite{bo-bo-ha-ma-we-ze-09} on the modular 
form\footnote[2]{$F(t)$ and $\, t$ are modular forms
on $\, \Gamma_0(6)$ which has four inequivalent
 cusps $\, \infty$, $\, 0$, $\, 1/2$, $\, 1/3$. $F(t)$
is a weight 2 modular form. } $F(t)$. This ODE
 corresponds to the linear differential
operator\footnote[4]{In Ap\'ery's proof of the irrationality 
of $\, \zeta(3)$ a crucial role is played
 by the linear differential operator (\ref{Aperyonemore}).}:
\begin{eqnarray}
\label{Aperyonemore}
&& (t^2 \, -34\, t \, +1)\cdot t^2 \cdot  D_t^3 \, \, 
+ \, (6\, t^2 \, -153\, t \, +3)\cdot t \cdot  D_t^2   \\
&& \qquad \qquad  \qquad \qquad \, 
+ \, (7\, t^2 -112\, t \, +1) \cdot   D_t 
\, \,  \,  + (t-5),  \nonumber 
\end{eqnarray}
that reads in terms of the homogeneous derivative 
$\theta \, =\,\, t \cdot d/dt$:
\begin{eqnarray}
\label{Aperytheta}
&& (t^2 \, -34\, t \, +1) \cdot \theta^3 \,  \,
+ \, (3\, t^2 \, -51\, t)\cdot \theta^2   \\
&& \qquad \qquad  \qquad \qquad \, 
+ \, (3\, t^2 -27\, t) \cdot  \theta 
\, \,  \,  + (t^2\, -5\, t),  \nonumber 
\end{eqnarray}
this operator\footnote[1]{Introducing the
 inhomogeneous order-two ODE corresponding to (\ref{Aperytheta}) 
with the very simple rhs $\, 6\, t$, and 
considering the ratio of solution
of  (\ref{Aperytheta}) and of this inhomogeneous  
order-two ODE, one can build~\cite{ModularForm}
 a modular form of weight 4, by performing the
 third order derivative with respect to $\tau$, the ratio
of two solutions of (\ref{Aperyonemore}).
} being linked to
 the {\em modularity of the algebraic variety}
\begin{eqnarray}
 x\, +\,{{1} \over {x}}\,   +\,y\, +\,{{1} \over {y}}\,  
 +\,z\, +\,{{1} \over {z}}\,   +\,w \, +\,{{1} \over {w}} 
 \, \, \,\,= \, \, \, \ \, 0,  
\nonumber
\end{eqnarray}
that is, to the one-parameter 
family of K3-surfaces\footnote[5]{The
 simplest example of Calabi-Yau manifolds
 are K3 surfaces. This Ap\'ery operator (\ref{Aperytheta}) was 
seen in~\cite{Asterisque}
(see also~\cite{bo-bo-ha-ma-we-ze-09}) to correspond
 to a {\em symmetric square} of a second order operator 
associated to a modular form (see also section (\ref{Z2})).
Along this line it is worth recalling that
some one-parameter families of K3 surfaces
can be obtained from the {\em square of families of elliptic curves}
(see the so-called {\em Shioda-Inose structures} and
 their Picard-Fuchs differential 
equations~\cite{Long} and see also relations (1.9) 
in~\cite{LianYau}).}~\cite{Peters}: 
\begin{eqnarray}
\label{XYZz}
1\,\,  - (1-XY)\cdot Z\,
  - \,\,  z \cdot X\, Y\, Z\cdot  (1-X)\cdot (1-Y)\cdot (1-Z)
\,\, = \,\,  \, \, \, 0.  \nonumber 
\end{eqnarray}
\vskip .1cm 
$\, \bullet$ Another example of linear differential
 equations, satisfied by modular
forms, can be found in page 18 of~\cite{ModularForm}: 
\begin{eqnarray}
t \, = \, \, 
 \Bigl({{\eta(2\tau)\eta(6\tau) } 
\over {\eta(\tau)\eta(3\tau)  }}\Bigr)^6,
 \qquad \quad 
F  \, = \, \, 
 \Bigl({{\eta(\tau)^2 \, \eta(3\tau)^2 } 
\over {\eta(2\tau)\, \eta(6\tau)  }}\Bigr)^2,
\end{eqnarray}
the third order ODE on $F(t)$ corresponding to the
linear differential operator
\begin{eqnarray}
\label{othertheta}
&& (64\, t^2 \, +20\, t \, +1) \cdot \theta^3 \,  \,
+ \, (192\, t^2 \, +30\, t)\cdot \theta^2   \\
&& \qquad \qquad  \qquad \qquad \, 
+ \, (192\, t^2 \, +18\, t) \cdot  \theta 
\, \,  \,  + (64\, t^2\, +4\, t),  \nonumber 
\end{eqnarray}

$\, \bullet$ A third example~\cite{ModularForm} is:
\begin{eqnarray}
t \, = \, \,  
\Bigl({{\eta(3\tau)\eta(6\tau) } 
\over {\eta(\tau)\eta(2\tau)  }}\Bigr)^4,
 \qquad \quad 
F  \, = \, \,  
 {{  (\eta(\tau)\eta(2\tau))^3 } 
\over {\eta(3\tau)\,  \eta(6\tau)  }},
\end{eqnarray}
with a third order ODE on the modular form $F(t)$.
This ODE corresponds to the
linear differential operator
\begin{eqnarray}
\label{othertheta}
&& (81\, t^2 \, +14\, t \, +1) \cdot \theta^3 \, 
+ \, (243\, t^2 \, +21\, t)\cdot \theta^2   \\
&& \qquad \qquad  \qquad \qquad \, 
+ \, (243\, t^2 \, +13\, t) \cdot  \theta 
\, \,  \,  + (81\, t^2\, +3\, t),  \nonumber 
\end{eqnarray}

\vskip .1cm

If one chooses two linearly independent 
solutions ($F_1$, $F_2$) of these last order-three 
linear differential operators
 appropriately~\cite{ModularForm},
then $\, t$ is a modular form of 
$\, \tau$ the ratio of these two solutions. The
 function $\, t(\tau)$ satisfies
a well-known  third order {\em non-linear} ODE
known as the {\em Schwarzian equation}
 in the litterature\footnote[2]{The simplest example 
of Schwarzian equation, associated with the complete elliptic
integral $\, K$, corresponds to a rhs in (\ref{Jacobi}) reading 
$\, -{{1} \over {2}} \cdot \,
{\frac {{t}^{2}-t+1}{{t}^{2} \cdot (t-1)^{2}}}$.
The study of the Schwarzian equation is, in general,
 complicated, 
but Halphen found a simpler equivalent
 system of differential equations~\cite{Harnad,Halphen}.}: 
\begin{eqnarray}
\label{Jacobi}
2 \, Q(t) \cdot \Bigl({{d t} \over {d \tau}}\Bigr)^2
  \,\,\, + \,\,\,  \{t, \, \tau\}
\, \,\,\, = \,  \,\, \,\, 0,
\end{eqnarray} 
where $\,  \{z,\, t \}$
 denotes the {\em Schwarzian derivative}
with respect to $\, \tau$:
\begin{eqnarray}
\label{Schwaderiv}
{{d} \over {d\tau}} \, = \, \, q \cdot {{d} \over {dq}},
 \qquad \qquad 
  \{z,\, \tau\}  \, \, = \, \, \, \, {{z^{(3)}} \over {z'}}\, 
 -{{3} \over {2}} \cdot \Bigl({{z''} \over {z'}}\Bigr)^2.
\end{eqnarray}
and where $\, Q(t)$ is a rational function that
 can be simply deduced~\cite{ModularForm} from the
coefficients of the $k$-th order (here $k\, = \, 3$)
linear ODE on $\, F(t)$.

\vskip .3cm

\section{Modular form solution of $\, F_3$, 
related to $h_6$, Ap\'ery and $\, Z_2$}
\label{soluF3}
\vskip .1cm

The order-three linear differential
 operator  $\, F_3$, occurring as a factor 
of the differential operator
 annihilating $\, \tilde{\chi}^{(5)}$, was rationaly
 reconstructed in~\cite{High}. It 
can be seen to be homomorphic to 
the {\em symmetric square} of a second order
operator. Similarly to what we had
 for $\, F_2$, the (analytical at $x=0$)
solution of $\, F_3$ corresponds
 to a series with {\em integer} coefficients,
suggesting, again, a modular form interpretation.

Actually the solutions of this second order operator can be expressed
in terms of quadratic expressions of Legendre 
 or $_2F_1$ hypergeometric functions 
  {\em with a rational pull-back}. 
The three solutions of $\, F_3$ can be expressed in terms of Legendre 
functions where the pull-back $\, P_1$, in these  Legendre 
functions, reads:
\begin{eqnarray}
\label{pullF3}
&& P_1(x)  \, = \, \, \, 
{\frac {1}{108}}\,{\frac { (1-2\,x)  \, (1+2\,x) 
 \, (1+ 32\,x^2)^2}{x^2}}  \\
\label{pullF3minusone}
&&\quad \quad \, = \, \,  
{\frac {1}{108}}\,{\frac { (1- 4\,x)^3 \cdot (1+ 4\,x)^3}{x^2}}
\,\quad  + \, \, 1.
\end{eqnarray}
The solutions can also be expressed as quadratic expressions 
 of hypergeometric functions:
\begin{eqnarray}
 _2F_1\Bigl([{{1} \over {6}},\, {{1} \over {6}}],
[{{1} \over {2}}]; \, P_1 \Bigr),
 \qquad  \quad  
\,\,\,  
  _2F_1\Bigl([{{2} \over {3}},\, {{2} \over {3}}],
[{{3} \over {2}}]; \, P_1 \Bigr).  
\end{eqnarray}

Do note that the pull-back (\ref{pullF3})
 is not unique. Another (Atkin-Lehner involution
related) pull-back works equally well:
\begin{eqnarray}
\label{otherpull}
&&P_2(x)  \, = \, \, \, 
P_1 \Bigl( {{1} \over {8\, x}}\Bigr)
 \,   \,  \,\, = \, \, \, \, \,\,
 -\, {\frac {1}{108}}\,{\frac { (1- 4\,x) 
 \, (1+ 4\,x)  \, (1+2\,{x}^2)^2}{{x}^4}}\,   \\
\label{otherpullminus1}
&& \qquad \quad \qquad   \,  \,  = \, \, \,  \, \,
{\frac {-1}{108}}\cdot {\frac { (1\, -2\,x)^3 \, (1+2\,x)^3}{{x}^{4}}}
\,\quad  + \, \, 1.
\end{eqnarray}
Note that these two rational pull-backs are functions of $\, x^2$.
These two pull-backs can be seen as a
 rational parametrization 
 $(a, \, b) \, = \, \, (P_1(x), \, P_2(x))$
of the $(a, \, b)$-symmetric genus zero curve:
\begin{eqnarray}
\label{HauptF3}
&&-625\,\, +525 \cdot (a\, +b) \,\,\,  +3\,ba\, +96\cdot (a^2\, +{b}^{2})
 \, \,   \nonumber \\
&&\qquad   -528 \cdot (b{a}^{2} \, +\,{b}^{2}a)\,\, 
  +4 \cdot ({a}^{3} +{b}^{3})
\,   +432 \cdot {a}^{2}\,{b}^{2}  
\,\,  \, \, = \, \,\, \,  \, 0. 
\end{eqnarray}

Keeping in mind\footnote[1]{And keeping in mind 
the well-known symmetry of many hypergeometric functions 
 changing the pull-back $\, p(x)$
into $\, q(x)  \, = \,\,  \, 1\, -p(x)$. 
The other well-known symmetry
 $P_1(x) \,\leftrightarrow\, 1/P_1(x)$ corresponds to 
$x^2\,\,   \longleftrightarrow \, \, (1\,-4\, x^2)/(1\,+32\, x^2)/4$.}
  (\ref{pullF3minusone})
 and (\ref{otherpullminus1}),
we could have considered the algebraic curve relating 
 $(A, \, B) \, = \, \, (1\, -P_1(x), \, 1\, -P_2(x))$, which reads:
\begin{eqnarray}
\label{curvpbis}
&&-432\, A^2 \, B^2  \,\, + 4\cdot (A^{3}+ B^{3})\, \, 
+336 \cdot (A^2 \, B \, + B^2\, A)  \\
&&\qquad   +381 \cdot A\,B \, -12 \cdot (A^{2}+B^2)  \, \,
 \,  +12 \cdot (A+B) \,\,-4  \, \, \,\,  = \, \,\,\,  \, 0, \nonumber 
\end{eqnarray}
which is rationally parametrized as
$(A, \, B)\, = \, \, (A(z), \, \, B(z))$, where $\, A(z)$ and $\, B(z)$
read respectively: 
\begin{eqnarray}
\label{AB}
A(z)\, \,   = \, \,   \, {\frac {1}{1728}}\,{\frac { (z+16)^{3}}{z}},
\quad \quad \, \,\, \, \,
B(z)\, \,  = \, \,   \, {\frac {1}{432}}\,{\frac { (z+64)^{3}}{{z}^{2}}}, 
\end{eqnarray}
where  $\, A(z)$ and $\, B(z)$ are
related by a {\em Atkin-Lehner involution}  $\, B(z)\, = \, \,A(2^{10}/z)$.
This rational parametrization
 is {\em extremely similar} to the  
parametrization (\ref{alphazbetaz}),
(\ref{paraX02})
of the {\em fundamental modular curve} (\ref{X02}).
One deduces, from (\ref{AB}), the rational  parametrization 
for the curve (\ref{HauptF3})
\begin{eqnarray}
\label{abpara}
a \, = \, \, -\,{\frac { (z+64)  \, (z-8)^2}{1728 \cdot \,  \,  z}}, 
\quad \, \quad  \,  \,   
b \, = \, \, -\,{\frac {   (z+16)  \, (z-128)^{2}}{432  \cdot \,  \,  {z}^2}}. 
\end{eqnarray}
where, again, $\, b(z) \, = \, \, a(2^{10}/z)$.
Within parametrization (\ref{abpara}), expressions
$\, (P_1(x), \, P_2(x))$ (see
(\ref{pullF3}), (\ref{otherpull}))
 correspond  to: $\,\, z \, = \, \,  -256 \cdot x^2 \, \, $
or $\,\, z \, = \, \,  -\, 4/x^2$.

From (\ref{AB}) it is thus tempting to interpret
 the {\em new}
 genus zero algebraic curve 
(\ref{HauptF3}), or (\ref{curvpbis}),
 as a {\em modular curve}\footnote[2]{Modular
 curves of genus 0, which
 are quite rare, turned out to be
 of major importance in relation with 
the monstrous moonshine 
conjectures~\cite{Ivanov,Tuite}. In general,
 a modular function field is a 
function field of a modular curve (or, 
occasionally, of some other moduli space 
that turns out to be an irreducible variety). Genus 0 
means that such a function field
 has a single transcendental function
 as generator: for example the $\, j$-function. 
The traditional name for such a generator, 
which is unique up to a M\"obius 
transformation and can be appropriately
 normalized, is a Hauptmodul 
(main or principal modular function).} relating 
two {\em  Hauptmoduls} corresponding 
to the two pull-backs (\ref{pullF3})
and (\ref{otherpull}), similarly 
to what was found (see subsection (\ref{Z2})) 
for the second order
operator $\, Z_2$ and its
 weight-1 modular form solutions. It was
seen to be related~\cite{bo-bo-ha-ma-we-ze-09} to a 
second order linear differential 
operator occurring in Apery's analysis of $\, \zeta(3)$:
\begin{eqnarray}
\label{Aperidiff}
&&4\,x \cdot \left( {x}^{2}-34\,x+1 \right) \cdot { D_x}^{2}
 \, \, \nonumber \\
&& \qquad \qquad  +4 \cdot  (1\, -51\,x\,  +2\,{x}^{2})
 \cdot D_x\, \,  \,  \, +x -10. 
\end{eqnarray} 
From its rational parametrization
 this {\em new curve} (\ref{curvpbis})
is extremely similar to (\ref{X02}), the {\em fundamental
 modular curve} $\, X_0(2)$. One has, of course, the well-known
(and slightly tautological) algebraic geometry statement 
that all the genus zero curves of the plane
 are birationaly equivalent. But refereing
to the ``second layer'' of parametrization (see (\ref{Ded2}) above)
can we say that this new curve is {\em also} a representation
 of $\, \tau \, \longleftrightarrow \, N \, \tau$ and thus ``truly'' 
a modular curve ?  

\subsection{ The new curve (\ref{curvpbis}) 
and the modular group $\, \Gamma_0(6)$}
\label{andGamma6}
Seeking for hypergeometric functions with pull-backs that 
{\em are not rational} functions anymore,
 but algebraic extensions,
we actually found another description of the solutions.
The second order operator (\ref{Aperidiff})
can be solved in terms of hypergeometric 
functions 
$\, _2F_1\Bigl([{{2} \over {3}},
\, {{2} \over {3}}],[{{3} \over {2}}]; \, P_{\pm} \Bigr)$
with the two possible algebraic (Galois conjugate) pull backs:
\begin{eqnarray}
\label{square}
&&P_{\pm} \, \,  = \, \,\, \,
 -{\frac {1}{216}}\,{\frac {{x}^{4}-23\,{x}^{3}-156\,{x}^{2}-23\,x+1}{{x}^2}} 
  \,\, \, \\
&& \qquad \qquad \pm \,  
{\frac {1}{216}}\,{\frac { (x-1)  \, ({x}^{2}-7\,x+1) }{{x}^{2}}}
\, \sqrt{1-34\, x \, +x^2}. \nonumber 
\end{eqnarray}
To get rid of the square root in (\ref{square}),
  we introduce a parametrization of 
the rational curve $\, y^2 \, = \, \, 1-34\, x \, +x^2$, 
namely:
\begin{eqnarray}
\label{ratpara}
x \, = \, \,  \,
 {{1} \over {2}} \,{\frac { (u+18)  \,  (u+16) }{u}}, 
\qquad \quad 
y \, = \, \, \, 
 {{1} \over {2}} \,{\frac {288-{u}^{2}}{u}}.
\end{eqnarray}
In terms of this rational parametrization (\ref{ratpara})
the  two possible pull-backs (\ref{square}) read respectively:
\begin{eqnarray}
\label{squarebis}
&&P_{+}(u)\, = \, \, \,
 -{\frac {1}{432}}\,{\frac { \left( u+24 \right)^{2} 
\left( {u}^{2}+12\,u-72 \right)^{2}}{u \cdot (u+16)  \, (u+18)^{2}}},
 \\
&&P_{-}(u)\, = \, \, \,
-{\frac {1}{216}}\,{\frac { \, (u+12)^{2}
 \, ({u}^{2}-48\,u-1152)^{2}}{{u}^{2} \, (u+16)^2 \, (u+18) }}
\,  \,= \, \, \,\,  \,
P_{+}\Bigl({{288 } \over {u}} \Bigr),
\nonumber 
\end{eqnarray}
or, more simply on 
$ \, (Q_{+}(u),\,Q_{-}(u)) \,\, = \, \,  \, (1\, -P_{+}(u),\, \,1\,- P_{-}(u))$:
\begin{eqnarray}
\label{QQ}
&&Q_{+}(u)\,\, = \, \, \,
 {\frac {1}{432}}\,{\frac { (u+12)^{6}}{u \cdot (u+16)  \, (u+18)^{2}}},
 \\
&&Q_{-}(u)\,\, = \, \, \,Q_{+}\Bigl({{288 } \over {u}} \Bigr)
 \,  \,= \, \, \,\,  \,
{\frac {1}{216}}\,{\frac { (u+24)^{6}}{{u}^{2} \cdot (u+16)^{2} \, (u+18) }}.
\nonumber
\end{eqnarray}

Noticeably, these two pull-backs (\ref{squarebis})
 {\em can actually be seen as another
 rational parametrization} $(a, \, b) \, = \, \, (P_{+}(u), \, P_{-}(u))$
{\em of the genus zero curve} (\ref{HauptF3}).
 The two pull-backs (\ref{QQ}) 
are another rational parametrization
of the ``new'' algebraic curve (\ref{curvpbis}), 
with $\, (A, \, B)  \, = \, \, (Q_{+}(u), \, Q_{-}(u))$.

Recalling the fact that the two pull-backs $\,P_1(x) $ and  $\,P_2(x)$
(see (\ref{pullF3}), (\ref{otherpull}))
were functions of $\, x^2$, 
the correspondence between $\,P_1(x) $ and  $\,P_2(x)$,
 and the two pull-backs (\ref{squarebis}),
corresponds to the following change of variables:
\begin{eqnarray}
\label{changeof}
x^2 \, = \, \, \, \, 
 -{\frac {1}{128}}\,{\frac { (u+16)\cdot u }{u+18}}.
\end{eqnarray}

The results (\ref{squarebis}) could have been
 obtained, alternatively, recalling 
the change of variable (\ref{cover}), namely 
$\, x \, = \, \, v \cdot (1 \, -9\, v)/(1-v)$, 
which transforms the linear differential operator (\ref{Aperidiff}) 
into (after rescaling by $\, (1-v)^{1/2}$):
\begin{eqnarray}
D_v^2 \,\, \,\, 
 +\, {{ 1-20\, v\, +27\, v^2}
 \over { (1-9\, v) \, (1-v)\, v }} \cdot D_v \, \, 
\,\, -  3 \cdot {{1-3\, v} \over { (1-9\, v) \, (1-v) \, v}},
\end{eqnarray}
which corresponds to the 
(staircase-polygon~\cite{Prell}) 
second order 
operator ${\cal Z}_3$ 
seen in Appendix A of~\cite{bo-bo-ha-ma-we-ze-09}
(see equation after (A.4)).
The variable $\, u$ is related to the previous 
variable  $\, v$  by $\, u \, = \, \, 18 \cdot (v-1)$.

However, recalling the weight-1 modular form
 interpretation of\footnote[2]{Or of the
 order-two operator ${\cal Z}_3$ corresponding to
 staircase polygons~\cite{Prell}.}
 operator $Z_2$ 
in~\cite{bo-bo-ha-ma-we-ze-09},
we had the occurrence of Hauptmoduls 
 ${\cal M}_z\, = \, \,  {{12^3} \over {j}}$
(resp. ${{12^3} \over {j'}}$) corresponding to 
 the (genus zero~\cite{Maier1}) modular curve\footnote[1]{ 
Which amounts to multiplying, 
or dividing, the ratio of the two periods of the elliptic
curve by $\, 6$.}
\begin{eqnarray}
\label{sixmodu} 
\Phi_6(j, \, j') \, \, \, = \, \, \, \, \Phi_6(j', \, j)
\,  \, \,= \, \, \, \, 0, 
\end{eqnarray}
obtained from the elimination of $\, z$ between\footnote[5]{
Here `` $\circ$ `` denotes 
the composition of two rational functions
i.e. $\, f(z) \circ g(z) \, = \, \, f(g(z))$.}:
\begin{eqnarray} 
\label{j6} 
&&j \, \, = \, \, j_6(z) \, = \, \, 
{\frac { (z+6)^{3} 
\, (z^3+18\,z^2+84\,z+24)^3}
{z \cdot (z+9)^{2} \, (z+8)^3}},\\
&& \quad \,\,  = \, \, \, 
j_2\Bigl( {\frac {z \cdot (z+8)^{3}}{z+9}}  \Bigr) 
\, \, = \, \, \,  {\frac { (z+16) ^{3}}{z}} \circ \, 
 {\frac {z \cdot (z+8)^{3}}{z+9}} , \nonumber  \\
&& \quad \,\,  = \, \, \, 
j_3\Bigl( {\frac {z \cdot (z+9)^{2}}{z+8}} \Bigr) 
\, \, = \, \, \, {\frac { (z+27)  \, (z+3)^{3}}{z}}
\circ \, {\frac {z \cdot (z+9)^{2}}{z+8}},
 \quad \quad \hbox{and:}
\nonumber  \\
&&j' \,  \, = \, \,j_6\Bigl( {{2^3 \cdot 3^2 } \over { z }}\Bigr) 
\, = \, \, \, {\frac { (15552+3888\,z+252\,{z}^{2}+{z}^{3})^{3}
 \, (z+12)^{3}}{{z}^{6} \, (z+8)^{2} \, (z+9)^{3}}} \nonumber \\
&& \quad  \,\,  = \, \, \,  
 j_2'\Bigl( {\frac {{z}^{3} \, (z+8) }{ (z+9)^3}} \Bigr) 
 \,\,  = \, \, \,
 {\frac { (z+256)^{3}}{{z}^2}}
 \circ \,   {\frac {{z}^{3} \cdot (z+8) }{ (z+9)^3}}  \\
&& \quad  \,\,  = \, \, \,  j_3'\Bigl( 
{\frac {{z}^{2} \cdot (z+9) }{ (z+8)^2}}\Bigr) 
\,\,  = \, \, \, 
{\frac { (z+27)  \, (z+243)^{3}}{{z}^{3}}} 
\circ \, {\frac {{z}^{2} \cdot (z+9) }{ (z+8)^2}}.
 \nonumber
\end{eqnarray}
with the covering~\cite{bo-bo-ha-ma-we-ze-09}
\begin{eqnarray} 
\label{changerz}
z \,\, = \, \,\, \,
{\frac {72  \, \,\, x}{ (1-x) 
 \, (1-4\,x) }},
\end{eqnarray}
which is a slight modification 
of (\ref{simplechange}).
Let us introduce the alternative covering $\, u \, = \, \, \,2 \, z$, 
the two Hauptmoduls  ${\cal M}_z\, = \, \,  {{12^3} \over {j}}$
(resp. ${{12^3} \over {j'}}$) read respectively:
\begin{eqnarray} 
\label{P1P2}
&&P_1^{(6)}(u) \, = \, \,\, \,
 {\frac { 110592 \cdot  u \cdot (u+16)^{3} \left( u+18 \right)^{2}}{
  (12+u)^{3} \cdot ( 192+336\,u+36\,{u}^{2}+{u}^{3})^{3}}}\\
&&\qquad  \, = \, \,\, \, 
 \Bigl( 12^3/\Bigl( {\frac { \left( u+32 \right)^{3}}{4 \, u}}\Bigr)\Bigr)
\circ \, 
\Bigl( {\frac {u \cdot (u+16)^{3}}{4  \cdot  (u+18)}} \Bigr),
 \nonumber \\
&&P_2^{(6)}(u)  \, = \,\,\, \,\, 
 P_1^{(6)}\Bigl( {{288 } \over {u}} \Bigr)  \\
&&\qquad \quad \, = \,\,\, \,\, 
\,{\frac { 3456  \cdot   {u}^{6} \cdot  (u+16)^{2} \, (u+18)^{3}}{ (u+24)^3
 \, (124416+15552\,u+504\,{u}^{2}+{u}^3)^{3}}} \nonumber \\
&&\qquad  \quad  \, = \, \,\, \, \, 
 \Bigl( 12^3/\Bigl({\frac { (u+512)^{3}}{2\, u^2}}
\Bigr)\Bigr)
\circ \, 
 {\frac {{u}^{3} \left( u+16 \right) }{ (u+18)^{3}}}.
\nonumber  
\end{eqnarray}
The relation between the Ap\'ery operator (\ref{Aperidiff})
and the $\, Z_2$ 
  weight-1 modular forms
 associated with (\ref{sixmodu}), 
seems to say that there should be some (at first sight totally 
unexpected) relation between hypergeometric functions 
with the pull-backs (\ref{squarebis}) 
and the hypergeometric functions
 with the pull-backs (\ref{P1P2}). {\em We have actually
been able to find such a ``quite non-trivial'' relation}
\begin{eqnarray} 
&& C_{6}(u) \cdot \,
 _2F_1\Bigl( [{{1} \over {12}}, \, {{5} \over {12}}],
[1]; \,\,  P_1^{(6)}(u)  \Bigr) 
 \nonumber \\
&& \quad  \quad = \, \, \,\, \, 
2^{1/2} \cdot \rho \cdot C_{+}(u) \cdot \,
 _2F_1\Bigl( [{{2} \over {3}}, \, {{2} \over {3}}],
[{{3} \over {2}}]; \, \, P_{+}(u) \Bigr)
 \, \,   \\
&&  \quad  \quad  \quad  \quad \quad 
  - \rho \cdot C_{-}(u) \cdot \, 
 _2F_1\Bigl( [{{2} \over {3}}, \, {{2} \over {3}}],
[{{3} \over {2}}]; \,\,  P_{-}(u)  \Bigr),  
\, \, \nonumber
\end{eqnarray}
where $\,C_{+}(u)$, $\,C_{-}(u)$, $\, C_{6}(u)$ 
and  $\, \rho$ read  respectively: 
\begin{eqnarray} 
 \Bigl( {{  (u+24)^2} \over {u}}\Bigr)^{1/2}  
 \cdot 
\Bigl({{ 16 \, u^2}  \over {(u \,+\,  16) \, (u \,+\,  18)^2 }}\Bigr)^{2/3}
\cdot  \Bigl( {{{u}^{2}+12\,u-72} \over {64 \, u}} \Bigr), 
 \nonumber 
\end{eqnarray}
\begin{eqnarray} 
 \, \Bigl( {{ 2 \cdot (u+12)^2} \over {u}}\Bigr)^{1/2}  
 \cdot 
\Bigl({{ 4 \, u}  \over {(u \,+\,  18) \, (u \,+\,  16)^2 }}\Bigr)^{2/3}
\cdot \Bigl( {{  u^2 -48\,u-1152}   \over {16 \, u}} \Bigr),  
 \nonumber 
\end{eqnarray}
\begin{eqnarray}
 \Bigl( {{144 \, u^2} \over {18^2 \cdot  (12 + u) \,
 (192+336\,u+36\,{u}^{2}+{u}^{3})  }}\Bigr)^{1/4},
 \quad \quad  \,  \, \,  
 {{2} \over {3}}\,
{\frac { \left( \Gamma  \left( 2/3 \right)  \right)^{3}}{{\pi }^{2}}}. 
 \nonumber 
\end{eqnarray}

Note however, that  $\,P_1^{(6)}(u)$ (or $\,P_2^{(6)}(u)$)
 {\em cannot be expressed} as a rational function of the two
pull-backs  $\, P_{\pm}(u)$ (see (\ref{squarebis})).
 The relation between  $\, a_6 \, = \, \, P_1^{(6)}(u)$ and
 $\, a \, = \, \, P_{+}(u)$ (resp.  $\, b_6 \, = \, \, P_2^{(6)}(u)$
and $\, b \, = \, \, P_{-}(u)$) reads a (necessarily genus zero) 
algebraic curve
\begin{eqnarray}
\label{rela}
&&(a-1)  \cdot (9\,a-25)^{3} \cdot  a_6^{2}\,  \\
&& \quad \quad +8\, \, (a-1)  \cdot (1458\,{a}^{2}-1215\,a+125) \cdot  a_{6}
 \,\, \,  + \, 16 \, \, \, \, \, = \, \, \,  \,\, \, 0, \nonumber 
\end{eqnarray}
and the same  genus zero algebraic curve where one replaces
$(a_6, \, a)\, \rightarrow \,(b_6, \, b)$. Note that
 the relation between   $\,P_1^{(6)}(u)$ and  $P_{-}(u)$
 (resp.  $\,P_2^{(6)}(u)$ and  $P_{+}(u)$)
 is much more involved. 

 Recalling (\ref{sixmodu}) one deduces, from the previous calculations,
that the (genus zero) modular curve
\begin{eqnarray}
\Phi_6\Bigl({{12^3} \over {a_6}}, \, \,   \,
 {{12^3} \over {b_6}} \Bigr) 
\, \,  \, \, = \, \,  \, \, \, 0, 
\end{eqnarray}
is {\em actually ``equivalent''  
to the genus zero algebraic curve} (\ref{HauptF3})
{\em up to the algebraic covering}  (\ref{rela}).   

\vskip .5cm 
\subsection{Dedekind parametrization of the new curve}
\label{Dednew}
\vskip .1cm

Let us revisit the rational curve (\ref{curvpbis})
that we rewrite (with a rescaling of $\, A$ and $\, B$
by $\, 1728$):
\begin{eqnarray}
\label{Xcurp}
&& -{y}^{2}{z}^{2}\,\, 
+16\, \, (y\, +z)  \, ({z}^{2}+83\,yz+{y}^{2}) 
\, \, \,  -82944 \cdot ({z}^{2}\, +{y}^{2}) \\
&& \quad \quad  +2633472\cdot yz\,\, +143327232\cdot (y+z)\, \,
 -82556485632 
 \,\,\,\, \, = \,\,\, \, \,\, 0.   \nonumber 
\end{eqnarray}
The curve (\ref{Xcurp}) is rationally parametrized by:
\begin{eqnarray}
\label{yzj2}
z \, = \, \,\, z(j_2) \, = \, \,
{\frac { \left( 256+j_2 \right)^3}{{ j_2}^2}}, \quad \quad \quad 
y \, = \, \,\, z\Bigl({{2^{14} } \over {j_2}} \Bigr)
\, = \, \,\, {\frac { \left( 64\, +j_2 \right)^3}{16 \cdot  j_2}}
. \nonumber 
\end{eqnarray}
The correspondence with the previous  rational
 parametrization is $\, j_2\, = \, \, 4 \cdot z$.
Similarly to what was done for the fundamental modular curve (\ref{X02}),
we can introduce a second ``layer'' of parametrization, writing $\, j_2$
as a ratio of Dedekind eta function (\ref{Dedekindeta})
\begin{eqnarray}
j_2 \, = \, \,\, j_2(q) \, = \, \,\, \,
4 \, {{\Delta(q)} \over {\Delta(q^2)}},
 \qquad \quad  A(j_2) \, = \, \,
{\frac { \left( 256+j_2 \right)^3}{{ j_2}^2}}, \nonumber 
\end{eqnarray}
which yields the following parametrization for (\ref{Xcurp})
($A(j_2)$ is the same function as in (\ref{paraj2}))
\begin{eqnarray}
\label{reinjecte}
z \, = \, \, A(j_2(q))\,\, = \, \,\, 
 A\Bigl( 4 \, {{\Delta(q)} \over {\Delta(q^2)}}  \Bigr), 
\qquad 
y \, = \, \,\, A\Bigl( {{j_2(q^2)} \over {4}}\Bigr)\, = \, \,
A\Bigl( {{\Delta(q^2)} \over {\Delta(q^4)}} \Bigr), \nonumber 
\end{eqnarray}
which makes clear that the algebraic curve (\ref{Xcurp})
 is a representation of $\, q \, \rightarrow \, q^2$.
The compatibility between the Atkin-Lehner involution
 $j_2 \, \leftrightarrow \, 2^{14}/j_2$ and the 
$\, q \, \rightarrow \, q^2$ transformation, corresponds to 
\begin{eqnarray}
A\Bigl( {{\Delta(q^2)} \over {\Delta(q^4)}} \Bigr) 
\,\,\, = \, \,\,  \,\, \,\,
A\Bigl(2^{14}/\Bigl( 4 \, {{\Delta(q)} \over {\Delta(q^2)}}  \Bigr) \Bigr)
\end{eqnarray}
which is nothing but (\ref{compatible}) corresponding to the 
functional equation (\ref{Ramanujan}). 

Matching $\,y/1728$ or $\,z/1728$ 
(see (\ref{pullF3}), (\ref{otherpull}))
 with $1\, -P_1$ or $1\, -P_2$,   
one gets:
\begin{eqnarray}
j_2 \, = \, \, - 1024 \, x^2 
\qquad \quad \hbox{or:} \qquad \quad 
j_2 \, = \, \, - {{16} \over { x^2}}. 
\end{eqnarray}
yielding a {\em straight interpretation of the
 $ \, x$ variable} in the $\, n$-fold integrals of the
Ising model {\em in terms of Dedekind eta function},
and more precisely, of the discriminant $\, \Delta$:
\begin{eqnarray}
 x^2  \, \, = \, \,  \,
  -{{1} \over {256}} \cdot {{\Delta(q)} \over {\Delta(q^2)}}, 
\quad \quad \hbox{or:} \quad  \quad  \quad 
 x^2 \, \, = \, \, \,    \, -\, 4 \cdot
{{\Delta(q^2)} \over {\Delta(q)}}.
\end{eqnarray}

\section{Modular form solution of $\, \tilde{L}_3$: 
is it associated with the new curve (\ref{curvpbis}) 
or with the fundamental modular curve $\, X_0(2)$ ? }
\label{L3tilde}
\vskip .1cm

In~\cite{Khi6} an order-three  linear differential 
operator $\, \tilde{L}_3$
was found as a factor of the minimal order operator
 for  $\, \tilde{\chi}^{(6)}$.
This order-three  operator
 is (as it should) globally nilpotent, and one can see that 
it is reducible to an order-two operator 
in the sense that it is 
homomorphic to the {\em symmetric square} of an order
 two linear differential operator:
\begin{eqnarray}
&&x \cdot  (1\, -16\,x)^2 \cdot (1\, - 4\,x)^{2} \cdot D_x^2 \\
&& \qquad \quad \,  
+ \, (1\, -24\,x)  \, (1 \, -16\,x)  \, (1\, -4\,x)^2 \cdot D_x 
\,\,\, +108\,{x}^{2},  \nonumber 
\end{eqnarray}
yielding  solutions in terms of hypergeometric functions like
 $ _2F_1([1/8, 3/8],[1/2];\, \, {\cal P}_1(x))$, with 
the pull-back $\, {\cal P}_1(x)$  reading
\begin{eqnarray}
\label{P1L3tilde}
{\cal P}_1(x) \,\, = \, \, \, \, \,
{\frac { (1\, -12\,x)^{2}}{ (1\, -16\,x) 
 \, (1-4\,x)^{2}}}, 
\end{eqnarray}
or more simply:
\begin{eqnarray}
1\, -{\cal P}_1(x) \,\, = \, \,\,  \,
-\,{\frac {256 \cdot {x}^{3}}{ (1 \, -16\,x)  \, (1\, -4\,x)^{2}}}.
\end{eqnarray}

\subsection{An ``uneducated'' guess}
A simple calculation shows that, some ``Atkin-Lehner'' 
transform of (\ref{P1L3tilde})
\begin{eqnarray}
{\cal P}_2(x) \, = \, \, \,\, 
 {\cal P}_1\Bigl( {{1} \over {64 \, x}}\Bigr)
 \,\, = \, \,\,  \, 
 - \,{\frac {4\cdot  (3-16\,x)^{2}\cdot x}{ (1-4\,x)
 \, (1\, -16\,x)^{2}}}, 
\end{eqnarray}
or more simply 
\begin{eqnarray}
1\, -\, {\cal P}_2(x)
 \, = \, \, \,\,\, 
{{1 } \over {(1\, -4\, x) \, (1\, -16 \, x)^2}}, 
\end{eqnarray}
provides {\em another rational parametrization} 
 $(a, \, b) \, = \, \, ({\cal P}_1(x) , \, {\cal P}_2(x) )$
{\em of the new genus zero curve} (\ref{HauptF3}).

Let us introduce the $\, x^2$-dependent transformation 
\begin{eqnarray}
 x \qquad \longrightarrow \qquad \,  \,
 R(x) \, \,\,  \, = \, \,  \,\,  \,
 {{1} \over {16 }} \cdot {{1\, -16 \, x^2} \over {1\, -4 \, x^2 }}, 
\end{eqnarray}
one actually finds a nice relation between the two pull-backs 
$ \, {\cal P}_1(x)$ and  this ``guessed candidate''  $ \, {\cal P}_2(x)$,
for the other pull-back of $\tilde{L}_3$ (if any ...):
\begin{eqnarray}
 {\cal P}_1\Bigl(R(x)\Bigr) \, = \, \, \, P_1(x),
\qquad \quad
 {\cal P}_2\Bigl(R(x)\Bigr) \, = \, \, \, P_2(x).
\end{eqnarray}
These two equalities are actually compatible
 with the two Atkin-Lehner involutions for 
$\, ({\cal P}_1, \, {\cal P}_2)$ and $\, (P_1, \, P_2)$, 
because of the nice functional relation on $\, R(x)$:
\begin{eqnarray}
R\Bigl({{1} \over { 8 \, x}}  \Bigr)
\, \,  \, = \, \,\, \,\, \,
 {{1} \over {64 \cdot R(x)}}. 
\end{eqnarray} 

It is, thus, extremely tempting to imagine that $\, \tilde{L}_3$
is, like $\, F_3$, related to the {\em new modular curve}
 (\ref{HauptF3}) or (\ref{curvpbis}). Furthermore,
 this would yield
some (deep  ...) relation between 
the singularities of the $\, \chi^{(2\,n)}$ 
and singularities of the $\, \chi^{(2\,n\, +1)}$, that is to say,
between the low and high temperature singularities of the 
susceptibility of the Ising model.
{\em This is not  the case}: $\, \tilde{L}_3$ is, in fact, related to the
fundamental modular curve $\, X_0(2)$.

\subsection{Modular form solution of $\, \tilde{L}_3$: the 
fundamental modular curve $\, X_0(2)$}
\label{L3tilde2}
\vskip .1cm

Actually, leaving, again, the framework
 of {\em rational} pull-backs, one gets
 the two (Galois-conjugate) {\em algebraic
 pull-backs} $\,\,\, {\cal P}_{\pm}[\tilde{L}_3]$: 
\begin{eqnarray}
&&3456 \, \cdot \, \, {\cal P}_{\pm}[\tilde{L}_3]
 \,\,\,\, \, \, \, \, =  \\
&& \quad \, \, \,   {\frac { \left( 40\,{x}^{2}-17\,x+1 \right) 
 \left( 400\,{x}^{4}-928\,{x}^{3}+297\,{x}^{2}-31\,x+1 \right) }{ x^6}} 
\nonumber  \\
&& \quad  \pm \,\,{\frac { (1\, - 12\,x) 
 \, (1\, - 4\,x)  \, (1- 7\,x)  \, (25\,{x}^{2}-17\,x+1) }{{x}^6}} 
\cdot \sqrt{1 \, -16\,x}. \nonumber 
\end{eqnarray}
The relation between these two Galois-conjugate pull-backs 
actually corresponds to (\ref{alphabeta}) 
which is {\em nothing} (up to a 1728 rescaling
factor, see (\ref{X02})) but
 the  {\em fundamental modular curve} $\, X_0(2)$.

To get rid of the square root singularity,
 we introduce the variable $\, y$:
\begin{eqnarray}
y^2 \,  =\,  \, \, 1 \, -16\,x, 
\qquad \quad \hbox{i.e.}  \qquad \quad 
 x \, = \, \, \, -{{1} \over {16}} \, \, (y^2 -1). 
\end{eqnarray}
The two previous algebraic pull-backs become respectively:
\begin{eqnarray}
 {{1} \over {27}} 
\,{\frac { \left( 5\,{y}^{3}-9\,{y}^{2}+15\,y-3 \right)^{3}}{
 \, (y+1)^{6} \, (y-1)^{3}}}, \quad \,\, \,  
 {{1} \over {27}} 
\,{\frac { \left( 5\,{y}^{3}+9\,{y}^{2}+15\,y+3 \right)^{3}}{
 \left( y+1 \right)^{3} \, (y-1)^{6}}}, 
\end{eqnarray}
which can be seen to be a rational 
parametrization of (\ref{alphabeta}). 
Recalling the previous parametrization (\ref{alphazbetaz}) 
one finds that the $\, z$ variable
in  (\ref{alphazbetaz}), must be equal to
 $\, z \, = \, \, 64 \cdot (y+1)^3/(y-1)^3$, 
 or $\, z \, = \, \, 64 \cdot (y-1)^3/(y+1)^3$.

\section{The puzzling $\, L_4$: preliminary results on $\, L_4$}
\label{L4}
\vskip .1cm

 Let us now focus on the order-four linear differential
operator $\, L_4$,
discovered as a factor of $\, \chi^{(6)}$, and that
 we were fortunate enough to get exactly by 
rational reconstruction\footnote[2]{For
 details on the rational reconstruction
see~\cite{col-enc-ratrecon}.}~\cite{Khi6}.
 Let us display a few results on $\, L_4$. 

\subsection{Negative results on $\, L_4$}
\label{negative}
Suppose that $\, L_4$ is equivalent (in the sense of 
equivalence of linear differential
 operators~\cite{Homomorphisms1,Homomorphisms})
to a symmetric cube
 of a second order linear differential
operator $\, L_2$.
Take a point $\,x\,=\, a$,  and 
suppose that the highest exponent of $\,\ln(x-a)$
that appears in the formal solutions of $\,L_2$,
 at $\,x\,=\,a$, equals $\,\rho$.
Then the highest exponent of $\,\ln(x-a)$
 that appears in the formal solutions
of $\, L_4$, at $x=\, a$, must be $\, 3\, \rho$.
Now look at the formal solutions of $\, L_4$,
 at $\, x=\,1/8$,
 one  gets a contradiction.
Similar reasoning (using both $\,x=\,1/8$
 and $x=\,0$) shows that $\, L_4$  {\em can also not
be related to the symmetric product of two second
order operators}. 

Our preliminary calculations 
show  that $\, L_4$  is not $\, _4F_3$-solvable
{\em if one restricts to rational pull-backs}, in the
sense that there is no $\, _4F_3$ differential 
operator that can be sent to $\, L_4$
(with a change of variables 
$\, x \, \rightarrow \,$  rational function in $\, x$,
followed by multiplying by
 $\exp(\int$(rational function in  $\, x$)),
   followed by Homomorphisms) to $\, L_4$.

In essence, if we allow the following functions in
 $ \mathbb{C}(x)$:  exp, log, and any $_pF_q$
 function, composed with any rational functions,
and anything one can form from those functions using
  addition, multiplication, derivatives, indefinite integral,
then we believe that 
$\, L_4$ is {\em not solvable in that class
of functions}.

\subsection{Positive results on $\, L_4$}
\label{positive}
The order-four linear differential operator $\, L_4$
exhibits, however, a set of nice properties. Let us display 
some of these  nice properties.

\vskip .2cm 

$\, \bullet$ An irreducible linear differential 
equation (resp. irreducible linear differential operator) 
 is said to be of Maximal Unipotent
 Monodromy (MUM) if all the indicial 
exponents at $\, x = \, 0$
 are zero (one Jordan block
to be Maximal). 
The formal solutions of an order-four MUM linear 
differential operator can be written as
\begin{eqnarray}
\label{fsolcalL4}
&&y_0 \, =\,\,\, y_0, \nonumber \\
&&y_1 \, =\,\,\, y_0 \, \cdot  \ln(x) \,+ \tilde{y}_1,
 \nonumber \\
&& y_2 \, =\,\,\, {1 \over 2} y_0 \cdot  \ln(x)^2\, 
 + \tilde{y}_1 \cdot  \ln(x)\,\,  + \, \tilde{y}_2, \\
&& y_3 \, =\,\,\, {1 \over 6} y_0  \cdot  \ln(x)^3\, 
 + {1 \over 2} \tilde{y}_1  \cdot  \ln(x)^2\,\, 
 + \tilde{y}_2 \, \ln(x)\, + \tilde{y}_3. \nonumber
\end{eqnarray}

The indicial exponents of  $\, L_4$ at $\, x\, = \, \, 0$, 
read $\, -6, \, -4, \, -4, \,$ and $\, 0$. 
Therefore, the order-four operator $\, L_4$  is {\em not} MUM,
however the formal solutions 
{\em can be cast exactly as for a MUM linear differential 
operator}.
The formal solutions of $\, L_4$ can be written as
\begin{eqnarray}
\label{fsolL4}
&& y_0 \, =\,\,\, y_0,\nonumber \\
&& y_1 \, =\,\,\, y_0 \cdot \ln(x)\,\, + \tilde{y}_{10}, 
\nonumber \\
&& y_2 \, =\,\,\, {1 \over 2} y_0  \cdot  \ln(x)^2\,
 + \tilde{y}_{21} \cdot  \ln(x)
\, + \tilde{y}_{20}, \\
&& y_3 \, =\,\,\, {1 \over 6}\, y_0 \cdot  \ln(x)^3\,
  + \tilde{y}_{32} \cdot  \ln(x)^2 \, 
+ \tilde{y}_{31}  \cdot  \ln(x)\,\, + \tilde{y}_{30} 
\nonumber
\end{eqnarray}

Such particular form for the set of series solutions 
is often obtained for irreducible operators.

There are four independent series that can be
 chosen as $y_0$, $\tilde{y}_{10}$, $\tilde{y}_{20}$
and $\tilde{y}_{30}$. The other series in front of the log's 
should depend on these four
chosen series.
For $L_4$, these series read
\begin{eqnarray}
\label{linfsolL4}
&& \tilde{y}_{21} \, =\,\, \, \, \,
\, \tilde{y}_{10}\,\,\,
 - {\frac{2854486264697}{459375 \cdot 10^6}}\, y_0, 
\nonumber \\
&& \tilde{y}_{32} \, =
\,\,\,\, {1 \over 2} \tilde{y}_{10} \,\,\,
+ {\frac{38100003421933}{11484375 \cdot 10^4}}\, y_0, 
\nonumber \\
&&  \tilde{y}_{31} \, \, =\, 
\,\,\, \, \tilde{y}_{20}\, \,\,
 + {85327 \over 128}\,  \tilde{y}_{10}\,\,
 - {\frac{103266884399422867}{504 \cdot 10^{11}}}\, y_0, 
 \nonumber
\end{eqnarray}
where we see that it is a matter of combination
 to cast the formal solutions
(\ref{fsolL4}) in the form (\ref{fsolcalL4}). Making
 the combination
$y_2\, -c_{21}\,y_1$ and 
$y_3\, -2 c_{32} y_2-(c_{312}\, -2 c_{21} c_{32}) y_1$,
(where $\,c_{21}$, resp.  $\,c_{32}$ and $\,c_{312}$,
are the coefficients in front of $\, y_0$
in $\, \tilde{y}_{21}$, resp. $\, \tilde{y}_{32}$ and
 $\, \tilde{y}_{31}$, in (\ref{linfsolL4})), 
 one obtains
the new set
\begin{eqnarray}
\label{lincast}
&& y_0 \, =\,\, y_0, \nonumber \\
&&  y_1 \, =\,\, y_0 \cdot  \ln(x)\,\,   + \tilde{y}_{10}, 
 \nonumber \\
&& y_2 \, =\,\, {1 \over 2} y_0 \cdot \ln(x)^2\, \, 
 + \tilde{y}_{10} \cdot  \ln(x)\, \, 
 + \tilde{y}_{20}\,\, -c_{21} \tilde{y}_{10},   \\
&& y_3 \, =\, \,{1 \over 6} y_0 \cdot \ln(x)^3\, \,
  +  {1 \over 2} \tilde{y}_{10} \cdot \ln(x)^2
 \nonumber \\
&& \qquad 
+ \left(\tilde{y}_{20}\, +c_{311} \tilde{y}_{10}\, 
-2c_{32} c_{10} \tilde{y}_{10} \right)
 \cdot \ln(x) \nonumber \\
&& \qquad \, + \, \tilde{y}_{30}\,
 -2 c_{32} \cdot \tilde{y}_{20} \, \,
 +(2 c_{32} c_{21} -c_{312} ) \cdot \tilde{y}_{10},   
  \nonumber
\end{eqnarray}
where $\, (\tilde{y}_{20}\, -c_{21} \tilde{y}_{10})$ 
identifies with
$\, (\tilde{y}_{20}\, +c_{311} \tilde{y}_{10}\, 
-2c_{32} c_{10} \tilde{y}_{10})$ 
since
$c_{21} =\, - c_{311}\,  +2c_{32} c_{10}$ 
for the actual values of the combination
coefficients.

The remaining difference between a MUM linear 
differential operator and the formal solutions
of $L_4$ is the beginning of the series
 of the non leading log's. For instance
the local exponent $-4$ being twice, one 
should have a series starting as $x^{-4}$
in front of the log's like $\, \tilde{y}_{10}$
\begin{eqnarray}
\fl \quad \tilde{y}_{10} \, =\,\,\,\,
{\frac {1}{84000 \,\, x^4}}\,+{\frac {11}{16800\,\, x^3}}\,\,
+{\frac {9329}{336000\,\, x^2}}\,\,
 +{\frac {8023}{8400 \,\, x}}\,\,
+{\frac {99922803261913}{3675000000000}}
\,\,+\, \,\cdots   \nonumber 
\end{eqnarray}
We may imagine that by acting by an intertwinner
 on the formal solutions, one
ends up with series starting at $x$, i.e. 
$L_4$ may be equivalent to a linear 
differential operator which is MUM.

\vskip .2cm 

\vskip .2cm 

$\, \bullet$ The order-four linear differential operator $\, L_4$
is of course {\em globally nilpotent}~\cite{bo-bo-ha-ma-we-ze-09}.
The $\, p$-curvature~\cite{Andre7} of the order-four
 globally nilpotent differential operator 
 $\, L_4$ can be put in a remarkably simple Jordan form:
\begin{eqnarray}
\label{jordan}
 \left[ \begin {array}{cccc} 
0&1&0&0\\ 
\noalign{\medskip}0&0&1&0\\ 
\noalign{\medskip}0&0&0&1\\ 
\noalign{\medskip}0&0&0&0
\end {array} \right]. 
\end{eqnarray}
Its characteristic and minimal polynomial is $\, T^4$. 
Such an operator {\em cannot be a symmetric cube
 of a second order} operator 
in $\, \mathbb{C}(x)[D_x]$. We, however, determined 
the differential Galois group~\cite{nFn,vdP} of 
this linear differential operator $\, L_4$
 and {\em actually found that it
 is\footnote[9]{$\, SP(4, \, \mathbb{C})$ 
 is not the monodromy group:
it is equal to the Zariski closure
 of the (countable) monodromy group, i.e.
the differential Galois group.} the symplectic group}
$\, SP(4,\, \mathbb{C})$. A crucial step to exhibit this symplectic
structure, amounts to calculating the {\em exterior square}
of $\, L_4$, and verify that this, at first sight order-six,
 exterior square either reduces to an  order-five operator,
 or is a direct sum of an order-five operator 
and an order-one operator with a rational solution. 
$\, L_4$ corresponds to this last scenario.

Denoting $\, y_i$ the four solutions of an order-four linear
differential operator, the exterior square of 
that operator is a linear differential operator that annihilates the
expressions
\begin{eqnarray}
\label{sixExt}
w_{i, j} \,=\,\,\,\,
 y_i \, {d y_j \over dx} \,  - y_j \, {d  y_i \over dx},
 \qquad \quad i \ne j =0, 1, 2, 3. 
\end{eqnarray}
It should, then, be (at first sight) of order six.

The six solutions (\ref{sixExt}) of the exterior square of 
a MUM order-four operator, 
contain log's with
degrees (at most) 1, 2, 3, 3, 4 and 5. There are then two solutions
 (with the same degree
in the log's) that can be equal
\begin{eqnarray}
y_0 \, {d \over dx} y_3 \, - y_3 \, {d \over dx} y_0 
\,\,  \,\,=\,\,\, \,\, \, \, 
y_1 \, {d \over dx} y_2 \, - y_2 \, {d \over dx} y_1.
\end{eqnarray}

When this happens the exterior square, annihilating
 five independent solutions, will be of order
five. This is equivalent (see proposition 2 in~\cite{Almkvist}),
 for the coefficients of the order-four
linear differential operator, to verify 
condition (\ref{condbis}) below.

Here, computing the exterior square of the linear 
differential operator $L_4$, one finds
an order-six linear differential operator with 
the {\em direct sum decomposition}
\begin{eqnarray}
{\rm ext}^{(2)} (L_4) 
\,\,\, = \,\,\,\,\, \tilde{N}_1 \oplus N_5, 
\end{eqnarray}
with
\begin{eqnarray}
\fl \qquad sol(\tilde{N}_1 )\,\, \,=\,\,\,\,\, {\frac{N(x)}{D(x)}}, \\
\fl \qquad N(x) \,=\,\,\,
-12\,+2548\,x\,-502593\,{x}^{2}\, +43407720\,{x}^{3}
-1959091320\,{x}^{4}
 \nonumber \\
\fl \qquad \qquad +52738591890\,{x}^{5}-904049598675\,{x}^{6}
+10126459925120\,{x}^{7}
 \nonumber \\
\fl \qquad \qquad -74115473257440\,{x}^{8}
+350453101085400\,{x}^{9}
-1133589089074624\,{x}^{10} 
\nonumber \\
\fl \qquad \qquad +4059589860750336\,{x}^{11}
-25595376023494656\,{x}^{12}
 \nonumber \\
\fl \qquad \qquad
+141123001405931520\,{x}^{13}
  -440315308230574080\,{x}^{14}
\nonumber \\
\fl \qquad \qquad
+705909942330064896\,{x}^{15}
 -496507256028790784\,{x}^{16}
\nonumber \\
\fl \qquad \qquad
+140082179425173504\,{x}^{17}, 
 \nonumber \\
\fl \qquad D(x)\,=\,\,\,
 {x}^{9} \cdot (1-16\,x)^{13} \cdot (1-4\,x) ^{2}
\, (8-252\,x+1678\,{x}^{2}-3607\,{x}^{3}-4352\,{x}^{4}). 
\nonumber
\end{eqnarray}
This decomposition induces on the six solutions (\ref{sixExt})
of the exterior square, the following
relation (up to a constant in $sol(\tilde{N}_1 )$)
\begin{eqnarray}
\fl \quad 9701589902493\,w_{0,1}\,  +609600054750928\,w_{0,2}
+91875 \cdot 10^7 \cdot  ( w_{1,2} - w_{0,3})
\,  = \, \, sol(\tilde{N}_1 ), 
\nonumber
\end{eqnarray}
which shows the occurrence of a 
relation between the four solutions $y_0$,
$y_1$, $y_2$ and $y_3$ of $L_4$ and their first derivative.
  
\vskip .4cm

 $\, \bullet$ Along this globally nilpotent line, the $\, L_4$ 
 operator is ``more'' 
than a {\em $\, G$-operator}~\cite{Andre5,Andre6}, 
with its associated $\, G$-series. 
The series solution  (analytical at $x=0$)  $\, sol(L_4)$
is a {\em series with integer coefficients}
 in the variable $\, y \, = \, \, x/2$: 
\begin{eqnarray}
 \label{solL4}
&&sol(L_4)\, = \, \, \, 175\, +34398\,y\,
 +4017125\,{y}^{2}\, +362935156\,{y}^{3}\, 
\nonumber \\
&&\,+28020752579\,{y}^{4} +1943802285620\,{y}^{5}
+124761498220195\,{y}^{6}\, 
\nonumber \\
&&\,+7549851868859190\,{y}^{7} +436341703365296321\,{y}^{8}\, 
\nonumber \\
&&\,+24309515324321362986\,{y}^{9} +1314618756208478845353\,{y}^{10}\, 
\nonumber \\
&&\, +69377289961823319909960\,{y}^{11}
+3588051829563766082490527\,{y}^{12}
\,\nonumber \\
&&\, +182471551181260556637299032\,{y}^{13}
\nonumber \\
&&\,  +9150139649421210256395488775\,{y}^{14}
\nonumber \\
&&\, +453470079520701103056020155546\,{y}^{15} 
 \\
&&\, +22252827613097363700809754930653\,{y}^{16}
\nonumber \\
&&\, +1083008337798028206538475233669454\,{y}^{17}
\nonumber \\
&&\, +52344841647844780032111214432202429\,{y}^{18}
\nonumber \\
&&\, +2515396349437801867561610046658122428\,{y}^{19}
\nonumber \\
&&\, +120295197044047707889910797105191140059\,{y}^{20}
\nonumber \\
&&\, +5729990034986443499765238359785037134524\,{y}^{21}
\nonumber \\
&&
\, +272033605883471055363216581302378024952171\,{y}^{22}
\nonumber \\
&&\, +12879727903873470148364481804530391226578654\,{y}^{23}
\, \, + \,\cdots 
\nonumber
\end{eqnarray}

\section{Various scenarios for $\, L_4$}
\label{scenarii}

The order-four linear differential 
operator $\, L_4$ is thus slightly puzzling.
It has clearly a lot of remarkable properties but {\em cannot be reduced}
in a simple, or even in an involved way (up to rational pull-backs, up 
to operator equivalence~\cite{Homomorphisms1,Homomorphisms}
 i.e. homomorphisms,
 and, up to symmetric powers or products),
to {\em elliptic curves or modular forms}.  Is this operator going 
to be a counter-example to our favourite ``mantra''
 that the Ising model
 is nothing but the theory of elliptic curves and other modular forms ?
 
Let us display a few possible scenarios\footnote[5]{Taking into account
the previous known results on $\, L_4$, and consequently having some
overlap.}.  

\subsection{Hypergeometric functions, modular forms, mirror maps}
\label{hypermodular}

Because of the globally nilpotent character of $\, L_4$, we have some 
``hypergeometric functions'' prejudice\footnote[2]{It has been conjectured
by Dwork~\cite{Dwork} 
that globally nilpotent second order operators are necessarily associated with
hypergeometric functions. This conjecture was
 ruled out by Krammer~\cite{Bouw,Dettweiler} for some examples
 corresponding to periods of abelian surfaces 
over a Shimura curve.}. Furthermore,
 we would also like to
 see some ``renormalization group'' symmetries acting on the
 solutions, may be in a more involved way than what was described, 
in subsection (\ref{recallmodumar}) and displayed in~\cite{Renorm},
as isogenies of elliptic curves. We seem to have obstructions with 
rational pull-backs on hypergeometric
 functions. We should therefore
consider generalizations to
 {\em algebraic pull-backs}, but we expect the 
pull-backs to be ``special'' possibly corresponding to
 modular curves. Furthermore the {\em integrality} property of
 the solution series (\ref{solL4}) 
(integer coefficients for the series)
 suggests to remain  close to concepts, and structures,
 like  modular forms, theta functions (which are modular forms)
 and mirror maps~\cite{mirror,Harnad}.

Along these lines, let us recall a selected hypergeometric function
closely linked with isogenies
 of elliptic curves, modular form and
 mirror maps. Let us recall
 (see comments after formula (6.2) in~\cite{Almkvist})
 that\footnote[1]{We have studied in some 
details the renormalization 
transformation $z \, \rightarrow \, -4z/(1-z)^2$
in~\cite{Renorm}.} 
\begin{eqnarray}
&&_3F_2\Bigl([{{1} \over {2}}, {{1} \over {2}}, {{1} \over {2}}],
[1, \, 1]; \,  z  \Bigr) \\
&&\qquad \, = \, \, \, \, \, 
(1-z)^{1/2} \cdot \, \, 
_3F_2\Bigl([{{1} \over {4}}, {{3} \over {4}}, {{1} \over {2}}],[1, \, 1];
 \,  {{-4\, z} \over {(1-z)^2}}  \Bigr),
\nonumber  
\end{eqnarray}
is a {\em  modular map}~\cite{Almkvist}. 
More generally,  one has remarkable
 quadratic relations (see (6.3) in~\cite{Almkvist}),
 where the {\em  Landen transformation} 
\begin{eqnarray}
z \, \quad \longrightarrow \quad \, \, {{4 \, z} \over {(1+z)^2}} 
\end{eqnarray}
clearly pops out.

This selected hypergeometric function satisfies the 
quadratic relation\footnote[3]{Compare this
 relation with the Bailey theorem of products~\cite{Bailey}:
 $ (_2F_1([{{1} \over {2}}, {{1} \over {2}}],[1]; t))^2$
$= \, _4F_3([{{1} \over {2}}, {{1} \over {2}}, \, {{1} \over {2}}, \, 1],
[1, \, 1, \, 1]; \,4 \, t \cdot (1-t))$.}:
\begin{eqnarray}
_3F_2\Bigl([{{1} \over {2}},\,  {{1} \over {2}},\,  {{1} \over {2}}],
[1, \, 1]; \,4 \, t \cdot (1-t) \Bigr)
\,\, \, \, = \, \, \, \, \,
\Bigl(\;_2F_1\Bigl([{{1} \over {2}}, {{1} \over {2}}],[1]; t\Bigr)\Bigr)^2. 
\end{eqnarray}
which yields relation~\cite{ModularForm,BorweinBorwein} 
 ($q$ denotes the nome):
\begin{eqnarray}
\label{3F2nice}
\theta_3^4(q) 
\,\, \, = \, \, \, \, \, \, 
_3F_2\Bigl([{{1} \over {2}}, {{1} \over {2}},
 {{1} \over {2}}],[1, \, 1]; \, 4 \,
 {{ \theta_ 2^4(q) } \over {\theta_ 3^4(q)  }}
 \cdot {{ \theta_ 4^4(q) } \over {\theta_ 3^4(q)  }}   \Bigr),  
\end{eqnarray}
which is naturally associated with a very simple {\em mirror map}
involving theta functions\footnote[8]{Which are, as we know,
 modular forms of half integer weight.}:
\begin{eqnarray}
z \, = \,  z(q) \, = \, \, \, 4 \,
 {{ \theta_ 2^4(q) } \over {\theta_ 3^4(q)  }} 
\cdot {{ \theta_ 4^4(q) } \over {\theta_ 3^4(q)  }}, \quad \quad 
\theta_ 3^4(q) \, = \, \,\, 
 {{q} \over {z \cdot \sqrt{1-z} }} \cdot {{ dz} \over {dq}}. 
\end{eqnarray}
Generalizations of this kind of relations
 are certainely the kind of relations we are
seeking for, for the solutions of $\, L_4$,
 but, unfortunately, this requires
a lot of ``guessing'' (of the selected hypergeometric function,
of the mirror map, of some well-suited ratio of theta functions, ...).

\subsection{Hadamard product}
\label{hada}
\vskip .1cm
 In our ``negative'' result section (\ref{negative}) 
we saw that $\, L_4$, which is a {\em globally nilpotent} operator,
cannot be reduced to $_2F_1$ hypergeometric functions associated
 with elliptic curves (modular forms), up to transformations
{\em compatible with the  global nilpotence}, namely equivalence
 (homomorphisms) of linear differential
 operators, symmetric powers, or symmetric products
 of linear differential operators
 (even up to {\em rational} pull-backs).  
Having in mind this idea of compatibility
 with the global nilpotence,
one more operation, a ``product'' operation, can still be 
introduced, namely the {\em Hadamard product}
which {\em quite 
canonically\footnote[3]{The fact that the global nilpotence is preserved
 by the Hadamard product is a consequence of the stability of the notion of
 $G$-connection under higher direct images 
for smooth morphisms~\cite{Baldassarri}.} builds 
globally nilpotent differential operators
 from globally nilpotent elementary ``bricks''}. 

It has been seen in 
 G. Almkvist and W. Zudilin~\cite{Almkvist},
 that one can build many (Calabi-Yau) order-four operators
 from the order-two elliptic curves operators, using the
{\em Hadamard product} of series expansions. The Hadamard
 product is\footnote[1]{Deligne's formula
 (simple application of the residue formula).}, 
just a convolution~\cite{Hadamard} ($|z| \, < \, |w|\,  <\, 1$):
\begin{eqnarray}
 f*g(z) \, \, \,  = \, \, \, \, \, \,  \, 
  {{1} \over {2 \, i \, \pi}} \cdot 
 \int_{\gamma}\, f(w) \cdot g(z/w) \cdot {{dw} \over {w}}. 
\end{eqnarray}

Let us recall the order-two (elliptic curve associated) operator
\begin{eqnarray}
\label{LE}
L_E \,\,\,   = \,\,  \, \, \, \, \,  
x \cdot  (1 -x) \cdot  D_x^{2} \, \,\,
 \,   + \, (1 -x) \cdot  D_x\,  \,\, +{{ 1} \over {4}}, 
\end{eqnarray}
which has $\, EllipticE(x^{1/2})$ as solution, 
 and let us consider the {\em Hadamard product} 
of (the series expansion, at $\, x\, = \, \, 0$,  of)
 $\, EllipticE(x^{1/2})$  with itself.
This Hadamard square of (the series of) EllipticE 
is actually (the series of) 
a selected $\,\, _4F_3$ hypergeometric function
\begin{eqnarray}
\label{EE}
 &&{{2 \cdot EllipticE} \over {\pi}} \star 
 {{2 \cdot EllipticE} \over {\pi}} \\
 && \qquad \qquad \qquad \, = \, \, \,  \,  \, 
 _4F_3 \Bigl( [-1/2,1/2,1/2,-1/2], [1,1,1]; \, \, x \Bigr),
 \nonumber  
\end{eqnarray}
which is a solution of the
 {\em globally nilpotent}\footnote[8]{The Hadamard
 product of two hypergeometric series
is of course a hypergeometric series. The
 minimal operator of a $\, G$-series
 is globally nilpotent, and the Hadamard
 product of two $\, G$-series is a $\, G$-series.}
 fourth order linear operator 
that we will write $\, Had^2(L_E)$:
\begin{eqnarray}
\label{Hada}
&&Had^2(L_E)\, = \, \, \,  \, 
-1 \, \, \, -8 \cdot (x-2) \cdot  D_x \, \, \,
+8\cdot (14 \, -13\,x) \cdot  x \cdot  D_x^2 
 \nonumber  \\
&&  \quad \quad \quad \,
 +\, 96  \cdot (1-x)  \cdot  {x}^{2} \cdot D_x^{3}  \,\,
\,\,  +\, 16 \cdot (1-x)  \cdot  x^{3} \cdot D_x^{4}.  
\end{eqnarray}

The Jordan form of the $\, p$-curvature~\cite{bo-bo-ha-ma-we-ze-09}
 of the  globally nilpotent fourth order linear operator
(\ref{Hada}) {\em actually 
identifies} with the $\, 4 \times 4$ matrix (\ref{jordan}),
of characteristic and minimal polynomial  $\, T^4$. 
Such a linear differential operator {\em cannot be a symmetric cube
 of a second order} operator 
in $\, \mathbb{C}(x)[D_x]$. We can, however, certainly 
say that the  globally nilpotent fourth
 order linear differential operator
(\ref{Hada}) is
 {\em ``associated with elliptic curves''}, and we will
also say, by abuse of language\footnote[1]{
The Maple command gfun[hadamardproduct]($eq_1$, $eq_2$)
returns the ODE that annihilates the termwise product of two
holonomic power series of ODEs, 
 $eq_1$ and $eq_2$.}, that the linear 
differential operator 
(\ref{Hada}) is the Hadamard
product (at $\, x=\, 0$) of  (\ref{LE}) 
with itself, or the Hadamard square of 
the linear differential operator (\ref{LE}),
and we will write  $\, Had^2(L_E)\, = \, \, \, L_E \star \, L_E$. 
Several examples of ``Hadamard powers''
 of the complete elliptic integral $\, K$ are
given in (\ref{generAppend}). 

In our miscellaneous analysis of various
 (large order globally nilpotent)
 linear differential operators, we try to decompose these (large)
 operators into products, and ideally
 direct-sums~\cite{bo-gu-ha-je-ma-ni-ze-08,High,Khi6,ze-bo-ha-ma-05b}, 
 of factors of smaller orders.  We then try, 
in order to understand their ``very
 nature'', to see if these {\em irreducible} factors  are, 
up to equivalence of linear differential operators,
 and up to pull-backs, symmetric products
 of operators\footnote[9]{That is
 simple products of the solutions.} of smaller 
 orders. Since the Hadamard product quite naturally builds 
globally nilpotent operators from globally nilpotent ones, 
 and since it already provided examples~\cite{TablesCalabi} of 
(Calabi-Yau) order-four operators for which the corresponding
{\em mirror symmetries are generalizations of Hauptmoduls} 
(basically products of elliptic curves 
 Hauptmoduls~\cite{ModularFormTwoVar}), we can 
see the  Hadamard product as a quite {\em canonical transformation 
to add  to the
symmetric product of linear differential
 operators}\footnote[2]{For operators, not necessarily irreducible,
 this amounts to considering five ``Grothendieckian'' operations:
three products, the products of the operators, the 
products of the solutions of the operators (symmetric product),
the Hadamard product (convolution, Fourier transform), as well as 
the operator equivalence, and the substitution
 (pull-back)~\cite{Andre5,Andre6,Andre7}.}.

We will say that an {\em irreducible} differential operator
is "associated with an elliptic curve"
if it can be shown to be equivalent, {\em up to pull-backs},
to a symmetric product, or a Hadamard product, of
second order hypergeometric differential operators
corresponding to elliptic curves
(see~\cite{Renorm}).
If the differential operator is {\em factorizable}, we will say that
it is "associated with an elliptic curve", if each factor
in the factorization is.

 Is $\, L_4$ in~\cite{Khi6} an
 operator ``associated with an elliptic curve''?
This looks like a quite systematic (almost algorithmic) approach. 
In practice, it remains, unfortunately, (computionally) very 
difficult\footnote[3]{The transformations
``Homomorphisms'' and ``Hadamard product''
 mess up each other quite badly so
a Hadamard product  becomes difficult to recognize
after  homomorphisms (i.e. gauge) transformations.} to recognize
 Hadamard products {\em up to  homomorphisms} transformations.

\subsection{Calabi-Yau and $\, SP(4,\, \mathbb{C})$. 
Recalling three-fold Calabi-Yau manifolds}
\label{calsub}
We have discovered a symplectic $\, SP(4, \, \mathbb{C})$
differential Galois group for $\, L_4$.
Many order-four operators (often obtained 
by Hadamard product of second order
operators) and corresponding to Calabi-Yau ODEs, were
 found to exhibit a symplectic $\, SP(4, \, \mathbb{C})$
 differential Galois
group, to such a large 
extend that it may be tempting, for order-four operators,
 to see the occurrence of a
 $\, SP(4, \, \mathbb{C})$ differential Galois
group as a strong~\cite{SP4}
 indication\footnote[8]{In fact Calabi-Yau ODEs
{\em do not} reduce to $\, SP(4, \, \mathbb{C})$ 
 differential Galois
group.} in favour of a Calabi-Yau ODE~\cite{SP4}. On the other
 hand, one may have
the prejudice that Calabi-Yau ODEs and manifolds, which 
are well-known in string theory, have no reason to
occur in (integrable) lattice statistical mechanics.
 This is no longer true after Guttmann's 
paper\footnote[2]{See~\cite{Guttmann}
 for the Fast Track Communication.}
 which showed very clearly~\cite{GoodGuttmann}
 the emergence of Calabi-Yau ODEs in lattice statistical 
mechanics.
 
At this step, let us recall the famous 
non-trivial\footnote[5]{Elliptic curves can be seen
as the simplest examples of Calabi-Yau manifolds.}
 example~\cite{Candelas} of Candelas et al.
of (three-fold) {\em Calabi-Yau manifold}. The order-four 
linear differential operator
(in terms of the homogeneous derivative 
$\, \theta \, = \, \, z \cdot d/dz$)
\begin{eqnarray}
\label{theta5}
\theta^4 \,\, -5\, z \cdot (5\, \theta\, +1) \cdot
 (5\, \theta\, +2) \cdot (5\, \theta\, +3) \cdot (5\, \theta\, +4), 
\end{eqnarray}
has the simple hypergeometric solution:
\begin{eqnarray}
\label{solQuint}
_4F_3\Bigl([{{1} \over {5}}, {{2} \over {5}}, {{3} \over {5}},
 {{4} \over {5}}],[1, \, 1, \, 1]; \, 5^5\, z  \Bigr), 
\end{eqnarray}
which is associated with the {\em three-fold}
 Calabi-Yau manifold~\cite{Candelas,Calabi2}:
\begin{eqnarray}
\label{threefold}
x_1^5 \, +x_2^5 \, +x_3^5 \, +x_4^5 \,+x_5^5
 \, \,\, -z^{-1/5} \cdot x_1\, x_2\, x_3\, x_4 \, x_5
\, \,\, \, \, = \, \, \, \,\, \,\, 0. 
\end{eqnarray}
Actually the hypergeometric solution (\ref{solQuint})
can be written as a multiple integral
with an algebraic integrand having (\ref{threefold})
as a divisor. Being  a multiple integral
with an {\em algebraic integrand} it is, in mathematical
 language~\cite{bo-bo-ha-ma-we-ze-09} ``a {\em period}'', and,
 consequently, the associated 
order-four linear differential operator (\ref{theta5})
is necessarily~\cite{bo-bo-ha-ma-we-ze-09} globally nilpotent. 

The differential Galois group
of (\ref{theta5}) is actually
 the symplectic~\cite{SP4,Yang} group
 $\, SP(4, \, \mathbb{C})$. More precisely, the Picard-Fuchs 
linear differential operator (\ref{theta5}), 
with its solution (\ref{solQuint}) ($x=z$), reads:
\begin{eqnarray}
&&{x}^{4} \cdot (1-3125\,x) \cdot   D_x^{4} \, 
+2\,{x}^{3} \cdot (3-12500\,x) \cdot D_x^{3} \,
 \nonumber \\
&&\qquad +{x}^{2} \cdot (7-45000\,x) \cdot   D_x^{2} \, \, \, 
+\, x \cdot (1-15000\,x) \cdot  D_x \,\, \,  \, + \, 120\,x, 
\nonumber 
\end{eqnarray}
which can be written in the  form (3.9) in~\cite{Candelas}, 
 when rescaling $\, z \, = \, 5^5 \cdot x$:
\begin{eqnarray}
&&{\frac {d^{4} F(z)}{d{z}^{4}}}  \, \, \, \, 
-2\,{\frac { \left( 4\,z-3 \right)  }
{ (1-z) \cdot z }} \cdot  {\frac {d^{3} F(z)}{d{z}^{3}}}
\, \, \, \, 
-{{1} \over {5}} \, \,{\frac { \left( 72\,z-35 \right) 
 }{ (1-z) \cdot z^2 }}
 \cdot {\frac {d^{2} F(z)}{d{z}^{2}}} \nonumber \\
&&\qquad \qquad - {{1} \over {5}}\,{\frac {
 \left( 24\,z-5 \right)  }{ 
 (1-z)\cdot z^3 }}\, \cdot  {\frac {d F(z)}{dz}}
 \, \, \, \, 
-{\frac {24}{625}}\,{\frac {F(z) }
{(1-z) \cdot z^3 }}, \nonumber 
\end{eqnarray}

Along this line, some list of Calabi-Yau ODEs and Calabi-Yau
linear differential operators have been obtained~\cite{TablesCalabi} by 
 G. Almkvist et al. seeking systematically for  order-four
 differential operators obtained from Hadamard product constructions
of second order operators, often 
within a symplectic and MUM framework.
Such long, and detailed, list of  Calabi-Yau
differential operators are precious, but, again, it is 
not straightforward to see if an  order-four operator,  
like $\, L_4$, reduces to one of the  Calabi-Yau
differential operators in such lists, up to homomorphisms, 
and {\em up to pull-backs}. 

\subsection{The $_4F_3$ scenario}
\label{scenario}
As far as order-four operators that cannot be
simply reduced to elliptic curves are concerned, we already 
saw~\cite{ze-bo-ha-ma-05}, in the Ising model, 
an example corresponding~\cite{Holonomy}
to the {\em form factors}
 $\, C^{(1)}(k, \, n)$, expressed in terms of 
a  $\, _4F_{3} $ hypergeometric function:
\begin{eqnarray}
\label{bkn}
&&b(k, n) \, 
= \, \, \, \, \,\, _4F_{3} \Bigl([{{1+k+n} \over {2}},
 \,{{1+k+n} \over {2}},\,{{2+k+n} \over {2}}, 
  \,{{2+k+n} \over {2}}],\nonumber  \\
&& \qquad \qquad \qquad \qquad \quad 
[1+k, \, 1+n, \, 1+k+n]; \,  16 \, x \Bigr). 
\end{eqnarray}
It is solution of an order-four 
linear differential  operator 
which can be written, in terms of the
 homogeneous derivative $\, \theta$
 (in the usual quasi-factorized form for
  $\, _nF_{n-1} $ hypergeometric function):
\begin{eqnarray}
\label{JJ}
&& J_{k,n} \,\, = \,\, \, \, \, \,
 16 \cdot x \cdot \Bigl(\theta \, +{{1+k+n} \over {2}}\Bigr)^2 
 \cdot \Bigl(\theta + \, {{2+k+n} \over {2}}\Bigr)^2   \\ 
&& \quad  \quad \quad  \quad \quad \quad \quad  \quad \quad \,
 - \, \, (\theta \, +k) \cdot (\theta \, +n)
 \cdot (\theta \, +k\, +n) \cdot \theta. 
 \nonumber 
\end{eqnarray}

All these operators (\ref{JJ}) are, in fact,
 homomorphic (see (\ref{homomor})). The linear 
 differential 
operator (\ref{JJ}) is not MUM
(except for $k\, = \, n\, = \, 0$),
however $\, b(k, n)$ is clearly a {\em Hadamard product}
 (see (G.1) in~\cite{ze-bo-ha-ma-05}) of two algebraic functions
for $\, k$ and $\, n$ integers (or an algebraic
function and a $_3F_2$ function otherwise). 

The  {\em  exterior square} of $J_{k,n}$
is a sixth order operator which is
 invariant by 
 $\, k\,\,  \leftrightarrow \,\,  n$. Noticeably,
 this {\em exterior square}
 of $\,J_{k,n}$  has a very simple {\em rational solution}:
\begin{eqnarray}
\label{defP}
{{1 } \over {P}} 
 \quad \qquad \hbox{where:} \qquad  \qquad
P \, = \, \, \, (1-16\,x) \cdot x^{k+n\, +1},
\end{eqnarray}
which shows that one actually has a {\em symplectic structure
 when $\, k+n$ is an integer number}. Actually, performing the
 direct-sum factorization\footnote[2]{
 DFactorLCLM in Maple.} of the
 {\em exterior square} of $J_{k,n}$,
gives (when $\, k \ne \pm n$), with $\, P$ given by (\ref{defP}) 
\begin{eqnarray}
&&Ext^2(J_{k,n}) \, = \, \, \,\, \, \, 
\Omega_1^{(R)} \, \oplus  \, \, \Omega_1^{(2)} \, \oplus \, 
\Bigl(Q_2^{(1)}  \cdot  Q_2^{(2)} \Bigr),
\qquad \quad \hbox{with:} \\
&&\Omega_1^{(R)} \, = \,\,  \, D_x \, + \, \, {{d} \over {dx}} \ln(P),
 \quad 
\Omega_1^{(2)} \, = \, \,
D_x \, + \, \,
 {{d} \over {dx}} \ln\Bigl({{x^N \cdot (1-16\, x)^M} 
\over {P_{k,n}}}\Bigr),
\nonumber 
\end{eqnarray}
where $P_{k,n}$ is a polynomial, where $\, N$ and $\, M$
  are integers depending of $\, k$ and $\, n$,
and where $\, Q_2^{(1)}$ and $\, Q_2^{(2)}$  
are equivalent, and homomorphic 
to $\, Q_2^{(1)}(k=1,n=0) $
\begin{eqnarray}
 D_x^{2} \, \, \,
 +2\,{\frac { (3-64\,x) }{ (1\, -16\,x) \cdot x }}
\cdot D_x \, \, \, +2\,{\frac {3-98\,x}{(1\,  -16\,x) \cdot x^2 }}, 
\end{eqnarray}
which 
solutions can be expressed in terms of hypergeometric functions:
\begin{eqnarray}
&&{{1} \over {x^2}} \cdot \, \, 
_2F_1\Bigl([{{3} \over {2}},{{3} \over {2}}],[2]; \, 1\, -16 \, x \Bigr),
 \nonumber \\
&&{{1} \over {x^2 \cdot (1-32\, x)^{3/2}}} \cdot \, \, 
_2F_1\Bigl([{{3} \over {4}},{{5} \over {4}}],[1];
 \, {{1} \over {(1-32\, x)^2}} \Bigr).
\end{eqnarray}

The integer $\, M$ 
is equal to $\, 0$ if $\, k-n$ is even,
 and is equal to $\, 1$ if $\, k-n$ is odd,
the integer $\, N$ and the degree of
 the polynomial reading respectively:
\begin{eqnarray} 
{{3} \over {2}} \cdot (n+k) \, + {{1} \over {2}} \cdot |n-k| \, +1,
 \quad
 {{n+k} \over {2}} \, + {{|n+k|} \over {2}}  \, \,
 -{{3} \over {2}} \, \, -{{1} \over {2}} \cdot (-1)^{n-k}. 
 \nonumber
\end{eqnarray}

 The symplectic form 
of  the {\em exterior square} of $J_{k,n}$ is singular 
if, and only if, $\, k \, = \, \pm \, n$. The exterior square 
$Ext^2(J_{k,n})$
has no direct sum factorization for $\, k \, = \, \pm \, n$.
It factorises in the product of an order-one, two order-two
and an  order-one operators.

Furthermore, the function 
\begin{eqnarray}
\label{akn}
a(k, \, n) \, \,  \,  = \, \, \, \, \,  \, \, 
 {k+n\choose k} \cdot b(k, \, n),  
\end{eqnarray}
which {\em corresponds to the form factor} $\, C^{(1)}(k, \, n)$, 
has a series expansion with {\em integer coefficients}:
\begin{eqnarray}
\label{54}
&&a(k, \, n) \, \, = \, \, \, {k+n\choose k} \, \, 
+ {\frac {  (k+n+1)\, (k+n+2)^{2} }{ (n+1)  
\, (k+1) }}\cdot {k+n\choose k} \cdot x \nonumber \\
&&\quad \quad \quad \, 
+\, {{\alpha_2(k, \, n)} \over {2}}    
\cdot {k+n\choose k} \cdot {x}^{2}\,
 \, \, \,  \, 
+ \,{{\alpha_3(k, \, n)} \over {6}} 
  \cdot {k+n\choose k} \cdot {x}^{3}\, \, \,  
 +  \, \,  \cdots 
\nonumber 
\end{eqnarray}
where $\alpha_2(k, \, n)$ and $\alpha_3(k, \, n)$ read respectively:
\begin{eqnarray}
&& {\frac { (k+n+1)  \, (k+n+2) 
  \, (k+n+3)^{2} \, (k+n+4)^{2} }{  
 (k+1) \, (k+2)  \, (n+1) (n+2) }}, \nonumber \\
&&
{\frac { (k+n+1)  \, (k+n+2) 
 \, (k+n+3) \, (k+n+4)^{2} \,
 \, (k+n+5)^{2}  \, (k+n+6)^{2}  }{ 
 (k+1) \, (k+2)   \, (k+3)  \, (n+1) \, (n+2)\, (n+3) 
  }}.  \nonumber 
\end{eqnarray}

\section{$\, L_4$ is Calabi-Yau}
\label{Calabi}

\subsection{Warm-up: discovering the proper 
algebraic extension for the pull-backs}
\label{warm}

The $\mbox{}_4 F_3$ function satisfies a linear differential operator
$L_{4,3}$ with three singularities $0, 1, \infty$.

The singularities of $L_4$ at $x=\, 1/16$, and at $x=\, \infty$,
 have exponents:
integer, integer, half-integer, half-integer, 
and have only one logarithm there.
This configuration is not compatible with any of the singularities of
$L_{4,3}$ under {\em rational} pullbacks.

The singularity at $x=\,1$ of $L_{4,3}$ has exponents
$0,\,1,\,2,\,\lambda$,  where $\lambda$ depends on the parameters of the
$\mbox{}_4 F_3$ function.  The exponents
 $0,\, 1,\,2$ correspond to solutions
without logarithms.  Thus, by choosing the $\mbox{}_4 F_3$ parameters
to set $\,\lambda$ to an integer (we take $\lambda =\, 1$) we get one
logarithm at $x=\,1$.  Then the $x=\,1$ singularity of $L_{4,3}$ has
the same number of logarithms as the $x=\, 1/16$
 and $x=\,\infty$ singularities
of $L_4$.  However, under rational pullbacks there is still no match
because the exponents of $L_{4,3}$ at $x=\,1$, which are
now $0,\,1,\,1,\,2$, do not match (modulo the integers) the exponents
of $L_4$ at $x=\, 1/16$ and $x=\, \infty$.

A pullback $x \mapsto (x-a)^2$ doubles the exponents at $x=\,a$, and
likewise, a field extension of degree 2 can divide the exponents in
half.  The exponents at $x=1$ of $L_{4,3}$ must be divided in half
to match (modulo the integers) the exponents of $L_4$
 at $x=\,1/16$ and $x=\,\infty$.
So this field extension must ramify at $x=\, 1/16$ and $x=\,\infty$,
and this tells us that the field extension must be
$\mathbb{C}(x) \subset \mathbb{C}(x, \sqrt{1 - 16x})$. We
 can write this latter field
as $\mathbb{C}(\xi)$ where $\xi =\, \sqrt{1-16x}$.
With a pullback in $\mathbb{C}(\xi)$,
 the $x=\,1$ singularity of $L_{4,3}$ can
be matched with the $x= \,1/16$ and $x= \,\infty$
 singularities of $L_4$.

A necessary (but not sufficient) condition for a homomorphism between
two operators\footnote[3]{Here the two operators are: $L_4$
 and a pullback of $L_{4,3}$, both viewed as elements
 of $\mathbb{C}(\xi)[D_\xi]$}
to exist, is that the exponents of the singularities
 must match modulo the integers,
and the number of logarithms must match at every singularity.
But once one knows that the pullback for $L_4$
 must be in $\mathbb{C}(\xi)$, it
suddenly becomes easy to find a pullback that meets this necessary
condition.  Once the pullback is found, we can check if a homomorphism
exists (and if so, find it) with DEtools[Homomorphisms] in Maple. 

\subsection{The $_4F_3$ result}
\label{result}

Seeking for $\, _4F_3$ hypergeometric functions up to homomorphisms,
and assuming an algebraic pull-back
 with the {\em square root extension},
$(1\, -16 \cdot  w^2)^{1/2}$, we actually found\footnote[1]{Details
 will be given in forthcoming publications.} 
that the solution of $\, L_4$ {\em can be expressed
 in terms of a selected} 
$\, _4F_3$ which
 is precisely the {\em Hadamard product of two elliptic functions}
\begin{eqnarray}
\label{4F3good}
 &&_4F_3\Bigl( [{{1} \over {2}},\, {{1} \over {2}},\,
 {{1} \over {2}},\, {{1} \over {2}}], [1,1,1]; \, \, z  \Bigr) \\
 && \qquad \, = \, \, \, \,  \, \,
 _2F_1\Bigl( [{{1} \over {2}},\, {{1} \over {2}}],
 [1]; \, \, z  \Bigr) \star \, 
 _2F_1\Bigl( [{{1} \over {2}},\, {{1} \over {2}}],
 [1]; \, \, \, z  \Bigr),  \nonumber 
\end{eqnarray}
where the pull-back $\, z$ {\em is nothing but}
 $\, s^8$ with $\, x \, = \, w^2,$ where $ \,   w$ is the 
natural variable for the $\tilde{\chi}^{(n)}$'s
 $\, n$-fold integrals~\cite{ze-bo-ha-ma-04,ze-bo-ha-ma-05}, 
$ \,   w\,=\,\, s/(2 \, \,  (1+s^2))$: 
\begin{eqnarray}
\label{pullsquare}
z \, \,\,  = \, \,\,\,  
 \Bigl({{ 1\, + \,  (1\, -16 \cdot  w^2)^{1/2}} 
\over { 1 \, - \,  (1\, -16 \cdot  w^2)^{1/2}}}  \Bigr)^4
 \,\,\,  = \, \,\,\,  s^8. 
\end{eqnarray}

Let us recall that $\, t\, = \, \, k^2 \, = \, \, s^4$ 
and that $EllipticK(k)$ in Maple 
is the integral with a $\, k^2$ in the square root, 
so $\, t^2\, = \, \, k^4 \, = \, \, s^8$
\begin{eqnarray}
{{2} \over {\pi}}\cdot EllipticK(y) \,\,  = \, \,\, 
  _2F_1\Bigl( [{{1} \over {2}},\, {{1} \over {2}}], [1]; \, \, y^2  \Bigr)
\end{eqnarray}
therefore  the solution of $\, L_4$ is expressed in terms of 
the Hadamard square of $\, EllipticK$,
yielding in Maple notations:
\begin{eqnarray}
&&_4F_3\Bigl( [{{1} \over {2}},\, {{1} \over {2}},\, {{1} \over {2}},
\, {{1} \over {2}}], [1,1,1]; \, \, t^2  \Bigr) \nonumber \\
&& \qquad \qquad 
\, = \, \, _2F_1\Bigl( [{{1} \over {2}},\, {{1} \over {2}}], 
[1]; \, \, t^2  \Bigr) \star \, 
 _2F_1\Bigl( [{{1} \over {2}},\, {{1} \over {2}}], [1]; \, \, t^2  \Bigr) 
\nonumber \\
&& \qquad \qquad  \, = \, \, 
{{2} \over {\pi}}\cdot EllipticK(t) \star
 \,{{2} \over {\pi}}\cdot EllipticK(t),  
\end{eqnarray}
extremely similar to the previously seen Hadamard product (\ref{EE}).

\vskip .1cm
\subsection{ The Calabi-Yau result}
This result could be seen as already achieving 
the connection with {\em elliptic curves}
we were seeking for.  In fact, looking at the Calabi-Yau list 
of fourth order operators obtained
 by Almkvist et al.~\cite{TablesCalabi}, 
one discovers that this selected $_4F_3$ hypergeometric function
 {\em actually corresponds to a 
Calabi-Yau ODE}. {\em This is  Calabi-Yau ODE
 number 3 in page 10 the Almkvist et al. list.}
(see Table A of Calabi-Yau equations page 10 in~\cite{TablesCalabi}).

\vskip .2cm

{\bf Remark 1:} The hypergeometric function (\ref{4F3good}) 
also corresponds to the 
{\em hyper body-centered cubic lattice
 Green function}~\cite{Guttmann,Guttmann2}:
\begin{eqnarray}
\label{body}
&&P(0, z) \, \, \, = \, \,  \\
&&\int_0^{\pi} \, \int_0^{\pi} \, \int_0^{\pi} \, \int_0^{\pi} \, 
 {{dk_1 dk_2 dk_3 dk_4} \over {
1\, - \, z \cdot \cos(k_1) \, \cos(k_2) \, \cos(k_3) \, \cos(k_4)
}}, \nonumber 
\end{eqnarray}

{\em It may well be,
following the ideas of Christoll~\cite{Chris,Christol}, 
that the Calabi-Yau three-fold corresponding to}
 (\ref{body}), (\ref{4F3good})
similar to  (\ref{threefold}), 
{\em is nothing but the denominator of the 
integrand of} (\ref{body}),
$(1\, - \, z \cdot \cos(k_1) \, \cos(k_2) \, \cos(k_3) \, \cos(k_4))$, 
{\em written in an algebraic way} ($z_i\, = \, \, \exp(i\, k_i)$):
\begin{eqnarray}
&&_4F_3\Bigl([{{1} \over {2}}, {{1} \over {2}}, {{1} \over {2}}, {{1} \over {2}}],
[1, \, 1, \, 1]; \,  z  \Bigr) \, 
\, \, \, \, \simeq   \\
&&\int \, \int \, \int \, \int \, 
 {{dz_1 dz_2 dz_3 dz_4} \over {
8 \, z_1\,z_2 \,z_3 \, z_4 \, \, \, 
-(1\, +\,z_1^2) \cdot  (1\, +\,z_2^2) \cdot  (1\, +\,z_3^2)
 \cdot  (1\, +\,z_4^2) \cdot  z
}}. \nonumber 
\end{eqnarray}
Along this line 
\begin{eqnarray}
\label{Calabimanifold}
8 \, z_1\,z_2 \,z_3 \, z_4 \, \, \, 
-(1\, +\,z_1^2) \cdot  (1\, +\,z_2^2)
 \cdot  (1\, +\,z_3^2) \cdot  (1\, +\,z_4^2) \cdot  z
\,\, = \, \, \,\, 0, \nonumber 
\end{eqnarray}
is a genus-one curve in $(z_1, \, z_2)$
which has to be seen on the same footing
 as the three-fold Calabi-Yau manifold (\ref{threefold}).

\vskip .2cm 
{\bf Remark 2:}
Note that the series expansion of (\ref{4F3good})
 for the inverse $\, 1/z$ of the pull-back, 
is a series with {\em integer} coefficients
 in\footnote[1]{This is  not true
 in $\, z$ or $\, s$.} the variable $\, w$:
\begin{eqnarray}
&& _4F_3\Bigl( [{{1} \over {2}},\, {{1} \over {2}},\,
 {{1} \over {2}},\, {{1} \over {2}}], [1,1,1]; \, \,
  \Bigl({{ 1\, - \,  (1\, -16 \cdot  w^2)^{1/2}} \over
 { 1 \, + \,  (1\, -16 \cdot  w^2)^{1/2}}}  \Bigr)^4 \Bigr) 
 \nonumber \\
&& \qquad
\, = \, \, \, \,
1\, +16\,{w}^{8}\, +512\,{w}^{10}\, +11264\,{w}^{12}+212992\,{w}^{14} 
\nonumber \\
&& \qquad +3728656
\,{w}^{16} \, +62473216\,{w}^{18}+1019222016\,{w}^{20} \\
&& \qquad \, +16350019584\,{w}^{22}
+259416207616\,{w}^{24}\,\nonumber \\
&& \qquad +4086140395520\,{w}^{26} \, + \, \cdots 
\nonumber 
\end{eqnarray}
We also have this {\em integrality property}
 for  $\, \,\,  256 \,  z \, $:
\begin{eqnarray}
&&_4F_3\Bigl( [{{1} \over {2}},\, {{1} \over {2}},\,
 {{1} \over {2}},\, {{1} \over {2}}], [1,1,1]; \, \,  256 \, z \Bigr)  
\,\, = \, \,\, 1\,\, +16\,z\,\,  +1296\,{z}^{2}\, 
+160000\,{z}^{3} \nonumber \\
&& \qquad+24010000\,{z}^{4}\,  +4032758016\,{z}^{5} 
 +728933458176\,{z}^{6}\,
\nonumber \\
&& \qquad+138735983333376\,{z}^{7}\, +27435582641610000\,{z}^{8}
 \nonumber \\
&& \qquad 
+5588044012339360000\,{z}^{9}\, 
+1165183173971324375296\,{z}^{10}\, \nonumber \\
&& \qquad +247639903129149250277376\,{z}^{11}\,   \\
&& \qquad +53472066459540320483696896\,{z}^{12}
 \, \, \, +  \,\cdots \nonumber 
\end{eqnarray}
The solution of $\, L_4$, had been seen
to be a series 
with {\em integer} coefficients (see (\ref{solL4})).  
Now that we know that $\, L_4$ has a {\em Calabi-Yau interpretation},
 this {\em integrality property} can be seen as associated with 
 {\em mirror maps and mirror symmetries}
 (see section (\ref{mirror}) below), 
{\em as well as inherited from the Hadamard square structure} 
(see (\ref{generAppend}) below).

\subsection{Speculations:  $\, \, \, _4F_3$ generalizations and beyond}
\label{firstset}
Let us consider a few generalizations of (\ref{4F3good}), the selected 
$_4F_3$ we discovered for $\, L_4$. 

More generally, the hypergeometric  function 
\begin{eqnarray}
\label{mnpqr}
_4F_{3} \Bigl([{{1} \over {2}}\, +\, q, \,{{1} \over {2}}\,
 +\, r,\,{{1} \over {2}}\, +\, s,   \,{{1} \over {2}}\, +\, t],
[n\, +1, \, m\, +1, \, p\, +\, 1]; \, \, \, x \Bigr), 
\nonumber 
\end{eqnarray}
corresponds to the linear differential operator:
\begin{eqnarray}
\label{nmpqrst}
&&\Omega_{n,\, m, \, p; \, q,\, r, \, s, \, t} 
\, \,\, \, \, \, =  
\nonumber   \\
&&   \, \,\Bigl(\theta \, + \,{{1} \over {2}} \, +\, q\Bigr)
\cdot \Bigl(\theta \, + \,{{1} \over {2}} \, +\, r\Bigr)
\cdot \Bigl(\theta \, + \,{{1} \over {2}} \, +\, s\Bigr)\cdot 
\Bigl(\theta \, + \,{{1} \over {2}} \, +\, t\Bigr)
  \nonumber \\
&& \quad  \quad \,   \, 
- {{1} \over {x}} \cdot  \Bigl(\theta \, +n \Bigr)
\cdot \Bigl(\theta \, +m \Bigr)\cdot  
\Bigl(\theta \, +p \Bigr)\cdot \theta. 
\end{eqnarray}
For any integer $\, n, \, m, \, p, \, q, \, r, \, s, \, t$ 
all these operators (\ref{nmpqrst}) 
are actually equivalent
 (homomorphic, see (\ref{homomor})). Therefore, 
all these linear differential 
operators (\ref{nmpqrst}) are homomorphic
to  (\ref{nmpqrst}) for 
 $\, n= \, m=  \, p=  \, q=  \, r=  \, s=  \, t= \, 0$,  
which is actually a Calabi-Yau equation.

\vskip .1cm

We have also encountered another kind of 
generalization of (\ref{4F3good}): 
the Hadamard powers generalizations (\ref{2nF2nm1})
(see (\ref{generAppend})).

It is thus quite natural to consider 
the linear differential operators
corresponding to generalizations like
 \begin{eqnarray}
\label{doublegeneral}
_{n}F_{n-1}\Bigl( [{{1} \over {2}}\, +\, p_1,\,\cdots, 
 {{1} \over {2}}\, +\, p_n], [1\, +\, q_1, \,\cdots,
 1 +\, q_{n-1}]; \, \, 4^n \cdot x  \Bigr),
 \nonumber 
\end{eqnarray}
where the $\, p_i$'s and $\, q_i$'s are integers
and see, if {\em up to homomorphisms and rational
or algebraic\footnote[5]{Not too involved
 in a first approach, just 
square roots extensions.} pull-backs} 
one can try to understand the remaining
quite large order operators $\, L_{12}$ and  $\, L_{23}$, 
in such a large enough framework. There is no conceptual 
obstruction to such calculations. The obstruction 
is just the ``size'' of the corresponding massive computer calculations
necessary to achieve this goal. 

\section{Mirror maps for the Calabi-Yau 
 $\, \, _4F_3([1/2,\,1/2,\, 1/2,\,1/2], \,[1,\,1,\, 1]; \, \, 256 \, x) $}
\label{mirror}
\vskip .1cm

An irreducible linear differential equation is said to be of 
Maximal Unipotent
 Monodromy (MUM) if all the exponents at 0 are zero (one Jordan block). 
This is the case for all the hypergeometric functions
\begin{eqnarray}
_nF_{n-1} \Bigl([{{1} \over {2}}, \,{{1} \over {2}},
 \,\cdots, \, \,  \,{{1} \over {2}}],
[1, \, 1, \, \cdots, \, \,  1]; \,  4^n \, x \Bigr). 
\nonumber 
\end{eqnarray}

The hypergeometric function 
\begin{eqnarray}
\label{yes}
_4F_{3} \Bigl([{{1} \over {2}}, \,{{1} \over {2}},
\,{{1} \over {2}},   \,{{1} \over {2}}],
[1, \, 1, \, 1]; \, \, 256 \, x \Bigr), 
\nonumber 
\end{eqnarray}
which is MUM, corresponds to the fourth order linear operator 
\begin{eqnarray}
&& x^4 \cdot (1-256\,x) \cdot   D_x^{4}\, 
+2\,{x}^{3} \cdot (3-1024\,x)\cdot  D_x^{3}\, \nonumber \\
&&\qquad 
+{x}^{2}\cdot  (7-3712\,x)\cdot  D_x^{2} \,\, 
 +x \cdot  (1\, - 1280\,x) \cdot D_x\, \, \, 
-16\,x,
\nonumber 
\end{eqnarray}
or, using the homogeneous derivative $\, \theta$:
\begin{eqnarray}
\label{theta4}
\theta^4 \,\,  \, \, 
-256 \cdot x \cdot  \Bigl(\theta \, + \, {{1} \over {2}}\Bigr)^4, 
\qquad \qquad 
 \theta \,= \, \, x \cdot {{d} \over {dx}}.
\end{eqnarray}

It verifies the symplectic condition~\cite{Almkvist}:
\begin{eqnarray}
\label{condbis}
a_1 \, = \, {{1} \over {2}} \cdot a_2  \cdot a_3 \,\,
 - {{1} \over {8}} \cdot a_3^3 \, 
\,\, + \, {{d a_2} \over {dx}} \, \, \,
- {{3} \over {4}} \cdot a_3 \cdot {{d a_3} \over {dx}} \,
\, - \, {{1} \over {2}} \cdot {{d^2 a_3} \over {dx^2}},
\end{eqnarray}
for the monic order-four operator:  
$\,\,\, D_x^{4} \,+ a_3 \cdot  D_x^3 \, \, \,
  + a_2 \cdot  D_x^2 \, \, \,+ a_1 \cdot  D_x \, \, + a_ 0$. 

Condition (\ref{condbis})
 is nothing but the condition
 for the {\em vanishing of the head coefficient} of $D_x^{6}$
of this exterior square 
 (see Proposition 3 of~\cite{Almkvist}).
 The exterior square of (\ref{theta4}) is 
an {\em irreducible order-five} operator, instead of the
order-six operator one expects at first sight. 

This opens  room for a {\em non-degenerate 
alternate 2-form invariant by the
(symplectic) group $\, SP(4, \, \mathbb{C})$}. 
Actually (\ref{theta4}) has a $\, SP(4, \, \mathbb{C})$
differential Galois group.

\vskip .3cm 

{\bf Remark: }
 Let us compare this situation 
with the one for a ``similar'' order-four operator
homomorphic to  (\ref{theta4}):
\begin{eqnarray}
\label{theta4screw}
\theta^4 \,\,  \, \, 
-256 \cdot x \cdot  \Bigl(\theta \, - \, {{1} \over {2}}\Bigr)^4, 
\end{eqnarray}
The exterior square of (\ref{theta4screw}) 
is an order-six operator which is {\em the direct sum 
of an order-five operator homomorphic
to the order-five  exterior square of (\ref{theta4})}
and an order-one operator 
\begin{eqnarray}
D_x \,\,\, + {{1} \over {(1 \, - 256\, x) \cdot x}} 
\end{eqnarray}
which has the simple rational solution $\, (1\, -256\,x)/x$.

These two operators, (\ref{theta4}) and (\ref{theta4screw}), have
the {\em exact same $\, SP(4, \, \mathbb{C})$ differential Galois
group},
 but, nevertheless, their corresponding exterior squares
 do not have the same order\footnote[3]{Since the log-degree
of these operators is equal to four, the order
of these exterior squares is at least five.}. The situation for
 (\ref{theta4}) can be thought as an ``evanescence'' 
of the rational solution. 

\vskip .2cm 

\subsection{Mirror maps in a MUM framework}
\label{mirrorMUM}
The solutions of (\ref{yes}) read:
\begin{eqnarray}
&&y_0 \, = \,\,\,\, \, \, 1 \, +16\,x \,
+1296\,{x}^{2}+160000\,{x}^{3}+24010000\,{x}^{4}
+4032758016\,{x}^{5} \nonumber \\
&&\qquad +728933458176\,{x}^{6}\, \, \, + \, \, \cdots, \nonumber \\
&&y_1 \, = \,\,\, \, y_0 \cdot \ln(x)\,\,   \,  +\,   \tilde{y}_1,
 \qquad \quad \quad \, \, \hbox{with:}
\nonumber \\
&&\tilde{y}_1 \, = \, \, 64\,x\,  +6048\,{x}^{2}\, 
+{\frac {2368000}{3}}\,{x}^{3}
+{\frac {365638000}{3}}\,{x}^{4}\, 
+{\frac {104147576064}{5}}\,{x}^{5} \nonumber \\
&&\qquad +{\frac {19045884743424}{5}}\,{x}^{6} \, 
+{\frac {25588111188676608}{35}}\,{x}^{7} \, \, + \,  \cdots
 \nonumber \\
&&y_2 \, = \,\,\, \, y_0 \cdot  {{\ln(x)^2} \over {2}} \,
  \,  +\, \tilde{y}_1\cdot \ln(x)
 \,\,  \,  + \,\,\, \,   \tilde{y}_2,
 \qquad \quad \quad \hbox{with:}\nonumber \\
&&\tilde{y}_2 \, = \,\,\, \,  \,32\,x\, \,  +5832\,{x}^{2}\,
 +{\frac {8182400}{9}}\,{x}^{3}\, 
+{\frac {1374099650}{9}}\,{x}^{4}\, \nonumber \\
&&\qquad \quad  
+{\frac {685097536032}{25}}\,{x}^{5}\, \,  
+{\frac {129379065232032}{25}}\,{x}^{6}\, \, + \, \cdots
 \nonumber
\end{eqnarray}
\begin{eqnarray}
&&y_3 \, = \,\,\, \,
 y_0 \cdot  {{\ln(x)^3} \over {6}} \, \,
 + \, \, \tilde{y}_1 \cdot  {{\ln(x)^2} \over {2}} \, \, 
+ \, \, \tilde{y}_2 \cdot  \ln(x) \, \, 
+ \,  \, \tilde{y}_3, 
 \nonumber \\
&& \tilde{y}_3  \, = \,\,\,\,  -\, 64\,x\, \,  -4296 \, x^2
\, \,  -{{10334080} \over {27}} \cdot x^3
-{{1110845155} \over {27}} \cdot x^4
\,\,\,  +\,  \cdots 
\nonumber 
\end{eqnarray}

Introducing the nome~\cite{mirror,nFn} $\, q$:
\begin{eqnarray}
\label{defnome}
q\, = \, \, \,\, \exp \Bigl({{y_1} \over {y_0}}   \Bigr)  \, 
  \, \, = \, \,\, \, \,
 x \cdot \exp \Bigl({{\tilde{y}_1} \over {y_0}}    \Bigr),  
\end{eqnarray}
one finds the expansion (with {\em integer} coefficients)
 of the nome $\, q$:
\begin{eqnarray}
\label{nome}
&&q\, = \, \, \,  x \,  \, + 64 \, x^2\,
  +7072\,{x}^{3} \,+991232\,{x}^{4} \,+158784976\,{x}^{5}
 \nonumber \\
&&+27706373120\,{x}^{6}+64\,{x}^{2}+5130309889536\,{x}^{7}
 \nonumber \\
&&+992321852604416\,{x}^{8} \nonumber \\
&&+198452570147492456\,{x}^{9}+40747727123371117056\,{x}^{10}
 \nonumber \\
&&+8546896113440681326848\,{x}^{11}
+1824550864289064534212608\,{x}^{12}
 \nonumber \\
&&+395291475348616441757137536\,{x}^{13} 
\nonumber \\
&&+86723581205125308226931367936\,{x}^{14}
 \nonumber \\
&&+19233461618939530038756686458880\,{x}^{15}
 \nonumber \\
&&+4305933457394032994320115176046592\,{x}^{16} 
\nonumber \\
&&+972002126960220578680860300103711764\,{x}^{17}
 \nonumber \\
&&+221026060926103071799983313019509871872\,{x}^{18}
 \, \, \,  + \,  \cdots 
\end{eqnarray}
and, conversely, the {\em mirror map}~\cite{mirror,LianYau,Almkvist}
 reads the following series with {\em integer} 
coefficients ($x\, = \, \, z(q(x))$): 
\begin{eqnarray}
\label{mir}
&&z(q) \, = \, \, q\,\,  -64\,{q}^{2}\, +1120\,{q}^{3}\,\, 
 -38912\,{q}^{4}\, -1536464\,{q}^{5}\nonumber \\
&& -177833984\,{q}^{6}\, 
-19069001216\,{q}^{7}-2183489257472\,{q}^{8}
\nonumber \\
&& -260277863245160
\,{q}^{9}-32040256686713856\,{q}^{10}\nonumber \\
&& -4047287910219320576\,{q}^{11} \nonumber \\
&& -522186970689013088256\,{q}^{12} \nonumber \\
&& -68573970045596462152576\,{q}^{13} \nonumber \\
&& -9140875458960295169327104\,{q}^{14} \nonumber \\
&& -1234198194801672701733531648\,{q}^{15} \nonumber \\
&& -168503147864931724540942221312\,{q}^{16}
 \nonumber \\
&& -23230205873245591254063032928212\,{q}^{17}
 \nonumber \\
&& -3230146419442584387013916457526784\,{q}^{18}  
\,\,\, +\, \cdots 
\end{eqnarray}

The {\em Yukawa coupling}~\cite{mirror,Almkvist}
\begin{eqnarray}
\label{Yukawa}
K(q) \, = \, \, 
 \Bigl( q \cdot {{d} \over {dq }} \Bigr)^2
 \Bigl(  {{y_2} \over {y_0}}\Bigr), 
\end{eqnarray}
has the following (integer coefficients) series expansion: 
\begin{eqnarray}
\label{Yuka}
&& K(q) \, = \, \,
 1\, +32\,q\,\, +4896\,{q}^{2} \, +702464\,{q}^{3} 
\nonumber \\
&&+102820640\,{q}^{4}+15296748032\,{q}^{5}
 \nonumber \\
&&+2302235670528\,{q}^{6}+349438855544832\,{q}^{7}
 \nonumber \\
&&+53378019187206944\,{q}^{8}+8194222260681725696\,{q}^{9} 
\nonumber \\
&&+1262906124008518928896\,{q}^{10}\, 
+195269267971549608656896\,{q}^{11} 
\nonumber \\
&&+30273112887215918307768320\,{q}^{12}
 \nonumber \\
&&+4703886698057200436126953472\,{q}^{13}
\nonumber \\
&& +732300206865552210649383895040\,{q}^{14}
 \nonumber \\
&&+114192897568357606610746318782464\,{q}^{15}
 \nonumber \\
&&+17832557144166657247747889907477280\,{q}^{16} \\
&&+2788280197510341680209147877101177216\,{q}^{17}
 \nonumber \\
&&+436459641692984506336508940737030913792\,{q}^{18}
\,\,\,  + \, \,\, \cdots  \nonumber 
\end{eqnarray}

The nome series (\ref{nome}) has a radius of convergence 
 $\, R\, = \, \, 1/256$, corresponding 
to the $\, z \, = \, \, 1/256$
 singularity. {\em The mirror map}
 (\ref{mir}), {\em as well as the Yukawa series} (\ref{Yuka}),
 have a radius\footnote[5]{Along this line of 
radius of convergence of the mirror
map and related Schneider-Lang transcendence
 criteria, see~\cite{radius}.} of convergence 
$\, R\, \simeq \, \,0.0062794754\, \cdots \,\,  $,
corresponding to the singularity 
$\, q_s\, \simeq \, \,0.0062794754\,\, \cdots \,\,  $
given by:
\begin{eqnarray}
&&q_s \, = \, \, \,  \exp\Bigl({{x_0} \over {x_1}} \Bigr),
 \qquad \qquad \hbox{with:}  \\
&& x_0 \, = \, \,  \,  
_4F_3\Bigl( [{{1} \over {2}}, \, {{1} \over {2}},\, {{1} \over {2}},\,
 {{1} \over {2}}], [1, \, 1, \, 1]; \, 1 \Bigr), 
\qquad \qquad  \hbox{and:} \nonumber \\
&& x_1 \, = \, \, \,  \sum_{n=0}^{\infty}\, 4 \cdot 
\Gamma(n+1/2)^4\cdot (\Psi(n+1/2) \,
 -\Psi(n+1))/\Gamma(n+1)^4/\pi^2.
\nonumber 
\end{eqnarray}

Introducing the rational function\footnote[2]{Which
 corresponds to take the values
 $\, r_0 \, = \, \, 16,\, $
$ r_2 \, = \, \,384 \, = \, \,(3/2) \cdot 256 ,$
$\, r_3 \, = \, \,512 $, $\, r_4 \, = \, \,256 $, 
in~\cite{mirror}.} 
\begin{eqnarray}
q_2 \, = \,  \,  \, \,
{{1} \over {2}} \,
{\frac {327680\,{z}^{2}-1792\,z+5}{z^2 \cdot  (1-256\,z) ^{2}}}, 
\end{eqnarray}
 one finds that the mirror map (\ref{mir}) 
{\em actually verifies}
a generalization of (\ref{Jacobi}), namely
 the so-called {\em quantum deformation of 
the Schwarzian equation} (see (4.24) in~\cite{mirror}): 
\begin{eqnarray}
\label{quantum}
 {{q_2} \over {5}} \cdot z'^2 \,\,  + \, \, \{z,\tau\}
 \, \, \, \,
=  \, \,\,\,\,\,\,
 {{2 } \over {5}} \cdot {{K"} \over {K}} 
 \, - \,{{1 } \over {2}} \cdot \Bigl({{K'} \over {K}} \Bigr)^2  \,  
\end{eqnarray}
where the derivatives are  
with respect to $\, \tau$, 
the ratio of the first two solutions 
 $\, \tau \, = \, \, y_1/y_0$ (log of the nome (\ref{defnome})),
and  $\,  \{z,\, \tau \}$
 denotes the {\em Schwarzian derivative} (\ref{Schwaderiv}).  

The rhs of (\ref{quantum}) generalizes the very simple
rational function rhs we had on the  Schwarzian equation
(\ref{Jacobi}) (see the related footnote). The
 rhs of (\ref{quantum})
 is basically a transcendental function depending
 on the Yukawa coupling function (\ref{Yukawa}). 
It is natural to try to obtain a (non-linear)
ODE bearing {\em only} on the mirror map $\, z(q)$,
 and {\em not} the Yukawa coupling function
 (\ref{Yukawa}) as well. 
This can be done (see (\ref{gettinghigher}))
 with a (complexity) price to pay,
namely that these {\em higher order Schwarzian} 
(non-linear) ODEs are {\em much more involved ODEs
of much larger order}. 

\subsection{Higher order Schwarzian ODEs on the mirror map}
\label{higher}
Actually, we have also obtained the {\em higher order Schwarzian 
(non-linear) ODE} (see (4.20) in~\cite{mirror}),
verified by the {\em mirror map} (\ref{mir}). It
 is an order-seven  non-linear ODE
given by the vanishing of a polynomial with integer coefficients 
 in $\, z, \, z', \, z'', \, \cdots, z^{(7)}$,
 having  1211 monomials in
$\, z, \, z', \, z'', \, \cdots, z^{(7)}$. This 
polynomial of degree 18 in $\, z$,
24 in $\, z'$, twelve in $\, z''$, six in $\, z^{(3)}$, 
four in $\, z^{(4)}$, three in $\, z^{(5)}$, two in $\, z^{(6)}$
and one in $\, z^{(7)}$
 can be downloaded in~\cite{down}
to check that (\ref{mir}) {\em actually verifies} 
 this higher order Schwarzian (non-linear) ODE.  

One can verify that these higher order 
Schwarzian ODEs, on the mirror map,
are {\em actually compatible}
 with the (renormalization group, isogenies~\cite{Renorm}, ...)
transformations $\, q \, \rightarrow \, q^n$,
 for any integer $\, n$. 
Changing $\, q \, \rightarrow \, q^n$ in the mirror map (\ref{mir})
 \begin{eqnarray}
\label{zqn}
&&z(q^n) \, \, = \, \, \, \, \,  
{q}^{n}\, \,-64\,{q}^{2\, n}\,\, +1120\,\,{q}^{3\, n}\,
-38912\,\,{q}^{4\, n}\,-1536464\,\,{q}^{5\, n}
 \nonumber \\
&& \qquad -177833984\,\,{q}^{6\, n}\,
 -19069001216\,\,{q}^{7\, n}\,
 \,-2183489257472\,\,{q}^{8\, n}
\nonumber \\
&& \qquad  -260277863245160\,\,{q}^{9\, n}\,
\, + \, \, \cdots 
\end{eqnarray}
one finds that this new function is
 {\em still a solution of the higher order Schwarzian ODE}. 

Conversely, one can consider the {\em reciprocal}
 higher order Schwarzian ODE 
 bearing on the log of the nome, 
$\, \tau\, = \, \ln(q)\, = \, \tau(z)$,
seen as a function of $\, z$. It is
 an  order-seven {\em non-linear} ODE given by the
 sum of 602 monomial terms:
\begin{eqnarray}
\label{Schwarz4t}
&&0 \, \, = \, \, \, z^6 \cdot (1-256 \cdot z)^2 \cdot 
P_{\tau}(z, \, \tau', \, \tau'', \, \cdots \, \tau^{(5)})
 \cdot \tau^{(7)}
 \,\,\,  + \,\,  \,  \cdots  \nonumber  \\
&&\quad \, + \, \, 
(1-256 \cdot z) \cdot
 Q_{\tau}(z,\,  \tau, \cdots,  \, \tau^{(6)}, \,\tau^{(7)}) 
 \nonumber  \\
&&\quad \quad \quad 
\, + \, \,
 R_t(z,\,  \tau,  \, \tau', \cdots,   \, \tau^{(5)}) \cdot \tau',
\end{eqnarray}
where the $\tau^{(m)}$'s are the $\, m$-th $z$-derivative of 
$\, \tau(z)$,
and where $P_{\tau}$ and $R_{\tau}$ 
are polynomials of the $z$ and the 
$\tau^{(m)}$'s derivatives (see (\ref{Rt}) below). 

Let us consider the Moebius transformation
(homographic transformation) on the  log of the nome
\begin{eqnarray}
\label{Moebius}
 \tau\,\quad  \longrightarrow \quad  \quad 
 {{ a \cdot \tau \, + \, b} \over {c \cdot \tau \, + \, d}},  
\end{eqnarray}
which transforms, as far as the $z$-derivatives are concerned,
 in an increasingly involved way 
with increasing orders of derivation:
\begin{eqnarray}
\label{deduced}
&&\tau' \,\quad  \longrightarrow \quad \, \, \,  
 {\frac {ad-cb}{ (c \cdot \tau\, +d)^{2}}} \cdot \tau', \\
&&\tau'' \,\quad  \longrightarrow \quad   \, \, \,  
 {\frac { (ad-cb) }{ (c \cdot \tau\, +d)^{2}}}
 \cdot  \tau'' \, \,  \,  \, 
-2\,{\frac { \left( ad-cb \right)
 \cdot c}{ (c \cdot \tau\, +d)^{3}}} \cdot  \tau'^2, 
\quad \quad \cdots \nonumber  
\end{eqnarray}
It is a straightforward  calculation to
 verify that the higher order Schwarzian ODE (\ref{Schwarz4t})
is {\em actually invariant}
 by the  Moebius transformation (\ref{Moebius})
and its deduced transformations 
on the derivatives (\ref{deduced}).
Do note that we do not impose   $\, ad \,-cb \, \,  =\,  1$:
 we are in $\, GL(2, \, \mathbb{Z})$ {\em not} in $\, SL(2, \, \mathbb{Z})$.
The previous symmetry $\, q \,  \rightarrow \, q^n$
(see (\ref{zqn})) of the higher order Schwarzian ODE
 corresponded to 
$\, \tau \,  \rightarrow \, n \cdot \tau$. 
We  have here a  $\, GL(2, \, \mathbb{Z})$ symmetry group of the 
higher order Schwarzian ODE, corresponding to the extension of the
well-known  {\em modular group} $\, SL(2, \, \mathbb{Z})$ by 
the  {\em isogenies} (exact representation 
of the  {\em renormalization group}~\cite{Renorm}) 
$\, \tau \,  \rightarrow \, n \cdot \tau$,
 which extend quite naturally 
the  modular group and  isogenies symmetries  encountered
 with elliptic curves~\cite{Renorm}. Even
 leaving the elliptic curves
or modular forms framework, for some natural generalizations
(Calabi-Yau are natural generalizations of Hauptmoduls)
 it was crucial to get mathematical 
structures with {\em canonical  exact representation 
of the renormalization group}~\cite{Renorm}. 

\vskip .1cm
\section{Late comments: the integrability behind the mirror.}
\label{late}

It is beyond the scope of this very down-to-earth 
paper to give a mathematical definition of mirror symmetries,
since mathematicians are still seeking for the proper
general framework to define them (mixed Hodge structures, 
flat connection underlying a variation of Hodge
 structures\footnote[2]{See~\cite{Griffiths}
for the introduction of the notion of
variation of Hodge structures with
 their Gauss-Manin connections.}
 in the Calabi-Yau case~\cite{Voisin,Morrison2,Gross},
T-duality~\cite{StromingerT}, 
 toric frameworks like in Batyrev's construction 
of mirror symmetry between
 hypersurfaces of toric Fano\footnote[3]{Mirror symmetry, 
in a class of models of toric varieties with zero first
 Chern class Calabi-Yau manifolds and positive first
 Chern class (Fano varieties) was proven by
 K. Hori and C. Vafa~\cite{Vafa}.} varieties~\cite{Batyrev},
{\em algebraic} Gauss-Manin connections~\cite{KO}, ...).

More familiar to physicists, in particular specialists
of integrability, is the notion of
 {\em Picard-Fuchs}~\cite{Picard-Fuchs} 
{\em equation}\footnote[1]{It is known that if 
a linear differential equation with coefficients in $\, \mathbb{Q}$
is of Picard-Fuchs type, then it also describes an abstract variation of
$\, \mathbb{Q}$-Hodge structures, and it 
is globally nilpotent~\cite{Golyshev}.}.
Along this  Picard-Fuchs line, 
the occurrence of Painlev\'e VI equations is well-known 
for the Ising model~\cite{Painleve,Fuchs} (see the Garnier
or Schlesinger systems~\cite{Schlesinger}). This Picard-Fuchs
notion is central in any "intuitive"
 understanding of mirror maps~\cite{Morrison}
and other Calabi-Yau manifolds (namely compact K\"ahler
 manifolds with  Ricci-flat K\"ahler metrics).

As far as the ``proper integrable framework'' of this paper is
concerned let us underline the following comments. 
The 2-D Ising model is a well-known free-fermion model with an elliptic 
parametrization. This elliptic parametrization is, of course, a straight
consequence of the Yang-Baxter integrability of the model
(here the star-triangle relation), and, therefore, it is not a surprise
to see elliptic functions in the integrals of correlation functions
(see for instance~\cite{Yamada,Martinez,Martinez2}).
Along this line, even Painlev\'e VI equations can be seen as a Gauss-Manin 
deformation of an elliptic function second 
order ODE~\cite{Painleve,Fuchs,Manin}. 

However, it is crucial to note that the elliptic parametrization
is not one-to-one related with a Yang-Baxter integrability: 
the sixteen vertex model, which, in general, is {\em not}
Yang-Baxter integrable, has a {\em canonical} (compulsary !) {\em elliptic
parametrization}~\cite{sixteen}, the elliptic
parametrization being a consequence of the  {\em integrability
of the  birational symmetries}~\cite{Baxterization}
 of the model, and of course not 
of a Yang-Baxter integrability that does not exist generically
for that very model.
The free-fermion character of the square
 Ising model is of course crucial in 
Wu et al.~\cite{wu-mc-tr-ba-76} (Pfaffian, Toeplitz, ...) 
calculations to write explicitly the $\chi^{(n)}$'s as integrals
of some integrand {\em algebraic} in some well-suited variables. 
However, Guttmann's paper~\cite{GoodGuttmann} makes crystal clear 
with miscellaneous examples of Green functions for many lattice statistical
mechanics, or enumerative combinatorics, problems in
 {\em arbitrary} lattice dimensions, that Calabi-Yau ODEs 
emerge in a lattice statistical mechanics framework which is
(at first sight)  {\em quite
remote from Yang-Baxter (tetrahedron, ...) integrability}, and
 even more {\em from free-fermion integrability}. 

If the occurrence of linear differential operators
{\em associated with elliptic curves} for square Ising
correlation functions, or form factors, is not 
a surprise~\cite{Holonomy},
the kind of linear differential operators that 
should emerge in quite involved {\em highly composite}
objects like the $n$-particle components $\chi ^{(n)}$
of the susceptibility of the square Ising model,
is far from clear. We just had a prejudice that they should be "special" 
and could possibly be  associated with elliptic curves. 
In fact, even if Yang-Baxter structures were the ``deus ex machina'' 
behind these "special" linear differential operators 
it seems impossible to use that property in any explicit calculation.

Moving away from Yang-Baxter integrability to 
other concepts of "integrability", we are actually using the
following ingredients: we have $n$-fold integrals   
of an integrand which is {\em algebraic} in the variables
of integration and in the other remaining variables.
 {\em This algebraicity is the crucial point}. As a consequence 
we know that these $n$-fold integrals can be interpreted as {\em "Periods"
of algebraic varieties} and verify globally nilpotent~\cite{bo-bo-ha-ma-we-ze-09}
 linear differential equations: they are~\cite{Andre5,Andre6,Andre7}
 "{\em Derived From Geometry}". 
However, inside this "Geometry" 
framework~\cite{Morrisson3} (in the sense of the
 mathematicians) theoretical physicists
 are exploring\footnote[3]{Sometimes,
without knowing it, like Monsieur Jourdain
 (Le bougeois gentilhomme).}  "Special Geometries". These
linear differential operators factorize in irreducible
operators that are also
necessarily globally nilpotent~\cite{bo-bo-ha-ma-we-ze-09}.
When one considers all the irreducible
globally nilpotent linear differential operators
of order $N$, that we have encountered,
(or the one's displayed by other authors
in an enumerative combinatorics 
framework~\cite{Guttmann,GoodGuttmann}, 
or in a more obvious Calabi-Yau framework~\cite{Batyrev,TablesCalabi}),
it apperas that their differential Galois group are not 
the $SL(N,\, \mathbb{C})$ or extensions of
 $SL(N,\, \mathbb{C})$, groups one could expect 
generically, but {\em selected} $\, SO(N)$,
 $\, SP(N,\, \mathbb{C})$, $\, G_2$, ...
differential Galois group~\cite{Katz}. These are, typically, 
classification problems in algebraic geometry and/or\footnote[8]{
Given a classification problem in algebraic geometry, using 
the mirror duality one can translate it into
 a problem in differential equations, solve this problem 
and translate the result back into geometry.} differential
geometry. 
Our linear differential operators are, in fact,
{\em "special" globally nilpotent} 
operators ($G$-operators). This paper can be seen as an attempt, through 
a fundamental model, the Ising model, to
try to characterise the ``additional structures and properties''
of these globally nilpotent operators. In the simple example
of $\, _pF_q$ generalised hypergeometric functions, only 
operators with $\, _nF_{n-1}$ solutions~\cite{nFn}
can be globally nilpotent. In a hypergeometric framework
 we are thus trying to see the emergence of
``special'' $\, _nF_{n-1}$ hypergeometric functions.

The last results, displayed in this paper,  
show clearly, with the $SP(4, \,\mathbb{C})$ differential Galois group
of $\, L_4$ for $\chi^{(6)}$,
 that these "special geometries" {\em already emerge}
 on the Ising model which, therefore, {\em does not restrict to the theory
of elliptic curves}~\cite{Kean}
 (and their associated elliptic functions and modular forms).
To define  these "special geometries" 
is still a work in progress\footnote[1]{For 
 the next $\chi^{(n)}$'s, $n \, \ge \, 7$, we just know that the 
corresponding differential operator are globally nilpotent. We
 can only conjecture the emergence of these ``special geometries'', 
as we already conjectured the integrality of the $\chi^{(n)}$'s
series in well-suited variables
 (see equation (8) in~\cite{High}).} but it
seems to be close to concepts like the concept of modularity 
(for instance, integrality of series)
 and other mirror symmetries~\cite{Lectures}.

\vskip .1cm
\section{Conclusion}
\label{conclu}
\vskip .1cm

All the massive calculations we have developped during several
years~\cite{bo-gu-ha-je-ma-ni-ze-08,High,bernie2010,Khi6,ze-bo-ha-ma-04,bo-ha-ma-ze-07b} 
 on the square lattice Ising model
give coherent exact results that do show a lot of remarkable and deep 
(algebraico-differential) structures. These structures
 all underline the {\em deep connection
between the analysis of the Ising model
 and the theory of elliptic functions}
(modular forms, selected 
hypergeometric functions~\cite{Renorm}, modular curves, ...).
In particular we have actually been able to understand 
almost all the factors obtained in the analysis of the
$\, {\tilde \chi}^{(n)}$'s, as linear differential operators
 ``{\em associated with elliptic curves}''. 
Some  linear differential operators have a very straightforward relation
 with elliptic curves: they are
 homomorphic to symmetric powers of $\, L_E$
 or $L_K$, the second order operators corresponding to 
 complete elliptic integrals $\, E$ or $\, K$. 
We showed, in this paper, that the solutions
of the second and third order 
operators 
$\, Z_2$, $\, F_2$, $\, F_3$,
  $\, {\tilde L}_3$ operators,
 {\em can actually\footnote[5]{At first sight it looks like a simple
problem that could be solved using utilities like the
 ``kovacicsols'' command~\cite{JAW} in maple13: it is not
even simple on a second order operator.} be interpreted as
 modular forms} of the elliptic curve of the Ising model.
These results are already quite a ``tour-de-force'' and 
their generalization to the much larger (and involved)
operators  $\, L_{12}$ and $\, L_{23}$, seems out of reach 
for some time. The understanding of the ``very nature''
of the globally nilpotent fourth order operator $\, L_4$
was, thus, clearly a very important
 challenge {\em to really understand the 
mathematical nature of the 
Ising model}. This has been achieved 
with the emergence of a {\em Calabi-Yau
equation}, corresponding to a selected $_4F_3$ hypergeometric function,
 which can also be seen as a {\em Hadamard product}  
of the complete elliptic integral\footnote[2]{Or the Hadamard product 
of $\, (1\, -\, 16 \, w^2)^{-1/2}$, the 
square root $\, (1\, -\, 16 \, w^2)^{1/2}$
 playing a fundamental role in the
 algebraic extension necessary to discover
the good pull-back (\ref{pullsquare}), which 
is a crucial step to find the solution (see (\ref{Calabi})). 
} $\, K$,
with a remarkably simple algebraic pull-back
 (square root extension (\ref{pullsquare})), 
 the corresponding Calabi-Yau fourth-order operator having 
 a {\em symplectic} differential Galois group
 $\, SP(4, \, \mathbb{C})$. The associated 
{\em mirror maps and higher order Schwarzian ODEs} present all 
 the nice physical and mathematical ingredients we had 
 with elliptic curves and modular forms, in particular 
 an exact (isogenies) representation of the generators of the 
renormalization group, extending the 
modular group $SL(2, \, \mathbb{Z})$ 
to a $GL(2, \, \mathbb{Z})$ symmetry group.

We are extremely close to achieve our journey
 ``from Onsager to Wiles'' (and now Calabi-Yau ...),
 where we will, finally, be able to say that the {\em Ising model is
 nothing but the theory of elliptic curves, modular forms and other
mirror maps and Calabi-Yau}. Do note that all the ideas, 
displayed here, {\em are
 not specific of the Ising
model}\footnote[8]{See~\cite{GoodGuttmann}.
We use the closed formulae for the $\,\chi^{(n)}$'s
as $n$-fold integrals with {\em algebraic integrands}, 
derived from Pfaffian methods~\cite{wu-mc-tr-ba-76}, 
but not directly the free-fermion character 
 of the Ising model.} and can be generalized to most of the problems 
 occurring in exact lattice statistical mechanics, enumerative
 combinatorics~\cite{Guttmann,Guttmann2}, particle physics,
 ...,  (the elliptic curves
 being replaced by more general algebraic varieties, and
the Hauptmoduls being
 replaced by the corresponding mirror symmetries
 generalizations~\cite{ModularFormTwoVar}). 

We do hope that these ideas will, eventually yield the emergence
 of a new Algebraic Statistical Mechanics classifying all the 
problems of theoretical physics\footnote[9]{Beyond lattice
statistical mechanics~\cite{Moore1,Moore2,Strominger}.}
 on a completely
 (effective) algebraic geometry basis. 

\ack
We would like to thank  Y. Andr\'e for
providing generously written notes on the compatibility
between $\, G$-operators and 
 Hadamard-product of operators.
We thank  D. Bertrand, A. Enge,
 M. Hindry,  D. Loeffler,  J. Nekovar, J. Oesterl\'e, 
 M. Watkins, for fruitfull discussions on modular curves,
modular forms and modular functions, D. Bertrand, L. Di Vizio, 
 for fruitfull discussions on differential Galois groups, C. Voisin
for fruitfull discussions on mirror symmetries,
and G. Moore
for stimulating exchanges on arithmetics and
complex multiplication. One of us (JMM)  thanks
 the Isaac Newton Institute 
and the Simons Center where part of this
work has been initiated, as well as the
 Center of Excellence in Melbourne 
for kind support. Alin Bostan was supported in part
 by the Microsoft Research-Inria Joint Centre.
As far as physicists authors 
are concerned, this work has been performed without
 any support
 of the ANR, the ERC or the MAE. 

\appendix

\section{Hadamard product of operators
 depends on the expansion point}
\label{Hadprod}
The Hadamard product of two operators depends on the
point around which the series expansions are performed,
 and, hence, the Hadamard 
product of the series-solutions.

Let us consider the complete elliptic integral $\, K$
which is already a Hadamard product
\begin{eqnarray}
{{2} \over {\pi}} \cdot EllipticK(4\cdot x^{1/2})
\,  \, \,\,=  \, \,  \,\, \,\,
  (1 \,-4 \,x)^{-1/2} \star (1 \,-4 \,x)^{-1/2}. 
\end{eqnarray}
and the order-two linear differential 
operator for $\,\, EllipticK(x^{1/2})$
\begin{eqnarray}
D_x^2 \,\, \, +{\frac { (1- 2\,x) }{(1-x) \cdot x}} \cdot D_x
 \,\, \,
-\, {{1} \over {4}}\,{\frac {1}{ (1 -x) \cdot x  }}, 
\end{eqnarray}

The Hadamard square of a linear differential operator
 {\em does depend on the point around which the
 series  are performed}. 
For $EllipticK$ with its three singularities $0, \, 1, \, \infty$, 
 the Hadamard square yields the {\em same} operator of order-four 
for the three expansion points $0, \, 1, \, \infty$, this
order four operator corresponding to 
$_4F_3\Bigl( [{{1} \over {2}},\, {{1} \over {2}},\,
 {{1} \over {2}},\, {{1} \over {2}}], [1,1,1]; \, \, z  \Bigr)$,
 namely:
\begin{eqnarray}
\label{orderfour}
&&D_x^{4}\,  \,\,  
 +2\,{\frac { (3 \, - 4\,x) }{(1-x) \cdot x }} \cdot D_x^3 \,\,
\, \, +\, {{1} \over {2}} \,
{\frac { (14-29\,x)}{ (1-x) \cdot x^2}} \cdot D_x^2\, 
\,  \nonumber  \\
&&\qquad +\, {\frac { (1 \, - 5\,x) }{(1-x) \cdot x^3 }}
 \cdot D_x
\,\, \, \,
 -{{1} \over {16}}\,{\frac {1}{(1 \, -x)\cdot x^3  }}, 
\end{eqnarray}
for the three expansion points $0, \, 1, \, \infty$.
However, for a generic expansion point, $\, x= \, c$
 (where $\, c \, \ne \, 0, \,1/2, \,  1, \, \infty$)
 one gets an {\em order-six} linear differential  operator.

\subsection{Hadamard powers generalizations}
\label{generAppend}

Let us denote
 $\, Had^n(F)\, = \, \, F \, \star F \, \star\, \cdots \, \star \,F$,
 the Hadamard product of $\, F$,  $\, n$-th time with itself. 
Relation (\ref{4F3good}) can straightforwardly 
be generalized to arbitrary Hadamard powers
of the complete elliptic integral $\, K$,
which, as we know, plays a crucial role in our analysis of 
the Ising model~\cite{Painleve,Fuchs}:
\begin{eqnarray}
\label{2nF2nm1}
&&_{2\, n}F_{2\, n-1}\Bigl( [{{1} \over {2}},\,\cdots, 
 {{1} \over {2}}], [1, \,\cdots, 1]; \, \, 16^n \cdot z  \Bigr) \\
&& \qquad \qquad \qquad \, \,  = \, \,\,  \, \, \, \, 
Had^n\Bigl(_2F_1\Bigl( [{{1} \over {2}},\, {{1} \over {2}}], [1];
 \, \, 16 \cdot z  \Bigr) \Bigr).
 \nonumber 
\end{eqnarray}
These hypergeometric functions~\cite{Goursat} are solutions 
of the $\, 2n$-th order linear differential operator:
\begin{eqnarray}
\theta^{2\, n} \, - \, \, 16^n \cdot z \cdot
 \Bigl(\theta \, +  \, {{1} \over {2}}  \Bigr)^{2\, n}.
\end{eqnarray}
These relations are a subcase of the (slightly)
 more general  Hadamard power
relations 
\begin{eqnarray}
\label{nFnm1}
_{n}F_{n-1}\Bigl( [{{1} \over {2}},\,\cdots, 
 {{1} \over {2}}], [1, \,\cdots, 1]; \, \, 4^n \cdot z  \Bigr)
\, \,  = \, \,\,  \, \, \, \, 
Had^n\Bigl(  {\frac {1}{\sqrt {1-4\,z}}}  \Bigr),
 \nonumber 
\end{eqnarray}
these last hypergeometric functions being solutions 
of the $\, n$-th order  linear differential operator:
\begin{eqnarray}
\theta^{ n} \, - \, \, 4^n \cdot z \cdot
 \Bigl(\theta \, +  \, {{1} \over {2}}  \Bigr)^n.
\end{eqnarray}
Do note that the corresponding
$\,_{(n)}F_{(n-1)}$ hypergeometric
 series are, actually, series with {\em integer coefficients}: 
\begin{eqnarray}
&&Had^n\Bigl(  {\frac {1}{\sqrt {1-4\,z}}}  \Bigr)
\,   \, \, =   \, \, \, \, \,
1\,  + \, \,  
\sum_{k=1}^{\infty} \, \Bigl(2 \cdot {2\, k-1\choose k-1} \Bigr)^n   \cdot z^k 
 \, \, \,  \, \, =  \nonumber \\
&& \qquad 1\, +2^n \cdot z\, +6^n \cdot {z}^{2}\, +20^n \cdot {z}^{3}\,
 +70^n \cdot {z}^{4}\, +252^n \cdot {z}^{5}\,
 +924^n \cdot {z}^{6}\,\nonumber \\
&& \qquad \qquad  +3432^n  \cdot{z}^{7} \,\, +12870^n \cdot z^8 
\,\,+ 48620^n \cdot z^9
 \, \, \, + \, \, \cdots  
\end{eqnarray}

\section{Equivalence of the $ J_{k,n}$; $\, \, \, $
Equivalence of the $\Omega_{n,\, m, \, p; \, q,\, r, \, s, \, t}$}
\label{homomor}
$\, \bullet$ 
All the operators $\, J_{k,n}$
defined by  (\ref{JJ}) are
 homomorphic. This can be seen recursively from the
 two operator equivalences
\begin{eqnarray}
\label{homoJJ}
 {\cal U}_1 \cdot J_{k,n} \,\, = \,\, \, \, \, \, J_{k,n+1} \cdot {\cal U}, 
\qquad \qquad  
{\cal V}_1 \cdot J_{k,n} \,\, = \,\, \, \, \, \,  J_{k+1,n} \cdot {\cal V}, 
 \nonumber 
\end{eqnarray}
where 
\begin{eqnarray}
&&{\cal U}  \,\, = \,\, \, \, \, 
{{4\, n} \over {x}} \cdot (1-16\,x)  \cdot \theta^3 \nonumber \\
&&\, -\, {{1} \over {x}} \cdot
(16\, (8\,n +6\,kn +{k}^{2} +5\,{n}^{2}) \cdot x\,
 - \, (k+5\,n) \, (k+n))  \cdot \theta^2
\nonumber \\
&&\, -\,  (16\, (k+n+1)  \, ({k}^{2}+3\,kn+5\,n+2\,{n}^{2}) \cdot  \theta\, \\
&&+{{k} \over {x}} \cdot (5\,{n}^{2} +2\,kn +{k}^{2})) \cdot \theta
\, -4 \cdot  ({k}^{2} +2\,kn +4\,n +{n}^{2})  \, (k+n+1)^{2}, 
\nonumber \\
&&{\cal U}_1  \, = \,\,\,  \,{\cal U} \,  \,\, \,
+{{4 \, n} \over {x}} \cdot (1\, -32 \, x)\cdot  \theta^2 \,
 \nonumber \\
&& \qquad \qquad 
\, -{{4} \over {x}} \cdot
 (8 \, (4\,kn+8\,n+3\,{n}^{2}+{k}^{2} )\, - k\, n)  \cdot \theta
\nonumber \\
&& \qquad \qquad -16\cdot  (k+n+2)  \, ({k}^{2} +2\,kn +4\,n +{n}^{2})
\end{eqnarray}
and where $\, {\cal V}$ (resp. ${\cal V}_1$) is
 $\, {\cal U}$ (resp. ${\cal U}_1$)
where $k$ and $\, n$ 
have been permuted. 

\vskip .2cm 

\vskip .2cm 

$\, \bullet$ Let us now show that the order-four
 linear differential operators
$\Omega_{n,\, m, \, p; \, q,\, r, \, s, \, t}$, 
corresponding to the hypergeometric  functions 
(\ref{mnpqr}),  are homomorphic.

 One has:
\begin{eqnarray}
&& \Omega_{n,\, m, \, p; \, q,\, r, \, s, \, t}
 \cdot  \Bigl(\theta\, +\, n \, +1 \Bigr)
\,  \,  \,\,= \,\, \,  \,\,
\Bigl(\theta\, +\, n \, +1 \Bigr) \cdot 
 \Omega_{n+1,\, m, \, p; \, q,\, r, \, s, \, t}, 
\nonumber   \\
&& \Omega_{n,\, m, \, p; \, q,\, r, \, s, \, t} \cdot
  \Bigl(\theta\, +\, m \, +1 \Bigr) 
\, \, = \, \, \,\,\,
 \Bigl(\theta\, +\, m \, +1 \Bigr) \cdot 
 \Omega_{n,\, m+1, \, p; \, q,\, r, \, s, \, t}, 
 \nonumber   \\
&& \Omega_{n,\, m, \, p; \, q,\, r, \, s, \, t} \cdot
  \Bigl(\theta\, +\, p \, +1 \Bigr)
\,  \,\, \, = \, \, \, \,
\,\Bigl(\theta\, +\, p \, +1 \Bigr) \cdot 
 \Omega_{n,\, m, \, p+1; \, q,\, r, \, s, \, t}.
  \nonumber  
\end{eqnarray}
and:
\begin{eqnarray}
&&\Bigl(\theta\, +\, q \, +{{3} \over {2}} \Bigr) 
 \cdot \Omega_{n,\, m, \, p; \, q,\, r, \, s, \, t} 
\,  \, \, \,= \,\,  \,\, \, \Omega_{n,\, m, \, p; \, q+1,\, r, \, s, \, t}
\cdot \Bigl(\theta\, +\, q \, +{{1} \over {2}} \Bigr),
 \nonumber   \\
&&  \Bigl(\theta\, +\, r \, +{{3} \over {2}} \Bigr) 
 \cdot \Omega_{n,\, m, \, p; \, q,\, r, \, s, \, t} 
\,  \, \, \,= \,\,  \, \,\,
 \Omega_{n,\, m, \, p; \, q,\, r+1, \, s, \, t}
\cdot \Bigl(\theta\, +\, r \, +{{1} \over {2}} \Bigr),
 \nonumber   \\
&& \Bigl(\theta\, +\, s \, +{{3} \over {2}} \Bigr) 
 \cdot \Omega_{n,\, m, \, p; \, q,\, r, \, s, \, t} 
\,  \, \, \,= \,\,  \, \,\,
 \Omega_{n,\, m, \, p; \, q,\, r, \, s+1, \, t}
\cdot \Bigl(\theta\, +\, s \, +{{1} \over {2}} \Bigr),
 \nonumber   \\
&&\Bigl(\theta\, +\, t \, +{{3} \over {2}} \Bigr) 
 \cdot \Omega_{n,\, m, \, p; \, q,\, r, \, s, \, t}
 \,  \, \, \,= \,\,  \, \,\,
 \Omega_{n,\, m, \, p; \, q,\, r, \, s, \, t+1}
\cdot \Bigl(\theta\, +\, t \, +{{1} \over {2}} \Bigr),
 \nonumber
\end{eqnarray}
A simple composition of all these elementary relations
show, by recursion, that the $\Omega_{n,\, m, \, p; \, q,\, r, \, s, \, t}$'s
 are all homomorphic.
\vskip .3cm 

\section{Getting the higher order Schwarzian ODEs}
\label{gettinghigher}

Let us call ${\cal L}_4$ the order four
linear  differential operator corresponding to
the Fuchsian ODE of the hypergeometric function
\begin{eqnarray}
 _4F_3([1/2,\,1/2,\, 1/2,\,1/2], \,[1,\,1,\, 1]; \, 256\,x ). 
\end{eqnarray}
The formal solutions of ${\cal L}_4$ are denoted
 $y_3$, $y_2$, $y_1$ and $y_0$,
where the subscript is for the higher exponent of the log.
The nome map reads (see (\ref{nome})) 
\begin{eqnarray}
q(x) \, = \,\,\,\, \exp \Bigl( {\frac{y_1}{y_0}} \Bigr)
\,\, \,=\,\,\,\,\,\,
x\,\, +64\,{x}^{2}\,\, +7072\,{x}^{3}\, +991232\,{x}^{4} \, 
\, \,\, +\,\, \cdots  \nonumber 
\end{eqnarray}
Our aim is to obtain the non linear ODE of $q(x)$.
 
Considering the log nome map as $y_1/y_0 = \, \ln(q)$,
 and, differentiating both sides, gives
\begin{eqnarray}
y_1' \, y_0\, - y_1 \, y_0'
 \,\,\,\, =\,\,\,\,\,\,
 y_0^2 \cdot  {\frac{q'}{q}}
\end{eqnarray}
where the left-hand side is the wronskian
 of the solutions $y_1$ and $y_0$.
This wronskian is solution of an order-five 
linear differential operator $ {\cal L}_5$,
which is the exterior square of $\, {\cal L}_4$.
We have then three equations
\begin{eqnarray}
{\cal L}_5 \Bigl( y_0^2 \cdot  {\frac{q'}{q}} \Bigr)
 \,\,  =  \, \, \, 0, 
\quad \quad 
{\cal L}_4 ( y_0 \cdot  \ln(q)) \,\,  =  \, \,\,  0, 
\quad \quad
{\cal L}_4 (y_0) \, = \,  \,  0,
\end{eqnarray}
to which we add the following two equations obtained by
 differentiating the last
two equations
\begin{eqnarray}
D_x \cdot {\cal L}_4 \, (y_0 \cdot \ln(q))  \, = \, \,   0,
 \quad \quad \quad \quad
D_x \cdot {\cal L}_4 \, (y_0)  \, = \, \,   0.
\end{eqnarray}
We solve this system of five equations in the unknowns
 $F^{(n)}\, =\,\, d^n y_0 /dx^n$,
to obtain
\begin{eqnarray}
F^{(n)} \,=\,  \, B_n \cdot F^{(0)}, 
\quad \quad \quad \quad
 n= \, 1, \,  2,  \, \cdots, \,  5
\end{eqnarray}
The $B_n$'s depend on $x$, $q(x)$ and its derivatives up to seven.
All these relations should be compatible, 
by which it is meant that the derivative
of one relation with $n$ gives the relation with $n+1$.
The result is then
\begin{eqnarray}
{\frac{d}{dx}} B_n\,\, + B_n \cdot B_1\,\, -B_{n+1} 
\, \, = \, \, \,  0, \qquad \quad
 n\, = \, \, 1, \,  2, \cdots, \,  4.
\end{eqnarray}
The gcd of these four nonlinear differential equation 
is the ODE of $q(x)$.
The nonlinear ODE of $q(x)$ contains the derivatives
$q'(x), \cdots, q^{(7)}(x)$ with degrees, respectively,
12, 16, 8, 5, 4, 3, 2 and 1.
The nonlinear differential equation 
of the mirror map (see (\ref{mir}))
\begin{eqnarray}
X(q) \,=\,\,\,\,
q\,\,\, -64\,{q}^{2}\, +1120\,{q}^{3}\, -38912\,{q}^{4}\, -1536464\,{q}^{5}\, 
 \, \,\, + \, \cdots  \nonumber
\end{eqnarray}
can be obtained by using $X \left( q(x) \right)= \, x$ and the ODE
of $q(x)$.
The nonlinear ODE of $X(q)$ involves the derivatives
$X^{(0)}(q), X^{(1)}(q), \cdots, X^{(7)}(q)$ with degrees, respectively,
18, 24, 12, 6, 4, 3, 2 and 1.

\section{Higher order Schwarzian ODEs 
for the hypergeometric function
$\,  _4F_3([1/2,\,1/2,\, 1/2,\,1/2], \,[1,\,1,\, 1]; \, 256 \, z) $}
\label{higherApp}

Actually the non-linear ODE on the mirror map (\ref{mir})
 is of the form 
\begin{eqnarray}
\label{Schwarz4zA}
&&0 \, \,  = \,  \, \, (1-256 \cdot z) \cdot 
S(z, \, z', \, z'' \, \cdots \, z^{(5)},\,  z^{(6)},\,  z^{(7)})
\nonumber \\
&& \qquad 
 \,\, + \, \,  z'^{12}  \cdot P(z, \, z',\cdots \, z^{(5)}), \\
&&P(z', \, z \, \cdots \, z^{(5)}) 
\,\, = \,\, \, -18\,z'^{3} \cdot ( 188416\,z'^{4} \, 
+6\,z'\, z^{(3)} \, -9\, z''^{2})\cdot z^{(5)} \,
\nonumber \\
&& \quad +1909301941633024\,z'^{12}  \, 
\, -695516266496\,z'^{9}\, z^{(3)} \, \nonumber \\
&& \quad +12582912\, \left( 82912\, z''^{2}\, 
-345 \, z^{(4)} \right)\cdot z'^{8} \, \, 
  -4\,(z^{(3)})^{2}) \, z'^{3} \,
 \nonumber \\
&& \quad  +26046627840\,
z'^{7}\, z'' \, z^{(3)}\, 
+2430\, z'\, z''^{4}\, z^{(3)}\,
 \nonumber \\
&& \quad -33914880\, 
(768\, z''^{3}\, -\, (z^{(3)})^{2}\, -\, z^{(4)} \,z'') \cdot z'^{6}\,
 \nonumber \\
&& \quad
-203489280\, z'^{5}\, z''^{2} \, z^{(3)} \, 
 +135\, (1130496\, z''^{4}\, +\, (z^{(4)})^{2})\, z'^{4}\,
\nonumber \\
&& \quad -180\,\, z^{(3)}\cdot  (3\,z^{(4)}\,z" \, 
 -1620\,z'^{2}\, z''^{2}\, (z^{(3)})^{2}\,
 -1215\, z''^{6}.
\nonumber 
\end{eqnarray}
This is a non-linear ODE where the derivatives $\, z^{(n)}$ 
are the $\, n$-th derivatives in $\, \tau  \, = \, \ln(q)$ 
of the mirror map $\, z(q)$.
 
Conversely, the ODE on the function
 $\, \tau \, = \, \ln(q)\, = \, \tau (z)$
is an order-seven non-linear ODE given by the
 sum of 602 monomial terms:
\begin{eqnarray}
\label{Schwarz4tA}
&&0 \, = \, \,\, z^6 \cdot (1-256 \cdot z)^2 \cdot 
P_\tau(z, \, \tau', \, \tau'', \, \cdots \, \tau^{(5)})
 \cdot \tau^{(7)}
 \,\,\, + \,\, \,  \cdots \nonumber  \\
&&\quad \quad \, + \,\, \, (1-256 \cdot z) \cdot
 Q_\tau(z,\,  \tau, \cdots,  \, \tau^{(6)}, \,\tau^{(7)})  
\nonumber  \\
&&
\quad \quad \quad \quad \quad   \, + \, \,
 R_\tau(z,\,  \tau,  \, \tau', \cdots,   \tau^{(5)})
 \cdot \tau',
\end{eqnarray}
where:
\begin{eqnarray}
\label{Rt}
&&R_\tau(\tau, \, \tau', \cdots,  \, \tau^{(5)}) 
\,\, \, = \, \, \,\, \, \,\,
 18\cdot  ( 9\,\tau''^{2}\, -6\,\tau^{(3)}\, \tau'
+188416\,\tau'^{2} )\cdot \tau^{(5)}
  \nonumber \\
&&\quad \quad -1043274399744\,\tau''^{2}\, \tau' \,
 +16957440\cdot  (256\,\tau'\, -\,\tau'')
\cdot \tau'\,  \cdot \tau^{(4)}\, 
\nonumber \\
&&\quad\quad   +13023313920\,\tau''^{3}\, +1909301941633024\,\tau'^{3}
+360\,(\tau^{(3)})^{3}  \nonumber \\
&& \quad\quad  +135\,\tau'\cdot (\tau^{(4)})^{2}
\quad   +4 \cdot  \Bigl(173879066624\, \tau'^{2}\,
-135\, \tau''\, \tau^{(4)}\,\nonumber \\
&& \quad \quad \quad \quad \quad 
  -4341104640\,\tau''\,\tau'
 +4239360\,\tau''^{2}\Bigr) \cdot \tau^{(3)}.
\nonumber
\end{eqnarray}
The non-linear ODE (\ref{Schwarz4tA})
 is {\em a homogeneous polynomial 
expression  of degree four in the seven derivatives 
$\, (\tau', \, \tau'', \, \cdots \, \tau^{(7)})$}.  

Rewritten in $\, q(z)$ the non-linear ODE is 
an order seven non-linear ODE given by the
 sum of  2471 monomial terms in
 $(z, \, q, \, q' \, q'' \, \cdots q^{(7)})$:
\begin{eqnarray}
\label{Schwarz4qA}
&&0 \, \, = \, \,\, \, z^6 \cdot q^6 \cdot(1-256 \cdot z)^2 \cdot 
P_q(z, \, q', \, q'', \, \cdots \, q^{(5)}) \cdot q^{(7)}
 \,\,\, + \, \,  \cdots  \nonumber \\
&&\qquad \,\, + \,\, \, (1-256 \cdot z) \cdot
 Q_q(z,\,  q, \cdots,  \, q^{(6)}, \,q^{(7)})  \\
&&\qquad \qquad \qquad 
\, + \, \, R_q(z,\,  q,  \, q', \cdots,  \,q^{(5)})
 \cdot q^6 \cdot q',
 \nonumber 
\end{eqnarray}
where $\,P_q$ and $\,Q_q$  are polynomials of
 $z, \, q', \, q'', \, \cdots \, q^{(5)}$
and $z, \, q', \, q'', \, \cdots \, q^{(7)}$ respectively,
and where  $\,R_q$ :
\begin{eqnarray}
&& R_q(z,\,  q,  \, q', \cdots,  \,q^{(5)}) \
\, \, \, =  \nonumber \\
&&\quad
18\,{q}^{4} \cdot \left( 188416\,{q}^{2}q'^{2}\,
 -6\,q'\,q^{(3)}\,{q}^{2}\,
+9\,q''^{2}{q}^{2}\, -3\,q'^{4} \right) \cdot q^{(5)}
 \nonumber \\
&& \quad  +4341104640\cdot  (q'^{2}\, {q}^{6}\, q^{(4)}\,
 +\,{q}^{4}q''\,q'^{4} \, -\,{q}^{3}q'^{6})    
\nonumber \\
&&\quad -1043274399744\,q'\,{q}^{6}q''^{2}\, 
-270\,q'^{5}\, q^{(4)}\,{q}^{3}\,
+135\,(q^{(4)})^{2}{q}^{6} q'\, 
 \nonumber \\
&&\quad +695516266496\,q'^{2}{q}^{6}q^{(3)}
+13023313920\,{q}^{6}q''^{3} \nonumber \\
&&\quad +347758133248\,q'^{5}{q}^{4}\,
 +1909301941633024\,q'^{3}{q}^{6}
\nonumber \\
&&\quad +13565952\,q'^{7}{q}^{2}\,
-17364418560\,q'\,{q}^{6}q^{(3)}\,q"\,  -36\,q'^{9} 
\nonumber \\
&&\quad +1215\,q''^{4}\, q'\, \,{q}^{4}\,
 +648\,q''^{2}\, q'^{5}\,{q}^{2}
-432\,q^{(3)}\,q'^{6}{q}^{2}\, +360\,{q^{(3)}}^{3}\, {q}^{6}\,
\nonumber \\
&&\quad +1620\,(q^{(3)}\,q''\,{q}^{3}q'^{4}\,\,-\,q''^{3}q'^{3}{q}^{3}\,
-\,q^{(3)}\,q''^{2}{q}^{4}q'^{2})\nonumber \\
&&\quad +540 \cdot (q'^{3}\, q''\,{q}^{4}\, q^{(4)} \,
 -\,q^{(3)}\,q'' \,{q}^{6}\, q^{(4)})
 \nonumber \\
&& \quad+16957440 \cdot \Bigl({q}^{6}\, q''^{2}\, q^{(3)}\,
-\,q'\,{q}^{6}\,  q^{(4)}\, q''\, \nonumber \\
&& \quad\quad \quad \quad  \quad \quad 
+\,q'^{4}{q}^{4}q^{(3)}\, -\,q'^{5}{q}^{3}q'' \,
 -\,q'^{3}{q}^{4}\, q''^{2}\Bigr)   \nonumber.
\nonumber 
\end{eqnarray}

One can verify that $\, q \, = \, Constant$ is a solution of
 (\ref{Schwarz4qA}). Furthermore
 the expansion (\ref{nome}) multiplied
by an arbitrary constant is still a solution of (\ref{Schwarz4qA}):
\begin{eqnarray}
q \, = \, \, C_0 \cdot (z +64\,{z}^{2} +7072\,{z}^{3} +991232\,{z}^{4}
+158784976\,{z}^{5} \, + \, \, \cdots) \nonumber 
\end{eqnarray}
which is natural since (\ref{Schwarz4qA}) is a linear ODE 
on derivatives of $\, \ln(q)$.
\vskip .3cm 

{\bf Remark.}
Such non-linear ODE is a ``machine'' to build series
 with {\em integer} coefficients.
For instance, if we explore the
 solutions of (\ref{Schwarz4qA}), 
of the form
$\, q \, = \, \,\, z^2 \, + \, \, \cdots\, \, \, $,  one gets: 
\begin{eqnarray}
&&q \, = \, \,\, {z}^{2}\,\,  +128\,{z}^{3}\, +18240\,{z}^{4}\,
 +2887680\,{z}^{5}\, +494460832\,{z}^{6}
 \nonumber \\
&&\quad +89757208576\,{z}^{7}+17035431116800\,{z}^{8}
+3347987811139584\,{z}^{9}
\nonumber \\
&&\quad +676624996390235600\,{z}^{10} 
+139902149755519715328\,{z}^{11}
\nonumber \\
&&\quad +29480532176870291252224\,{z}^{12}
+6312281522697932105646080\,{z}^{13}
\nonumber \\
&&\quad +1370123593804106822389706240\,{z}^{14}\,
 \nonumber \\
&&\quad 
+300913725420989219840662110208\,{z}^{15} 
\nonumber \\
&& \quad +66768780541145654061810373951488\,{z}^{16}
\,\, +\,\,   \cdots 
\end{eqnarray}
In fact, this new series is
 {\em nothing but the square of}  (\ref{nome}).
Similarly one easily verifies that the cube of (\ref{nome})
\begin{eqnarray}
&&q \, = \, \,\,
{z}^{3}\, \,+192\,{z}^{4}\,\, +33504\,{z}^{5} \,+5951488\,{z}^{6}
\nonumber \\
&&\qquad +1093928304\,{z}^{7}\, 
+207935296512\,{z}^{8} \nonumber \\
&&\qquad  +40712043902464\,{z}^{9} \, +8176029744758784\,{z}^{10}
\,\, +\,\, \cdots 
\end{eqnarray}
{\em is also a solution of} (\ref{Schwarz4qA}),
and this is also true for negative 
powers of the expansion (\ref{nome}),
for instance the inverse (in the sense of the multiplication),
\begin{eqnarray}
&&q \, = \, \,\,\, 
 \, {{1} \over {z}}\,\,\,  -64\, -2976\,z\,  
-348160\,{z}^{2}-52017616\,{z}^{3}
\, -8802913280\,{z}^{4}\nonumber \\
&& \qquad  \qquad -1608195557888\,{z}^{5} 
-309505032060928\,{z}^{6}
\, + \,\,  \cdots 
\end{eqnarray}
{\em is also a solution of} (\ref{Schwarz4qA}),
and similarly, (\ref{nome}) to the power $(-2)$ 
 \begin{eqnarray}
&&q \, = \,\,  \,{{1} \over {z^2}}\,\,  + {{128} \over {z}}\,
 \, -1856-315392\,z-50614176\,{z}^{2}-8875323392\,{z}^{3}
\nonumber \\
&& \qquad \qquad  -1658793979904\,{z}^{4}
\,\, +\,\,   \cdots 
\end{eqnarray}
{\em is also a solution of} (\ref{Schwarz4qA}).

This remarkable property, {\em expression of
 the renormalization group}\footnote[2]{We
 have an exact representation~\cite{Renorm} of
 $\, \tau \, \rightarrow \, N \cdot \tau$
 or $ q \, \rightarrow \,q^N$.},
is in fact a straight consequence of the fact that 
the non-linear ODE (\ref{Schwarz4tA}) is
 {\em a homogeneous polynomial 
 of degree four in the seven derivatives 
$\, (\tau', \, \tau", \, \cdots \, \tau^{(7)})$}.  

Conversely, in the ``mirror'', in
 the non-linear ODE (\ref{Schwarz4zA}),
one can change $\, q$ into $\, A \cdot q$
 since the derivatives are all log-derivatives of $\, q$ 
 \begin{eqnarray}
&&z \, = \, \,\,\,  A \cdot q\,\,\, 
 -64 \cdot A^2 \cdot {q}^{2}\,\,
+1120 \cdot A^3 \cdot {q}^{3}\,\,
-38912 \cdot A^4 \cdot {q}^{4}\,\nonumber \\
&& \qquad  -1536464 \cdot A^5 \cdot \,{q}^{5} \,
\,\,\, + \,\, \cdots 
\end{eqnarray}
is also solution of (\ref{Schwarz4zA}), and, 
of course, one can easily check the symmetry 
corresponding to change $\, q \, \rightarrow q^n$, 
the new mirror map series (\ref{zqn})
 \begin{eqnarray}
&&z(q^{n}) \, = \, \, {q}^{2\, n}\,\,\, -64\,{q}^{4\, n}\,\,\,
+1120\,{q}^{6\, n} \,\,\, -38912\,{q}^{8\, n} \, \,\, + \, \cdots 
\end{eqnarray}
being also solution of (\ref{Schwarz4zA}).

\vskip .8cm

\end{document}